# Université Paris XIII
# Institut Galilée

## Laboratoire de Physique des Lasers

**Habilitation à Diriger de Recherches**

Présenté par
**Athanasios Laliotis**

Sujet:

# Probing atoms and molecules close to macroscopic bodies

Soutenue le 27/06/2018

Jury compose de:

M.  **Dmitry Budker**
M.  **Jacques Robert**
M.  **Franck Pereira dos Santos**
M.  **Stefan Scheel**
M.  **Olivier Gorceix**
M.  **Daniel Bloch**

# Acknowledgements

Our group (SAI) is comprised of 4 permanent members, Daniel Bloch, Isabelle Maurin, Martial Ducloy and me. I would like to thank Daniel Bloch for being a constant comrade in the everyday battles to push the experiments forward. I also thank him for many intense scientific discussions and for bearing the 'responsibility' of the group and other scientific projects from which we have benefited a lot. I would like to thank Isabelle Maurin for interesting discussions and for doing a lot of 'tough work' in the lab that no one else in the group knows how to do. However, foremost I would like to thank her for being the glue that binds this group together. I am very thankful to Martial Ducloy for many scientific but also friendly discussions occasionally scattered all over the world (St. Petersburg, London, Novosibirsk, Shanghai, Singapore…). Although Martial is not present in the everyday activities of the group, our discussions remain extremely illuminating for me.

I do not want to forget the group's PhD students. I am grateful to Elias Moufarej, Philippe Ballin, Junior Lukusa Mudiayi, Thierry Passerat de Silans and last but not least, Joao Carlos de Aquino Carvalho (who will hopefully soon be a permanent member of staff in Brazil), for their hard work in the lab.

We also have a long list of external collaborators: I thank Horacio Failache, Arturo Lezama and the whole laser spectroscopy group for welcoming me in Montevideo and making me feel part of the team. Undoubtedly, this has been an excellent collaboration and I am glad that now, with the creation of the French-Uruguayan LIA, we get to be officially part of the same laboratory. After the Uruguayans come the Brazilians. I thank Martine Chevrollier, Pedro Chaves de Souza Segundo, Marcos Oria and Jose Rios Leite and for being a source of enthusiasm and new ideas. I also want to thank Ilya Zabkov for his hard and sometimes tedious work on simulating electromagnetic fields inside complicated structures and finally David Wilkowski for our productive discussions and collaboration and for inviting me twice in his lab.

Very special thanks should go to Paolo Pedri, Sean Tokunaga and Benoit Darquié with who I spent hours talking about physics. All three are great colleagues and friends.

I am also grateful to the last two lab-directors, Charles Desfrançois and Olivier Gorceix and the people in all the workshops, in particular Albert Kaladjian, Fabrice Wiotte and Thierry Billeton, and in the administration department, Solen Guezennec, Maryse Medina and Carole Grangier. Thanks to them, the LPL is a smoothly running machine with a nice working atmosphere.

Going back to my work at Imperial College London was probably one of the most interesting parts of the writing-up process of this manuscript. I would like to take this opportunity to thank Joe Cotter and Sam Pollock with who 'we expended a great volume of blood, sweat and tea' to make the pyramid project work. Finally, I thank Ed Hinds that supported my career with numerous recommendation letters. Ed's clarity of thought has been a great school of physics for me and I am very grateful I had the opportunity to work with him.

On a more personal note, I want to express my thanks to Elisa and Αριάδνη because they are my family, after all.

# Table of contents

# Curriculum Vitae

## Athanasios Laliotis

Nationality: Greek
Born 15/10/1975
Personal address: 53 rue Louis Blanc, 75010, Paris, France
Academic address: Université Paris13, Sorbonne Paris-Cité, Laboratoire de Physique des Lasers, 99 Avenue J. B. Clément, 93430 Villetaneuse France
Email: laliotis@univ-paris13.fr, athanasios.laliotis@univ-paris13.fr

## Current Position:

2008→….    *Associate Professor (Maître de Conferences)* at the *University of Paris13, Sorbonne Paris-Cité, Institut Galilée, Laboratoire de Physique des Lasers.*

## Academic Appointments:

2017-2018    *CNRS Research Fellow (délégation CNRS), half-time.*

2016-2017    *CNRS Research Fellow (délégation CNRS), half-time.*

2015-2016    *Half-year sabbatical leave (CRCT).*

10/14-11/14  *Invited Fellow* at the *Nanyang Technological University of Singapore.* Invitation by David Wilkowski and Nikolay Zheludev for a period of one month during which I worked in the Centre of Disruptive Photonic Technologies (CDPT) in the field of Hybrid Systems (atoms close to metamaterials).

06/12-09/12  *Invited Fellow* at the *Universidad de la Republica, IFFI Montevideo, Uruguay.* Three months in Montevideo, working in the group of Horacio Failache and Arturo Lezama. During this period, we studied the confinement of alkali atoms in random porous media.

2011-2012    *CNRS Research Fellow (délégation CNRS), half-time.*

2010-2011    *CNRS Research Fellow (délégation CNRS), half-time.*

2006-2008    *Research Associate* at *Imperial College London, UK.* Two-year post-doctoral training under the supervision of Ed Hinds at the Centre for Cold Matter. My research was focused on trapping cold atoms on silicon based chips (atom-chips).

2005-2006    *Research Associate* at the *University of Paris13.* One-year post-doctoral training funded by the European network FASTnet. I worked with Daniel Bloch and Martial Ducloy on the subject of atom-surface interactions in the nanometric range.

**EDUCATION**

2001-2005    *Philosophy Doctorate (PhD) at Imperial College London in the Department of Electrical & Electronic Engineering.*
I worked under the supervision of Eric Yeatman on the subject of integrated optics. My thesis was entitled *'Analysis and fabrication of homogeneous, high concentration erbium doped waveguide amplifiers'*.

1998-2000    *Masters of Science and Engineering at Princeton University* in the *Department of Electrical Engineering.*
My studies at Princeton focused mainly on the field of optics and optical communications as well as on the field of semiconductor physics.

1994-1998    *Bachelors in Physics* (BSc) from the *University of Athens.*
A four-year degree in general physics.

**TEACHING EXPERIENCE**

2008→….    As an Associate Professor at the *University of Paris13* I have taught a number of courses at the graduate and undergraduate level. For the Master's program on 'Photonics and Nanotechnologies', I have created the courses of Microsystems (MEMS) and Nanophotonics, which I taught for a period of 6 years (2008-2014). Since 2014, I teach the course of Photonic Properties of Matter for $1^{st}$ year graduate students. Since 2011 I am responsible for the course on Electromagnetism for $2^{nd}$ year undergraduate students (formation SPI-L2PC). I have also taught the courses of Statistical Physics ($3^{rd}$ year undergraduate level), Optics ($1^{st}$ year undergraduate level), Sensors ($3^{rd}$ year undergraduate level).

2000-2008    During my PhD at *Imperial College London*, I taught the Optical Communications Laboratory to students at the Masters level. Later on, during my post-doc I was a tutor of the course of Statistical Physics for Bachelors students.

**CO-SUPERVISION OF PhD STUDENTS**

During my years as a post-doc at Imperial College London and subsequently as an Associate Professor in the University of Paris I have actively participated in the mentoring (co-supervision) of PhD students. Here I give a list of the PhD students I have worked with, trying give a brief description of my personal contribution to their supervision.

*Samuel Pollock*, Imperial College London (thesis defence 2010), (Ed Hinds: 50%, Athanasios Laliotis: 25%, Joe Cotter: 25%)
The principal contributors to Sam's supervision were Ed Hinds (PhD supervisor), myself and Joe Cotter. During the two years I spent as a post-doc at Imperial College London I interacted with Sam on a daily basis, working very closely with him on a cold atom experiment aiming to make integrated MOT's on a chip using pyramidal structures. Fabrication of the integrated pyramids turned out to be a very important part of his thesis for which the group relied on my expertise. Thanks to a collaborative project between Paris13 and Imperial College London, I continued participating in the pyramid project even after I left London. This allowed me to continue interacting with Sam until the end of his thesis, including the writing-up stage. We have three joint peer-reviewed journal publications. Unfortunately, Sam has left academia after the end of his thesis and has been working in the private sector.
*Thierry Passerat de Silans*, University of Paris13 (thesis defence 2009).
The work for this thesis was performed partly at the University of Paris13 with Daniel Bloch as the supervisor and partly at the Federal University of Paraiba in Joao Pessoa under the supervision of Martine Chevrollier. I interacted with Thierry during the first month of his arrival in Paris13 guiding him through the selective reflection experiment on the third resonance line of Cs, whose main goal was to demonstrate the temperature dependence of the Casimir-Polder interaction in the near-field. I also interacted with him during the year 2008-2009 at the end of his thesis. I consider my contribution to his supervision to be about 5%. Isabelle Maurin and Marie-Pascale Gorza in Paris13 as well as Marcos Oria in Joao Pessoa also contributed in Thierry's supervision. We have three joint peer-reviewed journal publications. Thierry is an Assistant Professor at the Federal University of Paraiba in Joao Pessoa

*Philippe Ballin*, University of Paris13 (thesis defence 2012). (Athanasios Laliotis: 20 %, Daniel Bloch: 40%, Isabelle Maurin: 40%)

Philippe Ballin's thesis focused on the spectroscopy of atomic vapour confined inside an opal. I supervised Philippe in his attempts to put arrays of nanospheres (opals) inside a Cs vapour cell and during the building of the experimental selective reflection set-up. I also interacted with him during the writing up of his thesis, in particular his studies on the nature of vapour confinement inside a complex structure like the opal. We have one joint peer-reviewed journal article and one joint article published in conference proceedings. Philippe has not pursued an academic career and is working as a programmer in a private company.

*Elias Moufarej*, University of Paris13 (thesis defence 2014). (Athanasios Laliotis: 45%, Daniel Bloch: 45%, Isabelle Maurin: 10%).

Elias's thesis was also in the subject of vapour confinement in photonic crystals fabricated with soft chemistry methods. Elias was my student during his undergraduate and graduate studies and he also did his Master's thesis under my supervision. I was very strongly involved in his supervision at all levels, experiment, theory and writing-up. We have jointly published 2 peer-reviewed journal papers and we also have a publication in conference proceedings. Elias is still doing research in atomic and molecular physics as a post-doc.

*Joao Carlos de Aquino Carvalho* University of Paris13 (Athanasios Laliotis: 40%, Daniel Bloch: 40%, Isabelle Maurin: 20%).

Joao started his thesis in September 2014 on the subject of the effects of near field thermal emission on atomic level shifts and radiative properties. His work is a direct continuation of my work on Casimir-Polder interaction in the presence of thermally excited surface modes. I guide him daily in his experimental work but I also try to mentor him through his efforts to learn Casimir-Polder theory. Joao has a four-year funding of the Brazilian government and is expected to finish his thesis in July 2018.

*Junior Lukusa Mudiayi* University of Paris13.

Junior joined our group as a PhD student in September 2016 working on the subject of selective reflection on rovibrational molecular transitions. As I am the principal investigator in this relatively new research topic, I participate very actively in his supervision (~50-60%). I also supervised Junior's Master's thesis, bibliography project and I taught him courses at the graduate and undergraduate level. Daniel Bloch is Junior's PhD supervisor and interacts with him regularly. Sean Tokunaga, Isabelle Maurin and Benoit Darquie also work and interact with Junior.

*Santiago Villalba*, Engineering Faculty of Montevideo (thesis defence 2014).

Santiago's thesis was on the spectroscopy of Rb atoms confined inside the pores of random, porous media. I worked with him while I was an invited fellow at Montevideo for a period of 2,5 months and during various shorter visits. Although we have only interacted for a short period, the work we produced while I was a visitor at Montevideo is a very important part of Santiago's thesis. I estimate my contribution to his supervision to about 5-10%, the rest being probably split between his PhD supervisors Horacio Failache and Arturo Lezama. We have 2 joint peer-review publications. Santiago is a researcher at the University of Montevideo, although his position is not yet permanent.

*Ilya Zabkov,* Lebedev Institut (PhD student of Vassily Klimov).

Me and Daniel Bloch have also co-supervised Ilya Zabkov for the period of two months (October 2012 and October 2013) that he spent in the University of Paris13 as a visiting student. During this time, we guided Ilya on his efforts to simulate the electric field inside an opal of nanospheres. The reflection and transmission spectra of opals obtained by Ilya's simulation were also compared to experiments performed by Elias Moufarej and myself. We have one joint peer-reviewed journal publication. Ilya obtained his PhD degree in 2017 in Moscow.

## OTHER PROFESSIONAL EXPERIENCE

2001-2003 **"Assistant Editor" of the "International Journal of Electronics"**
- Participation in the selection process of the articles.

2000-2001 **Demokritos Research Institute, Athens**
**Research Assistant**
- Participation in the development of an e-learning network.

## AWARDS-PRIZES

- ECOS-SUD project: Collaboration with IFFI at Montevideo, Uruguay.
- Invited Fellow at NTU, Singapore, October-November 2014.
- CREI Funding: Collaboration with IFFI at Montevideo, Uruguay.
- Invited Fellow at IFFI Montevideo Uruguay, June-August 2012.
- Royal Society (CNRS Royal Society joint projects, 2010): Collaboration with Imperial College London.
- European funding FASTnet: Research Assistant 2005.
- The Rank Prize Funds 2005
- Nortel Networks Funding: PhD 2001-2004.

## LANGUAGES

**Languages** Greek, English, French

## PEER REVIEWED PUBLICATIONS

1. J.C. De Aquino Carvalho, P. Pedri, M. Ducloy, **A. Laliotis**, 'Retardation effects in spectroscopic measurements of the Casimir-Polder interaction', *Physical Review A*, **97**, 023806 (2018).
2. E. A. Chang, S. Abdullah Aljunid, G. Adamo, **A. Laliotis**, M. Ducloy, D. Wilkowski 'Tuning Casimir-Polder interactions in atom-metamaterial hybrid devices', *Science Advances*. **4**, eaao4223 (2018).
3. J.C. De Aquino Carvalho, **A. Laliotis**, M. Chevrollier, M. Oria, D. Bloch 'Sub-Doppler Fluorescence from an optically thick atomic vapour', *Physical Review A*, **96**, 034405 (2017).
4. **A. Laliotis** and M. Ducloy, 'Casimir Polder interaction with thermally excited surfaces' *Physical Review A*, 91, 052506, (2015).
5. I. Maurin, E. Moufarej, **A. Laliotis**, D.Bloch, 'Infiltrating a thin or single layer opal with an atomic vapour: Sun-Doppler signals and crystal optics', *JOSAB*, 32, 1761-1772 (2015).
6. E. Moufarej, I. Maurin, I. Zabkov, **A. Laliotis**, P. Ballin, V. Klimov, D. Bloch, 'Infiltrating a thin or single layer opal with an atomic vapour: Sub-Doppler signals and crystal optics', *European Physics Letters*, 108, 17008 (2014).
7. **A. Laliotis**, T. Passerat de Silans, I. Maurin, M. Ducloy, D. Bloch, 'Casimir-Polder interactions in the presence of thermally excited surface modes', *Nature Communications* 5, 5364, (2014).
8. T. Passerat de Silans, **A. Laliotis**, I. Maurin, M-P. Gorza, P. Chaves de Souza Segundo, M. Ducloy, D. Bloch, 'Experimental observations of thermal effects in the near field regime of the Casimir-Polder interaction', *Lasers Physics* 24, 074009 (2014).

## CONFERENCE PROCEEDINGS

8. **A. Laliotis,** E. M. Yeatman 'Compact planar waveguide amplifiers using sol-gel glass on silicon', IEE 1$^{st}$ International Conference on Photonic Access Technologies, pp.4/1-4/5, December 2002.

## INVITED CONFERENCES (invited talks delivered by myself are highlighted)

1. **A. Laliotis**, J. C. de Aquino Carvalho, I. Maurin, M. Ducloy, D. Bloch, 'Spectroscopic measurements of the Casimir-Polder interaction with atoms and molecules', Casimir and Van der Waals physics: Progress and Prospects, Hong Kong SAR China (2016). (presented by A. Laliotis)
2. S. Villalba, L. Lenci, *A. Laliotis*, A. Lezama and H. Failache 'Nonlinear atomic spectroscopy in a random porous medium', Latin America Optics and Photonics Conference, Medellin, Colombie, (2016). (presented by H. Failache)
3. **A. Laliotis**, J. C. de Aquino Carvalho, P. Chaves de Souza Segoundo, T. Passerat de Silans, I. Maurin, M. Ducloy, D. Bloch S. Tokunaga, B. Darquié, 'Spectroscopic measurements of the Casimir-Polder interaction with atoms or molecules', International Workshop on Quantum Manipulation of Atoms and Photons, Shanghai, October 2015. (presented by A. Laliotis)
4. **A. Laliotis**, T. Passerat de Silans, J. C. de Aquino Carvalho, P. Chaves de Souza Segoundo, I. Maurin, J. R. Rios Leite, M. Ducloy, D. Bloch, 'Effets de la température sur l'interaction Casimir-Polder en champs proche', COLOQ 14, Rennes July 2015. (presented by A. Laliotis)
5. *A. Laliotis*, J. C.de Aquino Carvalho, T. Passerat De Silans, P. Chaves de Souza Segundo, I. Maurin, M. Ducloy, D. Bloch, 'Atom in front of a hot surface: Thermal dependence of the Casimir-Polder interaction', EGAS 47 (47th European Group for Atomic Spectroscopy), Riga (2015). (presented by D. Bloch)
6. **A. Laliotis**, T. Passerat de Silans, I. Maurin, M. Ducloy, D. Bloch, 'Casimir-Polder forces in the presence of surface polariton modes', PSAS' 2014 Precision Physics of Simple Atomic Systems, Rio de Janeiro, May 2014. (presented by A. Laliotis)
7. *A. Laliotis*, T. Passerat De Silans, I. Maurin, M. Ducloy, D. Bloch, 'Casimir-Polder in the Near-Field van der Waals regime Experimental Observation of Temperature effects for Cs*/sapphire',Casimir Physics, Les Houches, France, (2014). (presented by D. Bloch)
8. I. Maurin, E. Moufarej P. Ballin, *A. Laliotis*, D. Bloch, 'Infiltrating an artificial opal with an atomic vapour: observation of sub-Doppler signals in linear spectroscopy', META14 (5th International Conference on Metamaterials, Photonic Crystals and Plasmonics), Singapore, Singapore, (2014). (presented by D. Bloch)
9. **A. Laliotis**, T. Passerat de Silans, I. Maurin, M-P. Gorza, M. Ducloy, D. Bloch, 'Experimental Observations of temperature effects in the near-field regime of the Casimir-Polder interaction', MPLP Laser Physics International Symposium, Novosibirsk 2013. (presented by A. Laliotis)
10. *A. Laliotis*, I. Maurin, E. Moufarej, M. Ducloy, D. Bloch, 'Optical probing of atoms with a subwavelength confinement', QMAP 2013 , 3rd workshop of the Sino-French Research Network on "Quantum Manipulation of Atoms and Photons" (GDRI QMAP), Palaiseau, France, (2013). (presented by D. Bloch)
11. D. Bloch, I. Maurin, *A. Laliotis*, P. Ballin, E. Moufarej, 'Transitions optiques dans des vapeurs confinées de façon mésoscopique', J3N Journée nationale nanosciences et nanotechnologies,Bordeaux, France, (2012). (presented by D. Bloch)
12. P. Ballin, E. Moufarej, I. Maurin, *A. Laliotis*, D. Bloch, 'Sub-Doppler optical resolution by confining a vapour in a nanostructure', XVII IQSE (seventeenth international school on quantum electronics: laser physics and applications), Nessebar, Bulgarie, (2012). (presented by D. Bloch)
13. P. Ballin, I. Maurin, *A. Laliotis*, E. Moufarej, M-P Gorza, M. Ducloy, D. Bloch, 'Laser spectroscopy of atoms with a sub-wavelength confinement',Second France Japan workshop on Nanophotonics,Toba, Japon, (2011). (presented by D. Bloch)
14. D. Bloch, P. Ballin, I. Maurin, *A. Laliotis*, M-P Gorza, M. Ducloy ,'Atomic spectroscopy and nano-optics',France-Japan workhop on nanophotonics, Villetaneuse, France, (2010). (presented by D. Bloch)
15. **A. Laliotis**, T. Passerat de Silans, I. Maurin, M. Ducloy, D. Bloch, 'Temperature dependence of vdW forces between an atom and a dispersive surface' Laser Physics 2010, Ashtarak,

Armenia. (presented by A. Laliotis)

16. I. Maurin, **A. Laliotis**, T. Passerat De Silans, G.Dutier, M.-P. Gorza, D.Bloch, M. Ducloy, D. Sarkisyan, " Vapour spectroscopy close to an interface: confined optical frequency references ", RF-YS2008 (Russian French Young Scientist symposium, St. Petersburg, 22-27 sept 2008. (presented by A. Laliotis)

17. *A. Laliotis*, I. Maurin, P. Todorov, I. Hamdi, G. Dutier, A. Yarovitski, S. Saltiel, M-P Gorza, M. Fichet, M. Ducloy, D. Bloch, 'Testing the distance-dependence of the van der Waals interaction between an atom and a surface through spectroscopy in a vapour nanocell', XIV international School on Quantum Electronics and Lasers Physics, Sunny Beach, Bulgaria, (2006). (presented by D. Bloch)

18. **A. Laliotis**, 'Erbium Doped Waveguide Amplifiers', The Rank Prize Funds, 2005 Symposium. (presented by A. Laliotis)

## CONFERENCES (National and international)

1. J. Lukusa Mudiayi, B. Darquié, S. Tokunaga, P. Chaves de Souza Segundo, I Maurin, J. R. Rios Leite, D. Bloch, **A. Laliotis**, 'High-resolution linear spectroscopy on a micrometric layer of molecular vapor', EQEC and CLEO' Europe, Munich, Allemagne, (2017).(**oral**)

2. J. Lukusa Mudiayi, J. C. de Aquino Carvalho, B. Darquié, I. Maurin, S. Tokunaga, P. Chaves de Souza Segundo,  J. R. Rios Leite, M. Ducloy, D. Bloch, **A. Laliotis**, 'Towards miniaturized molecular spectroscopy and measurements of the Casimir-Polder interaction', Kick-off meeting du DIM SIRT, Palaiseau, France, (2017).(**poster**)

3. J. C. de Aquino Carvalho, **A. Laliotis**, I. Maurin, D. De Sousa Meneses, P. Echegut, M. Ducloy, D. Bloch, 'Atom as a short range probe of thermally excited surface modes', C'nano 2017,Lyon, France, (2017).

4. J. C. de Aquino Carvalho, **A. Laliotis**, P. Chaves de Souza Segundo, I. Maurin, D. De Sousa Meneses, P. Echegut, M. Ducloy, D. Bloch, 'Atom probing of thermally excited surface polaritons', EQEC and CLEO' Europe, Munich, Allemagne, (2017).(**poster**)

5. J. C. de Aquino Carvalho, **A. Laliotis**, P. Chaves de Souza Segundo, I. Maurin, M. Ducloy, D. Bloch, 'Atom probing of thermally populated surface polaritons', EGAS49 (49th Conference of the European Group on Atomic Spectroscopy), Durham, UK, (2017).(**poster**)

6. J. C. de Aquino Carvalho, **A. Laliotis**, M. Chevrollier, M. Oria, D. Bloch, 'Sub-Doppler features in the backward-emitted fluorescence of a dense vapor and analogies with thin-cell spectroscopy', EGAS49 (49th Conference of the European Group on Atomic Spectroscopy), Durham, UK, (2017).(**poster**)

7. J. C. de Aquino Carvalho, **A. Laliotis**, P. Chaves de Souza Segundo, I. Maurin, M. Ducloy, D. Bloch, 'Atom probing of thermally populated surface polaritons', ICOLS 2017 (International Conference on laser Spectroscopy), Arcachon, France, (2017).(**poster**)

8. J. C. de Aquino Carvalho, **A. Laliotis**, P. Chaves de Souza Segundo, I. Maurin, D. De Sousa Meneses, P. Echegut, M. Ducloy, D. Bloch, 'Direct energy transfer from thermally excited polaritons to atoms: A quantum analogue to near field heat transfer', NANOMETA2017 (6th International Topical Meeting on  Nanophotonics and Metamaterials), Seefeld, Austria, (2017) (**poster**).

9. I. Maurin, E. Moufarej, **A. Laliotis**, I. Zabkov, V. Klimov, D. Bloch, 'Réseau bidimensionnel ou film mince organisé de nanosphères : expériences et modélisation', Nouveaux systèmes périodiques / Instrumentation diffraction, Toulouse, France, (2016) (**oral**).

10. J. C. de Aquino Carvalho, **A. Laliotis**, P. Chaves de Souza Segundo, I. Maurin, D. De Sousa Meneses, P. Echegut, M. Ducloy, D. Bloch, ' Sonder un mode polaritonique de surface avec l'interaction Casimir-Polder', Journée "Photonique de précision", Paris, France, (2016) (**poster**).

11. J. C. de Aquino Carvalho, **A. Laliotis**, T. Passerat De Silans, M. Chevrollier, M. Oria, D. Bloch D., 'Sub-Doppler retrofluorescence from an optically thick atomic vapor', Latin America Optics and Photonics Conference, Medellin, Colombie, (2016) (**poster**).

12. J. C. de Aquino Carvalho, **A. Laliotis**, M. Chevrollier, M. Oria, D. Bloch, 'Contribution sub-Doppler en rétrofluorescence pour une vapeur atomique optiquement épaisse', COLOQ15 - Optique Bordeaux 2016, Talence, France, (2016) (**poster**).

# Chapter 1: Introduction

The present document describes my research after graduating with a PhD degree from the Electrical Engineering Department of Imperial College London at 2005. After my graduation, I spent one year as a post-doc in Paris13 (in the group SAI) before moving to the Physics Department of Imperial College London for another post-doctoral position in the group of Ed Hinds (2006-2008). Since 2008, I am an Associate Professor in Paris13 and part of the SAI group.

I start the document by describing my work on the subject of cold atom trapping on silicon fabricated microchips in the group of Ed Hinds (Chapter 2: Integrated Magneto Optical Traps on silicon chips). I then describe my works here in Paris13, as well as my collaborations with the group of atomic spectroscopy in the Engineering Faculty of Montevideo, Uruguay and with the group of David Wilkowski in the Nanyang Technical University (NTU) of Singapore. Chapter 3 (Spectroscopy of three dimensionally confined atomic vapour) describes spectroscopic experiments on atomic vapours under confinement, either in opals (Paris13 project) or in porous media (collaboration with Montevideo). The experiments of atoms confined in opals revealed the signature of a sub-Doppler structure in linear spectroscopy whose origins are still to be explained. Chapter 4 (Casimir-Polder interactions at finite temperature: theory) and Chapter 5 (Casimir-Polder interactions at finite temperature: the experiments) describe works towards understanding and experimentally demonstrating the near field temperature dependence of the Casimir-Polder interaction in thermal equilibrium due to thermal excitation of surface-polariton modes. Chapter 4 focuses on a theoretical description of temperature effects, described as the interaction of an atom with near field thermal emission. Chapter 5 details the spectroscopic experiments that eventually demonstrated near field thermal effects in the Casimir-Polder interaction, trying not to neglect experimental details, or experimental failures that do not always appear in publications. Chapter 6 (Beyond the van der Waals approximation: retardation effects and metasurfaces) examines the effects of retardation on spectroscopic experiments of the Casimir-Polder interaction. Chapter 6 also briefly explains experimental attempts to go beyond flat surfaces (collaboration with NTU) and eventually tune the Casimir-Polder interaction by engineering the polariton excitation of metasurfaces. Finally, Chapter 7 (Beyond atoms: selective reflection on a molecular vapour) is focused on a relatively new project that aims at probing micrometric layers of molecular gases with selective reflection and thin cell spectroscopy. Experiments with molecular gases close to surfaces pave the way towards fabricating compact high-resolution molecular frequency references and towards performing measurements of the Casimir-Polder interaction with molecules.

The chapters might appear disjoint. This is because I have tried to some extent to diversify my works as much as possible, and pursue different projects. There is, however, an Ariadne's thread that makes it easier for the reader to walk through the text. The underlying theme is the study of the behaviour of quantum objects (atoms or molecules) close to macroscopic (classical) bodies (surfaces). This englobes miniaturisation of quantum systems using technologies such as atom-chips or hybrid systems with the aim of producing a new generation of quantum devices, as well as more fundamental issues such as the spectroscopic effects of confinement and the nature of vacuum and thermal fluctuations and their effects on quantum matter. This theme also stands at the interface between 'clean' QED quantum physics (where sometimes theory elegantly meets the experiments) and the much more difficult area of surface physics were things could still be almost empirical, calculations are difficult or cumbersome and experimental conditions are hard to control.

The subject that is most dear to me and that I spent most time thinking about is the Casimir-Polder interaction. In this perspective, the experimental confirmation of temperature effects in the near field of the Casimir-Polder interaction described in Chapter 5 and the physical interpretation of the phenomenon described in Chapter 4 are central in this text. Furthermore, the demonstration of near field energy transfer from a thermally excited polariton to an atom and the pursuit of a spectroscopic experimental measurement of the Casimir-Polder molecule-surface interaction emerge as dominant future projects.

# Chapter 2: Integrated Magneto Optical Traps on silicon chips

I joined the Centre for Cold Matter at Imperial College London in October 2006, working as a research fellow with Ed Hinds. At the time, half of the group was working on cold atom experiments integrated on atom-chips. The possibility of combining the physics of cold atoms with the technology of microfabrication was a fascinating prospect offering not only the opportunity to reduce the voluminous size of cold atom experiments (lab on a chip) but also to explore new fundamental physics. Integration and miniaturisation technologies offered for example a way to reduce the size of optical cavities, which provided an alternative way to achieve strong atom-cavity coupling and fabricating integrated quantum information devices [1]. In addition, placing atoms at close proximity to current wires offered a way to create magnetic traps with steep gradients and eventually explore BEC interferometry [2] and the physics of low dimensional quantum gases. These perspective gains made atom-chips very attractive and many research groups devoted significant efforts on overcoming the technical difficulties that came in tandem with the advantages of microfabrication.

Loading the atoms on the chip is an extremely crucial process in atom chip experiments, but also more generally for every experiment involving atomic microtraps. In the case of atom chips the standard protocol for loading atoms on the chip was introduced by T.W Hansch and colleagues [3]. Experiments would start by creating a surface-MOT (2 counter-propagating beams reflected on the surface of the chip and 2 counter-propagating beams parallel to the surface of the chip), using several independent laser beams and a set of external coils. The atoms would then be transferred to a MOT that used chip coils/wires (U-shaped wires), and eventually to a magnetic trap (usually Z-shaped wire), after which they could be dispatched to desired locations on the chip. The idea was successful mostly due to good overlap between the magnetic field of the U-shaped and Z-shaped wires that facilitates the transfer of the atomic cloud between successive traps. Despite its success, the process was inconvenient, complicated and somewhat incompatible with the lab on a chip idea. Trapping atoms directly on the chip (here directly refers to the direct use of magnetic fields generated on the chip), in a simple and reliable way, was at the time, a research goal of several groups working on atom-chips.

In the group of J. Schmiedmayer the proposed scheme for direct trapping on the chip involved the use of especially designed U-wires [4]. The magnetic fields generated in these experiments are a much better approximation of a real quadrupole field. As such, a surface MOT can be made without the use of external coils allowing on-chip trapping directly from a room temperature vapour. The wires used in these experiments are not microfabricated. They are macroscopic wires placed in proximity to the chip. Additionally, the set-up requires several laser beams in order to slow the atoms down. T. Pfau and colleagues proposed a scheme where a grid of macroscopic wires is placed underneath a reflective surface [5]. By tuning the direction and strength of the current at each wire, one can create MOTs at different locations of the chip surface. Here again the atoms are slowed down by 4 laser beams in the surface MOT geometry. In a different experiment, a transparent (high transmission) chip with permanent magnetisation was used to trap atoms [6]. The chip can be placed inside a conventional 6 laser-beam MOT geometry (one pair of counter-propagating beams at each axis). Using the permanent magnetisation and an external bias magnetic field, MOT's can be created simultaneously at many parts of the chip. This is an elegant and practical idea although it would probably be hard to integrate this device with other atom-chip functions such as wires or optical components like waveguides.

In the group of Ed Hinds we used pyramid hollows fabricated on a silicon wafer to directly trap atoms on the surface of a chip. The pyramid MOT is a simple way to magneto-optically trap atoms [7]. The basic principle consists of illuminating a macroscopic pyramid of highly reflective surfaces and an apex angle of 90° with a single circularly polarised laser light, while using external coils to create a magnetic quadrupole field whose zero is at the centre of the pyramid. Multiple reflections of the laser light on the pyramid walls automatically create a set of beams whose polarisation is appropriate for atom-trapping. The scheme was implemented with macroscopic structures in order to trap large atom numbers with relatively small laser power [7].

Micro-hollows of pyramidal shape can be fabricated on silicon chips by anisotropic wet etching with a KOH solution, which primarily etches the <100> planes of silicon while exposing the <111> planes, thus creating pyramidal hollows with an apex angle of 70,1°, defined by crystallography. The possibility of operating 70,1° pyramids as magneto-optical traps was investigated in [8]. It was shown that successful MOT operation is not possible when the pyramids are covered with a high reflectivity coating such as gold (more than 95% reflectivity for the relevant wavelengths), due to the fact that beams that are incident close to the pyramid corners undergo three, instead of two reflections, before exiting the pyramid. This changes the polarisation state of the beams and impedes the trapping of cold atoms in the pyramid centre. In the same article [8] it was also shown that 70,1° pyramids can work as MOTs if the gold coating is removed close to the corners of the pyramid (a so-called flower design) or alternatively by coating the pyramid with a metal, such as aluminium, whose reflectivity is lower, (~70% at 780nm), therefore reducing the relative importance of the beams that undergo three, instead of two reflections inside the pyramid. Using these techniques the group created a MOT of ~$10^6$ Rb atoms [8] inside a macroscopic 70,1° pyramid whose opening was ~16.3x16.3mm$^2$ (the atom number reported in [8] was actually overestimated).

Following the demonstration of successful MOT operation in 70,1° pyramids the group embarked on the ambitious goal of fabricating micro-MOT's on silicon chips, in collaboration with the group of Michael Kraft of the University of Southampton. The chip designed for this experiment was indeed a marvellous example of MEMS technology. In Fig.2.1 we show a microscope picture of the chip taken after fabrication. It contains multiple rows of pyramid hollows of different opening sizes ranging from 200um up to 1.2 mm. The opening size of the pyramid is limited by the thickness of the chip, in this case 1mm, allowing for a maximum opening of 1.3mm. The pyramids are surrounded by current carrying wires whose thickness depends on the pyramid size. The magnetic field of the wires in addition to an external bias field can create a magnetic quadrupole inside the pyramid hollows. The chip is mounted on a package with pins that serve as electrical contacts to the outside world (see fig.1 caption for further details). Eventually the entire structure is put in a vacuum chamber with an electrical feedthrough that allows driving current in the chip wires.

It is worth outlining the basic fabrication steps, described in more detail in [9]. The process starts by PECVD deposition of SiN on a silicon wafer, which is then patterned with square openings. SiN serves as a mask for the KOH etching process that creates the pyramid hollows inside the chip. After pyramid fabrication, the chip is sputter-coated with gold layer of few tens of nanometres and the gold is patterned using a wet etching step to create the flower patterns inside the pyramids. Subsequently the base of the wires (wire tracks) is patterned on the same gold layer. Finally, the wires are electroplated to their full thickness of about 3µm. This is a delicate multistep process that requires the use of several photolithographic masks. The chip was fabricated in the University of Southampton while characterisation was performed at Imperial College London.

Miniaturisation and integration of pyramidal hollows is also associated with a reduction of the available trapping volume and the capture velocity, and subsequently reduction of the number of atoms in the magneto-optical trap. Additionally, the proximity of atoms to hard room-temperature walls can also cause an additional atom loss mechanism in the MOT [10]. In reference [9] we give a crude estimate of the scaling law governing the number of atoms in a pyramid MOT as a function of its opening size, based on numerical simulations first introduced by C. Wieman and colleagues [11]. Our calculation showed that the number of atoms N scales down as ~$L^{3.6}$, where L is the pyramid opening. This gave an estimate of N=6000 trapped atoms inside the 1mm pyramid. A major drawback of the model reported in [11] is that it neglects saturation effects when calculating the scattering rate of the atom inside the multiple beam MOT. This approximation can be reasonable for large MOTs with large Doppler detuning but during the course of our experiment, we found that it breaks down much faster than expected in the case of micro-MOTs.

The pyramid chip, shown in Fig 2.1 was put inside a vacuum chamber with a base pressure of $10^{-9}$-$10^{-8}$ mbarr and Rb atoms were supplied by controlling the current of a dispenser. Repeated efforts were made to detect a MOT inside the pyramids using fluorescence imaging but proved unsuccessful. Additionally, it became evident that the minimum detectable atom number in a pyramid micro-MOT was severely limited by the strong scattering of laser light from the pyramid faces.

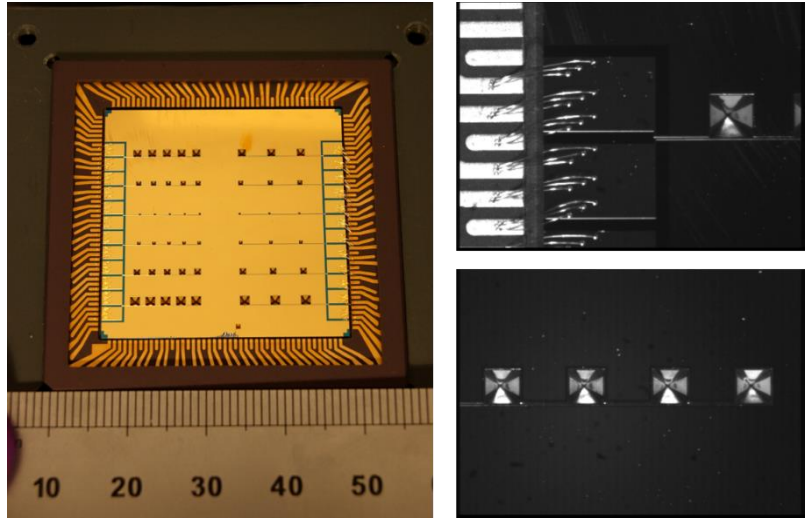

**Fig 2.1** Photographs of the pyramid chip mounted on its package. The chip contains rows of pyramids of 200, 400, 600, 800, 1000 and 1200 μm surrounded by wires of 25 μm or 50 μm (for pyramids larger than 800 μm). The integrated wires end in large pads linked to the electrical contacts (pins) of the package by a set of bonded wires. The pyramids and most of the chip's surface is coated with gold. The gold is removed close to the corners of each pyramid, making a flower-design that allows the pyramid to operate as MOTs.

## Atom number scaling as a function of pyramid size

In order to test the scaling of the number of atoms, we went back to the macroscopic glass pyramid [8] and measured the number of atoms as a function of pyramid size using a circular aperture (for alignment issues) placed outside the chamber. This reduces the laser beam size as well as the effective pyramid opening. We performed the experiments with a gold-coated pyramid with a flower-design, an aluminium coated pyramid and an aluminium coated pyramid with a flower design. All three coatings showed very similar behaviour. The results for an aluminium coated pyramids are shown in Fig 2.2 (from ref. [10]). We explore an effective pyramid size from ~9mm to about 4mm, below which measurements became exceedingly difficult as the number of trapped atoms became sensitive to imperfections of the macroscopic pyramid in particular close to the apex. Inspection of Fig. 2.2 reveals that for small pyramid sizes (until 6-7mm) the atom number scales as $L^6$, where L is the square root of the pyramid opening and that the $L^{3.6}$ scaling, suggested in [11] is only valid for larger pyramid openings. One can extrapolate, using the data of Fig 2.2 that L=1mm wide pyramids would trap approximately one atom. Despite our elaborate imaging techniques, which will be briefly explained in the following section, this number remained off detection limits.

Our scaling law data find a rather simple physical interpretation, outlined here without going into extreme detail. The number of atoms in a MOT is a balance between the loading rate of atoms ,R, in the trap and the rate with which atoms are lost ($1/\tau$), governed primarily by collisions with energetic Rb atoms of the background vapour. The loading rate depends on the flux of atoms, whose velocity is below the capture velocity $u_C$ (velocity below which, atoms can be efficiently slowed down and trapped) upon the surface that delimits the MOT trapping volume (depending on geometry). In the case of pyramids $R \sim L^2 u_C^4$ and the steady state, number of atoms in the MOT is $N_\infty \sim \frac{L^2 u_C^4}{\tau}$ [10]. In order to find the exact capture velocity one has to resort to numerical simulations of atom movement inside the optical molasses but as a first approximation the friction force exerted on the atoms is given by the difference of the scattering rates between the two beams of opposing directions, multiplied by the photon recoil momentum $\hbar k$.

$$F = \hbar k \frac{\Gamma}{2} \left[ \frac{s}{1 + s + 4\left(\frac{\delta - ku}{\Gamma}\right)} - \frac{s}{1 + s + 4\left(\frac{\delta + ku}{\Gamma}\right)} \right] \quad (2.1)$$

Here $\Gamma$ is the natural linewidth, s the saturation parameter, k the wavenumber and u the atomic velocity. For simplicity, the distance dependent detuning due to the magnetic field has been ignored here. More details can be found in refs [12-14] as well as many atomic physics textbooks. A plot of the friction force as a function of velocity can be seen in the inset of Fig.2.2. For small velocities ($|u| \leq \frac{\delta}{k}$) the force is linear with velocity ($F = -\alpha u$) which implies that the capture velocity (the ability of the molasses to slow atoms down to rest), is proportional to the trap size $u_C \sim L$ . The above arguments lead directly to the scaling law $N_\infty \sim L^6$. As the trap size increases, the MOT becomes less efficient, since the slowing force no longer increases linearly with velocity. In this regime, the scaling law is given by $N_\infty \sim L^{3.6}$. For large MOTs (typically above ~1cm), one also has to account for more complex phenomena such as collisions between trapped atoms as well as radiation trapping inside the MOT [12].

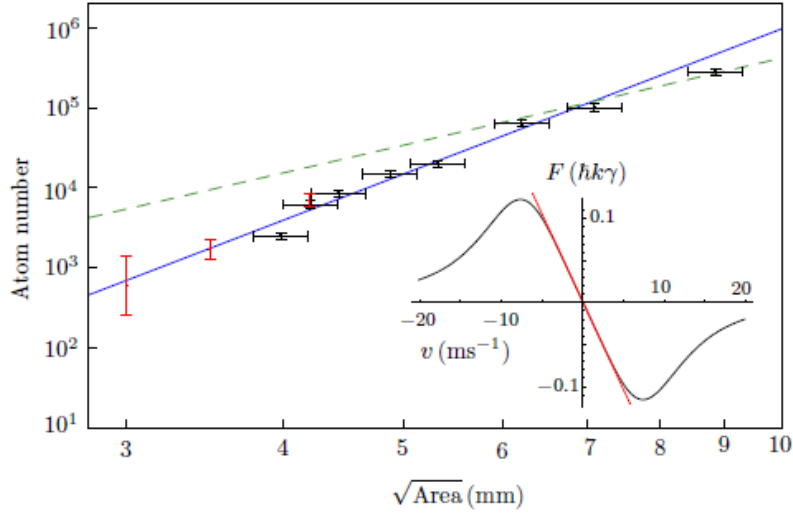

**Fig 2.2** (from ref.[10]) Number of atoms as a function of the square root of the pyramid opening area (L) for a macroscopic pyramid with a circular aperture (black points) and for the integrated pyramids of refs [14] (red points). The scaling law $L^6$ and $L^{3.6}$ is represented with a blue solid and green dashed lines respectively. For pyramid openings smaller than 6-7mm the atom number clearly follows the $L^6$ power low. The inset shows a plot of the friction force as given by equation (2.1) as a function of velocity for a detuning $\Delta$=-2$\Gamma$.

In our experiments, we use typically a detuning of $\delta \sim -2\Gamma = -2\pi \, 10MHz$ whereas the wavelength of the D2 Rb is 780nm. This means that the slowing force is linear for velocities smaller than $\frac{\delta}{k} \sim 7.8$m/sec. In order to estimate experimentally the capture velocity in our pyramids, we performed extensive measurements of the number of atoms in the MOT as a function of time for different values of Rb background pressure. This allows us to measure the capture and loss rates and as such estimate the capture velocity of the MOT, which is found to be $u_C \sim 4.6$m/sec for a MOT of 4.2mm, and ~7.8m/sec for a pyramid with an opening of L~7mm. As such, for pyramid openings larger than L~7mm, the capture velocity is no longer expected to linearly increase with size and the number of atoms in the MOT should scale as $N_\infty \sim L^{3.6}$. Although the argument followed here puts many complicated MOT physics under the carpet, it adequately explains the experimental observations. For comparison we mention here that simulations performed using the model described in [11] by C. Wieman and colleagues estimate that the capture velocity is 7.8m/sec for pyramid sizes smaller than 1mm. As previously mentioned the model fails to take into account the combined saturation effects of all the interacting beams and therefore overestimates the friction force on the atoms. More elaborate simulations, analysed in Sam Pollock's thesis, that take into account the combined saturation of all beams are in much better agreement with the experimental results.

Finally, we should mention that the loss of atoms due to collisions with the walls was also investigated in the course of this project. We found that these are negligible when the MOT is more than a few hundred microns away from the pyramid sides. This loss mechanism would be relevant only for pyramids smaller than 1mm, which as discussed above cannot trap any atoms, anyway.

## Imaging the MOT

We detected the MOT fluorescence using a camera with an attached lens that was placed outside the vacuum chamber. This limits the numerical aperture of the system but makes it significantly more flexible and easy to use. In most experiments, the MOT is in proximity to the chip's reflective surface that scatters light from the incident MOT beam(s) due to imperfections or surface roughness. Scattered photons create a background signal, which is the main source of noise in our detection. In order to increase the signal to noise ratio of our measurements we modulate the number of atoms in the MOT (switch the MOT on and off) while keeping the background light as constant as possible. The background is then removed by image subtraction. Since the images are shot-noise limited we improve the MOT visibility by averaging many images. This is the equivalent of a synchronous detection implemented with a CCD camera, whose frame repletion rate was ~400Hz (depending on the exact operation conditions). Additionally we used a spatial filtering, which consisted of convolving the detected images with a Gaussian profile of radius similar to that of the MOT (cross-correlated the raw image with the expected image of the MOT). This 'smoothing' technique further increased the SNR of the images. It's worth mentioning that in our experiments the MOT size was limited by temperature [12,13] and was about 100μm radius, independent of the number of atoms.

To modulate the MOT fluorescence we attempted to modulate the frequency of either the MOT or the repump beam, but we found that this significantly destabilises the MOT operation. Alternatively, we tried displacing the MOT using an external bias field. We were able to shift the MOT's position at frequencies ~100Hz and eventually achieved sensitivity to approximately ~100 atoms. Unfortunately, this technique was only applicable to big pyramids. For smaller pyramids, we just applied a magnetic field bias that did not allow MOT operation inside the pyramid hollow. As such, the MOT on/off sequence was much slower (a few seconds), dominated essentially by the MOT loading time. Nevertheless, the atom number sensitivity remained typically as low as 100 atoms.

## Integrated pyramid MOTs

Our experiments with the macroscopic pyramid made abundantly clear that the only way forward was to integrate bigger pyramid hollows on our atom chips. To do this we used 3mm thick silicon wafers, that allowed fabrication of 4.2 mm wide pyramids. Although the size of the pyramids is macroscopic, they were entirely fabricated using MEMS technology and in principle they can be integrated with most (if not all) atom chip components (wires, waveguides, fibres…). The first of the new generation chips was fabricated by me at the University of Southampton. Our design included simply the fabrication of pyramids, without integrating the current carrying wires. The anisotropic KOH etch lasted approximately 2 days (50 hours). During its course, many things went wrong, but finally, after a week's work in unremarkable Southampton, I made it back to Imperial College London with more than a few intact pyramid hollows (Fig. 2.3). Due to the extremely long etching times, the pyramid walls are very rough, as can be seen in Fig 2.3 (a,d), and the pyramids never worked as MOTs. A second chip was fabricated, at the London Centre of Nanotechnology by Joe Cotter, using an improved etching method (oxygen bubbles were funnelled into the KOH solution) that somewhat reduced the wall roughness (Fig. 2.3). Nevertheless, these pyramids still did not work as MOTs.

We then tried to smooth the pyramids by post-processing the wafers. At first, we tried depositing a silicon dioxide layer (by thermal deposition) or a photoresist layer (by spin or spray coating) but the results were disappointing. We then tried to perform an isotropic maskless etch either by HNA (wet etching) or by ICP (Inductively Coupled Plasma), a processing step which we used to smooth the walls of micro-fabricated optical cavities [14]. We discovered that the HNA etch was very aggressive and seriously affected the shape of the pyramids, turning them into very smooth spheres. The isotropic ICP etch turned out to be the best approach reducing significantly the roughness while preserving the pyramid shape (Fig. 2.3 c,f). In December 2008, after a few years of efforts, ~7000 atoms were detected in the silicon pyramids with an opening of 4.2mm. Consequently, a new pyramid chip, with pyramid sizes ranging from 3.5mm to 2.5mm was made and mounted on a PEEK holder with build-in copper wires. We observed 2000 and about 600 atoms in the 3.5mm and 3mm pyramids respectively in consistence with the $L^6$ scaling law and the data taken in the macropyramids as can be seen in Fig 2.2. The first observation of a MOT in an integrated pyramid was reported in [15]. A detailed experimental study of these micro-MOT's, including our findings on the scaling law was published later in [10].

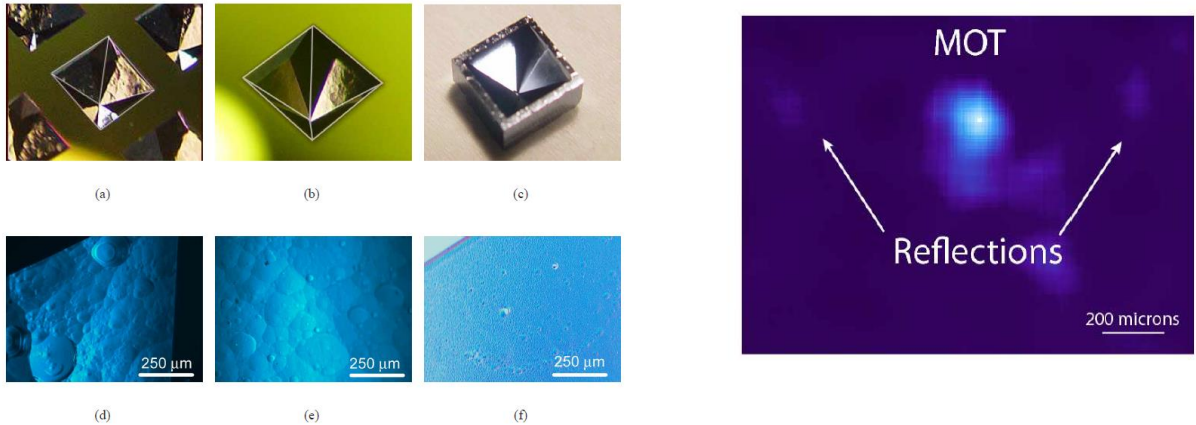

**Fig 2.3** (from ref. [14]). Photographs and SEM pictures of the micro fabricated pyramids fabricated with a conventional KOH etch bath (a), (d), a KOH bath funnelled with oxygen bubbles. The micropyramids shown in (c), (f) have also been post-processed with an ICP isotropic plasma etch for 30 min. The beneficial effects of ICP polishing are quite evident. In Fig 2.3 (g) we can see an image of 2000 Rb atoms in a micropyramid with a 3.5mm opening size.

## Integrated pyramids: The aftermath

The possibility of miniaturising cold atoms experiments, starting from a MOT, for various types of applications such as portable atomic clocks or accelerometers is indeed an ongoing subject of research. Compact cold-atom instruments do not necessarily require atom chip technology. The group of J. Kitching at JILA, for example are pursuing the miniaturisation of conventional 3D MOTs by a simple reduction of beam sizes [16]. Exploring conventional geometries seems to be more favourable in terms of scaling law. Although one has to carefully define the characteristic MOT size, it seems that the scaling $L^6$ is applicable for MOT smaller than ~3mm rather than ~7mm reported in the case of pyramids. Nevertheless, miniaturisation of 6-beam set-up is non-trivial and has not been demonstrated as of yet.

The group of M. D. Himsworth in the University of Southampton is now working towards the fabrication of miniaturized, compact MOTs using chip technology. The group has recently demonstrated a so-called switch-MOT [17]. This is essentially the AC version of a mirror-MOT that requires only one viewport in the chamber. Although this is an elegant way of making a trap, the benefits for integration seem minor. The group of E. Riis and A. Arnold at the University of Glasgow has also explored the possibility of trapping atoms in tetrahedron instead of pyramid shaped hollows [18]. Using this geometry on can trap atoms even by placing three tetrahedron-like mirrors outside the vacuum chamber. A major drawback of this geometry is that it cannot be easily integrated on chip scale. An alternative approach, pioneered by the same group in Glasgow in collaboration with J. Cotter and Ed Hinds (Imperial College London) was to etch reflective nano-grating on the surface of silicon chips. In this case, a single laser beam is diffracted on the grating, creating the necessary conditions for a MOT [19]. The grating-chip is outside of the vacuum chamber without any integrated wires, which means that the chip cannot be used for further atom manipulations.

More than 15 years after the initial atom-chip revolution, the community has somewhat condensed to two categories of research: a) experiments on low dimensional BEC's and b) cavity QED experiments. Atom chip technology has also reached a level of maturity and technological developments become more difficult and scarce. The major difficulty has so far been the integration of light carrying devices (integrated waveguides) with current carrying wires. Significant steps have been made in that direction with a waveguide chip demonstrated in the group of Ed Hinds [20], albeit without integrated wires, as well as in the groups of J. Scmiedmayer and J. Reichel using optical fibres [1,21] that are placed on the chip after the fabrication of the wires. Lattices, a very important step is the study of cold atoms have also been demonstrated on the surface of a chip [22].

Another issue that is closely linked to the integration of light carrying devices is atom trapping in nanometric distances close to dielectric surfaces, which represents a paramount step towards strong atom-light coupling and integrated quantum devices [23]. So far, atom trapping at submicron distances away from the surface of a chip has only been demonstrated in the group of V. Vuletic [24]. This remains up to date the only on-chip experiment where atoms have 'sensed' the Casimir-Polder force. In this respect, light carrying devices such as tapered optical fibres (in the groups of A. Rachebautel, J. Kimble and J. Laurat) and photonic bandgap waveguides (M. Lukin, V. Vuletic and J. Kimble) have performed much better, demonstrating atom trapping [25] and strong light-atom coupling [26] a few hundred nanometers away from surfaces. Hybrid systems, aiming to interface atoms with nanofabricated devices, represent now a new technological challenge in the field of atomic physics.

## Publications resulting from this work and personal contribution

My work at Imperial College London gave overall 4 peer reviewed publications. The first, [G. N. Lewis, Z. Moktdir, C. Gollasch, M. Kraft, S. Pollock, F. Ramirez-Martinez, J. Ashmore, A. Laliotis, M. Trupke, and E. A. Hinds, "Fabrication of Magnetooptical Atom Traps on a Chip," J. MEMS **18**, 347 (2009)] describes the fabrication and characterization of the first pyramid chip. The design of the chip was done at CCM before my arrival by Michael Trupke, Ed Hinds and Fernando Ramirez-Martinez who was the first PhD student on the pyramid project. Chip fabrication was entirely done at the University of Southampton. I participated in the characterization of the chip and in the numerical modelling of atom trapping in micro-MOTs. The paper was written by myself, Michael Trupke and Ed Hinds. Subsequent experiments showed that the pyramid hollows in this chip could not trap any atoms due to the small trapping volume

The demonstration of an integrated pyramidal micro-MOTs and a systematic study of the scaling law that governs the number of atoms trapped in micro-MOTs was studied in two publications [S. Pollock, J. P. Cotter, A. Laliotis, E. A. Hinds, 'Integrated magneto-optical trap on a chip using silicon pyramid structures' *Opt. Express* **17** 14109–14 (2009)] and [S. Pollock, J. P. Cotter, A. Laliotis, F. Ramirez-Martinez, E.A. Hinds, 'Characteristics of integrated magnetooptical traps for atom chips', *New J. Phys.* **13** 043029 (2011)]. This work was a hard and collective effort of Sam Pollock (the second and last student on the pyramid project), Joe Cotter (post-doc), myself and Ed Hinds.

Finally, a collaborative effort between CCM and Southampton describing fabrication methods for smoothing integrated cavities (or any integrated reflective surface) was also published [A Laliotis, M Trupke, JP Cotter, G Lewis, M Kraft, EA Hinds, 'ICP polishing of silicon for high-quality optical resonators on a chip', *Journal of Micromechanics and Microengineering,* **22** 125011 (2012)]. Although most of the experiments were performed during my post-doc years, a lot of the data analysis was done much later thanks to a collaborative project between Imperial College London and the University of Paris13 funded by the Royal Society which was led by Ed Hinds and Joe Cotter on the English side and myself and Benoit Darquié on the French side.

# Chapter 3: Spectroscopy of three dimensionally confined atomic vapour

Confining atoms to small spaces represents an essential step towards the fabrication of integrated and miniaturised atomic clocks, frequency references and sensors. Towards this end, microfabricated atomic vapour cells (Cs and Rb) were pioneered by J. Kitching, L. Holleberg and colleagues at JILA. These tiny sealed cells (a few $cm^3$ in size) [1] were fabricated using silicon technology and were used to demonstrate portable atomic clocks [2] and magnetometers [3]. Very recently, microfabricated cells with paraffin-coated walls that increase the coherence of atomic clocks and magnetometers [4] were also demonstrated. In addition, the group of Atomic Spectroscopy in Montevideo has studied the use of thin cells as atomic clocks [5], although in this case the experiment is not miniaturised and remains macroscopic. An alternative route towards fabrication of compact cells was proposed by F. Benabid and colleagues at the University of Bath, using hollow core fibres filled with molecular gases [6]. Although fibres are not a completely miniaturised system, since one dimension (the fibre length) remains macroscopic, the compactness and flexibility of fibre systems is unparalleled. A disadvantage of hollow fibres lies in the difficulty of filling them with alkali vapours that have strong dipole couplings and are very popular for many applications (clocks, magnetometers…). Alkali atoms have been probed inside hollow fibres [7] but the fabrication of a fully sealed alkali cell based on hollow core fibres remains elusive.

In all the aforementioned experiments, the atomic confinement is macroscopic compared to the wavelength of optical excitation. Pushing confinement to its extremes can naturally carry technological importance as one explores the ultimate limits of miniaturisation but also reveals new fundamental physics. The physics of atomic confinement was pioneered in the 50's by R. H. Dicke who predicted a narrowing of spectral lines due to the effects of collisions that change the velocity without affecting the internal state of the emitter [8]. Experiments have demonstrated this type of Dicke-narrowing in microwave or Coherent Population Trapping spectroscopy using collisions with buffer gas molecules. Around the same time, R. H. Dicke and R. H. Romer showed that narrowing can be observed when emitters are confined within the walls of a cell with thickness less than $\lambda/2$ (where $\lambda$ is the wavelength of excitation) [9]. Here the collisions are 'hard' in the sense that they interrupt the interaction of the emitters (atoms or molecules) with the electric field. The experimentally observed spectral narrowing is due to an enhanced contribution of slow atoms (in the direction of propagation of the excitatory field), that are less affected by the collisions with the walls. The first experiments demonstrated the effect in cm size cells using rotational transitions of ammonia [9]. Optical transitions were investigated many years later [10] using more demanding nanocell technology. Confinement can also be beneficial when studying collective effects, such as supper-radiance, also pioneered by R. H. Dicke, or the related cooperative Lamb shift [11] , as well as blockade effects such as the Rydberg blockade [12], [13] (and other related many-body phenomena such as correlated growth of Rydberg aggregates [14]).

Modern fabrication technologies allow the fabrication of periodic dielectric structures, photonic crystals, of a periodicity comparable to optical wavelengths. Tailoring the modes of the electromagnetic field inside photonic crystals is a booming field of nanophotonics with various applications, including engineering the properties of quantum emitters. For this purpose, atom confinement in such periodic structures is a challenge in modern physics promising the possibility of a new generation of quantum devices [15]. Beyond the well-ordered structure of crystals, the study of randomness and disorder is also of fundamental and technological importance. Anderson localisation, random lasers and Levy statistics of photons are some prime examples of fundamental physics strongly linked to disorder. Vapour confinement has also been studied in random porous media for increasing the optical path and the absorption of molecular vapours [16]. Additionally, spectroscopy of atoms or molecules under confinement allows the study of collisions [17] and provides a new tool for estimating the pore-sizes of random media that find numerous applications (vacuum pumping, dispensers, catalysers…) [18].

# Confining atomic vapour inside opals

In Paris13 we confined atomic caesium vapour inside the interstitial regions of an opal, which is an assembly of identical nanospheres. The spheres are usually made of glass and their size can be chosen between ~100nm up to a maximum of a few microns. Assembling the nanospheres can be done either by sedimentation (a liquid solution containing a concentration of nanospheres is left to slowly evaporate, leaving a deposition of a relatively well-order crystal of nanospheres) or by a Langmuir-Blodgett layer-by-layer deposition of nanospheres on a substrate (usually in our case a glass window). Opals fabricated by sedimentation have less periodicity defects but their size is not easily controlled and the deposition can take significantly long time. In contrast, Langmuir-Blodgett deposition can be faster with a well-controlled number of layers but opals fabricated this way have many periodicity defects occurring primarily due to a dispersion in the sphere diameter. The main advantage of opals against other possible systems of confinement such as photonic crystals made with MEMS technologies is that they can be fabricated easily by soft chemistry, making it possible to perform experiments with many different samples. Compared to porous media, opals allow a much better control of the size and geometry of the interstitial regions due to a relatively good control of the sphere size. On the other hand, the choice of sphere diameter is rather limited and opals have many periodicity defects making them a poor quality photonic crystal.

The first sample we explored was a big opal (~1cm$^3$) of 200nm diameter spheres, made by sedimentation. The opal was sandwiched between two glass windows and subsequently put in a sealed cell containing caesium, made by well-known glass blowing techniques. The cell was made by F. Thibout at the Laboratoire Kastler Brossel (LKB). Upon cell fabrication, the opal lost its white, milky coloration, characteristic of a transparent but strongly scattering medium, and turned black due to the clustering/ condensation of caesium inside the interstices. Naturally, we tried heating the cell in an attempt to evaporate the caesium atoms condensed inside the opal. We found that extremely high temperatures ~300ºC were required for the opal to return almost to its natural milky colour. At these temperatures, one risks a caesium chemical attack, which gives a brown coloration to the glass and makes it opaque. At intermediate temperatures, around 200 C the opal became blue/green, probably due to caesium aggregates. For these reasons, the idea of a big opal was abandoned very quickly and the group turned into the study of opals made by the Langmuir-Blodgett technique in the group of S. Ravaine in the University of Bordeaux. Opals of different sphere diameters and different number of layers were deposited on glass windows, some of which were used for optical characterisation of experiments, while others were subsequently inserted in a caesium sealed cell. Although caesium condensation could not be avoided at room temperature, almost all opals retrieved their milky colour at temperatures of ~150 C, comfortable enough to perform spectroscopic experiments.

## Reflection spectroscopy of Cs vapour inside the opals

Probing atoms inside the interstitial regions of the opal is difficult since laser transmission is usually negligible and in our case contaminated by a large parasitic atomic signal, originating from the free atoms inside the Cs cell (not confined in the opal). In the course of preliminary characterisation experiments, we noticed, much to our surprise, that a reflected beam was observable from the window/opal interface.

The observation of a reflected beam, despite the corrugated nature of the opal interface, opened up the possibility of performing reflexion spectroscopy of atoms inside (or in the vicinity) of the opals, a method, analogous to selective reflection spectroscopy performed at a flat vapour/window interface. At normal incidence, frequency modulated selective reflection (FMSR) is a linear sub-Doppler technique that probes atoms close to the cell window (for a flat surface typically ~$\lambda/2\pi$, where $\lambda$ is the wavelength of optical excitation), commonly used in our group for probing atom-surface interactions. At oblique incidence, selective reflection on a flat interface is broadened by a residual Doppler effect [19], making usually it less attractive for spectroscopic experiments.

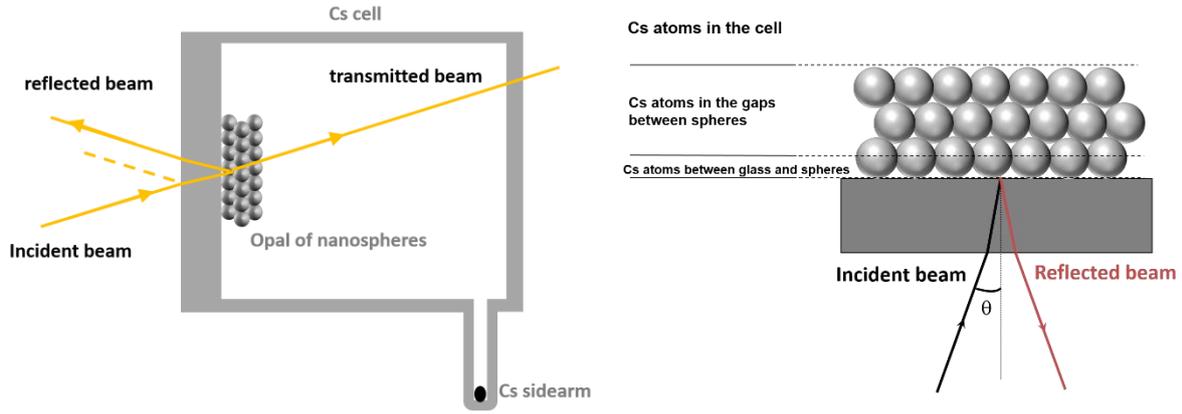

**Fig. 3.1** (a) Schematic of the sealed Cs cell containing the opal, deposited on one of the cell windows. The opal is typically heated to ~160°C while the Cs density in the cell is determined by the temperature of sidearm ~120°C. An atomic signal is measured on the beam reflected from the window/opal interface. The transmitted beam is weak and carries an atomic signal from free atoms inside the cell volume (b) A schematic (not on scale) of the opal deposited on a glass window. The lines define three distinct region where Cs vapour is under different confinement conditions.

The reflexion experiment implemented for probing atoms inside the opals is shown in Fig 3.1. A laser beam is incident on the opal/window interface at various angles of incidence varying from 0° to ~60°. The laser is frequency modulated (FM excursion of 10-20MHz and a frequency of 10kHz) and scanned around an atomic transition frequency. The reflected beam is focused on a photodiode and the output is demodulated using a lock-in amplifier. The fist experiments were conducted on the first two resonant lines of Cs (D1 and D2) at 894nm and 852nm respectively. We used two cells for our measurements containing opals of 10 and 20 layers of nanospheres whose diameter was ~1µm.

At normal incidence (small angles), a sub-Doppler signal was observed reminiscent of selective reflection signals on a flat window/vapour interface. As we increased the angle of incidence (from 10° to 30°), the reflection signals were significantly broadened due to the residual Doppler contribution. For angles ranging from 30° to 60° a sub-Doppler contribution was superimposed to the Doppler broadened structure. The spectral lineshapes depend strongly on polarisation and angle of incidence while the signal remains linear as a function of power. No crossover resonances were observed on the D2 line ($6S_{1/2} \rightarrow 6P_{3/2}$) thus excluding any possible interpretation of the sub-Doppler resonances as a residual saturated absorption signal. In Fig. 3.2 we show the reflection spectra on the D1 line ($6S_{1/2} \rightarrow 6P_{1/2}$) for angles between 30 and 55 degrees for both TE and TM polarisations. The sub-Doppler contribution is more prominent for the TM polarisation. Both cells, 10 and 20 layer opals, yielded similar results [20].

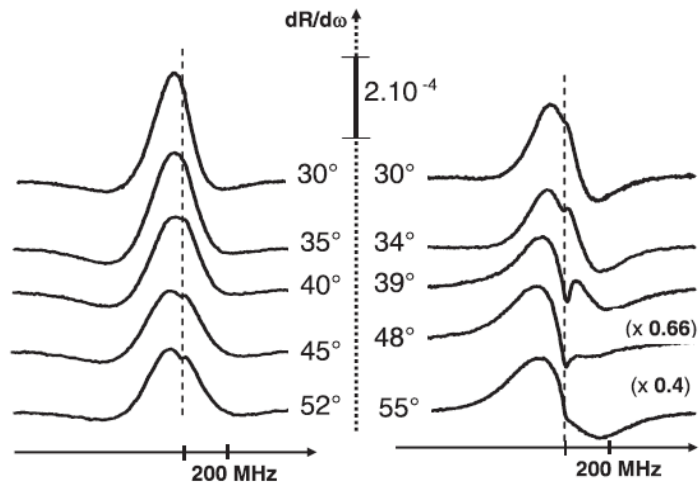

**Fig. 3.2** (from ref [20]) Reflection spectra from the opal/glass substrate interface for TE and TM polarization states and varying angles of incidence. Due to the FM demodulation, our signals represent the derivative of the reflectivity as a function of frequency. A frequency marker of 200MHz (Doppler FWHM is about 400MHz) is shown. In this range of incidences a sub-Doppler contribution is superimposed on an otherwise Doppler broadened spectra. The effect is more prominent for TM polarisations.

Identifying the origin of the atomic signal, and in particular the sub-Doppler contribution at large angles of incidence, is a major difficulty in reflection measurements. Since the opal transmission is weak and the atomic signal does not depend on the number of layers (between 10 and 20) one can safely assume that there is no contribution from the free atoms inside the cell. As can be seen in Fig 3.1 (b), Cs atoms between the window and the opal do not experience the same type of confinement as the atoms that are inside the opal. The atomic confinement inside the opal is complicated as the interstitial regions are interconnected, sometimes forming long empty tubes. Nevertheless, one can assume that inside the opal the confinement is more or less three-dimensional in interstices of a size that is fraction of the sphere diameter d (the value of d/2 was assumed in [20]), while between the window and the opal the atomic confinement is essentially uni or bidimensional.

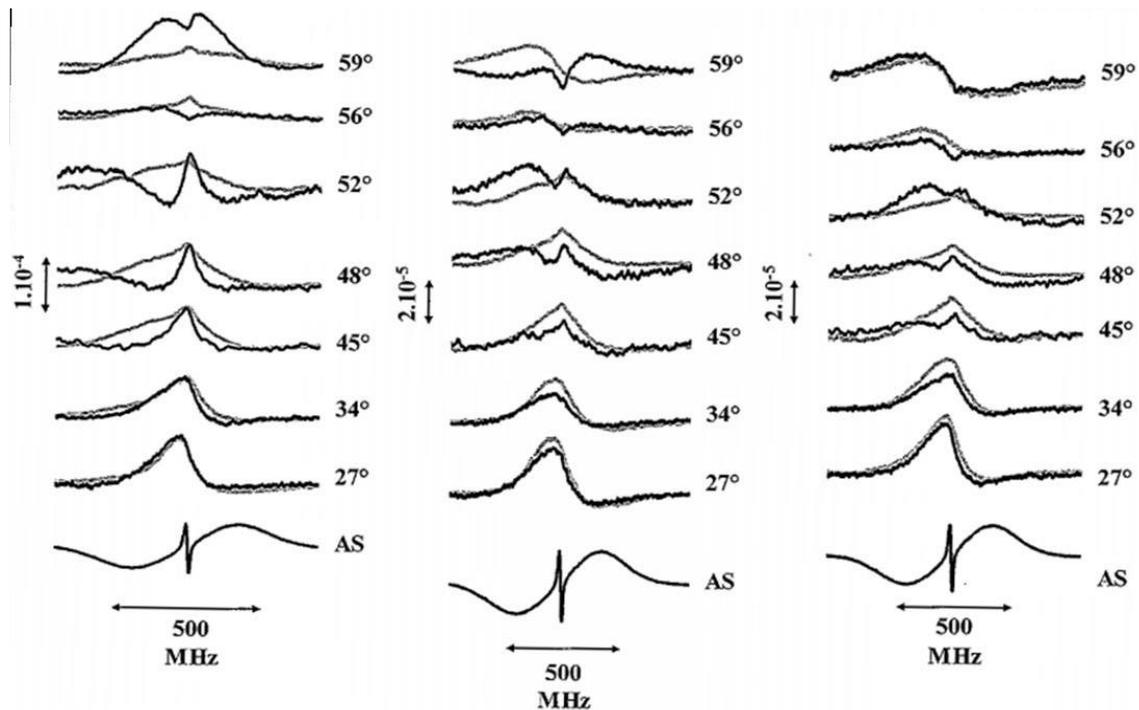

**Fig. 3.3** Reflection spectra of Cs infiltrated opals of 1, 2, 3 layers of 1μm diameter spheres for TM (black) and TE (grey) incident polarisations. The FM characteristics were similar to the ones used for obtaining the data of Fig. 3.2 (10 and 20 layers of 1μm diameter spheres). A sub-Doppler contribution is also present for 1 layer opals having similar characteristics as the one observed for multi-layered opals, i.e more predominant for TM polarisations with lineshapes that rapidly change with the angle of incidence.(from the thesis of Elias Moufarej)

As previously mentioned, the opal is a low quality photonic crystal that retains a periodicity especially in the z-direction (direction perpendicular to the window). This periodicity leads to the observation of Bragg peaks in the reflected beam for angles given by a modified Bragg condition that accounts for the inhomogeneity of the medium [21]. At a Bragg incidence, the contributions of the inner planes of the opal are in phase and add constructively increasing the opal reflectivity, suggesting also that the contribution of the atoms inside the opals should be enhanced. Here the second order Bragg peak of the opals, made of 1μm diameter spheres, is expected around ~50°, which is not incompatible with the range of incidence angles for which the sub-Doppler contribution is observed (30°-55°) [22]. This, along with the fact that the signal lineshapes seem to be largely dependent on the angle of incidence, suggesting a dependence on light propagation inside the opals, led us to attribute the sub-Doppler contribution to atoms that are there-dimensionally confined inside the opals. Subsequent optical characterisation of the opals (measurement of the reflectivity and transmission at 894nm, away from the atomic resonance) revealed that a second order Bragg peak indeed exists for an angle of incidence around 60° (see Fig. 3.5). The evidence presented in [20] indicated that the sub-Doppler contribution originated from atoms confined in the inner regions of the opal. However, a theoretical model of the expected reflection spectra would be indispensable for proving the above claims. The difficulty of developing such a model stems from the combination of a hard atomic physics problem with an equally hard nanoptics problem. In our experiments, the atoms are always in a transient regime of interaction with the electric field, confined in interstitial regions of very complex geometry, in which the electric field can be only numerically simulated after significant time and effort. Although some steps towards a better theoretical understanding of the opal experiment were made in the course of this project a theoretical model capable of reproducing the experimentally measured spectra has not yet been developed.

Significant steps for understanding the effects of atomic confinement inside opals were done during the PhD thesis of Elias Moufarej, whose project consisted in a systematic experimental study of atomic reflexion spectra inside different opal structures. An interesting experiment performed by Elias Moufarej was the study of atomic confinement in opals with different number of layers. In particular, we explored opals of 1,2,3,4 layers of 1μm, as well as 730nm diameter nanospheres for both incident polarisations, TE and TM [23]. Although, the principal aim of the experiment was to observe and explore the gradual build-up of the sub-Doppler structure as the number of layers increases, we observed that strong sub-Doppler resonances were also present for 1-layer opals. In Fig. 3.4 we show a collection of experimental reflection spectra at the Cs D1 line (894nm) for incidences ranging between $27^o$ to $59^o$ for opals of 1,2,3 layers of 1μm diameter spheres. As in the case of 10 and 20 layer opals, shown if Fig. 3.3, sub-Doppler structures predominantly, but not exclusively, appear for TM polarisations for angles ranging from $35^o$ to $59^o$. The spectra vary significantly with the number of layers and as a function of angle. Investigation of opals with 730nm diameter spheres revealed that a sub-Doppler structure appears predominantly for 1-layer opals for TE polarisations. Experiments were also performed with a 10-layer opal of 400nm diameter spheres, showing again sub-Doppler structures but for a different range of angles of incidence. Furthermore, in an attempt to explore the influence of the $\lambda/D$ ration (where $\lambda$ is the optical wavelength and D the sphere diameter), we performed experiments on the second resonance of Cs ($6S_{1/2} \rightarrow 7P_{3/2}$) at 455nm. These experiments were not very conclusive mostly due to the more complicated hyperfine structure of the transitions (three hyperfine transitions inside the Doppler range) but also partly because of the degradation of the signal to noise ratio compared to the more comfortable D1 line at 894nm.

The results presented here provide no definite answer about the origin of the sub-Doppler signal observed in reflection experiments on an opal infiltrated with Cs atoms. In our case, atoms are confined in between 'hard walls' and it is improbable to have a Dicke narrowing effect similar to the one observed with buffer gas collisions [8], as this requires collisions that only change the velocity of the atoms while not perturbing the interaction with the electric field. Having a selection of slow velocities due to confinement [9], an extension of the Dicke narrowing of thin cells, is a possibility. However, in the case of a three-dimensional confinement inside an interstitial region of a characteristic size of ~500nm, the proportion of atoms slow enough to avoid any wall collision within their excitation lifetime is approximately $2 \times 10^{-4}$ instead of $7 \times 10^{-2}$ in the 1D case [24]. Even though the FM detection enhances a narrow contribution compared to a Doppler-broadened signal one would still expect the narrow sub-Doppler contribution to remain a very small addition to the broad signal. In our experiments, the sub-Doppler contribution is, at its peak (Fig. 3.2, $39^o$ with TM polarisation), not much smaller than the Doppler broadened signal. The above arguments as well as the experimental results on the 1-layer opals suggest that the sub-Doppler contribution might not originate from the inner layers of the opal. It is possible that the sub-Doppler signals originate from a bi-dimensional confinement in long tubes either inside the opal or on the interface opal/glass. In this case, the proportion of atoms that are slow on the plane perpendicular to the tube axis is more favourable ~$4 \times 10^{-3}$. Alternatively, there could be two distinct mechanisms responsible for a sub-Doppler signal, one originating from the inner layers (giving a sub-Doppler signal for opals of many layers) and a different one giving raise to the sub-Doppler signals observed for the 1-layer opals.

Finally, it should be noted that pump-probe experiments were also performed inside opals. For these experiments, we used a strong pump beam resonant with the Cs D1 line and a probe beam resonant with the Cs D2 line. The experiments were performed in the vapour cells containing 10 layers of 1μm diameter spheres and they are described in more details in [25]. These measurements also revealed sub-Doppler features on the probe signal exclusively when the probe beam was at large angles of incidence. A noteworthy observation is that when the pump and probe beams are incident at opposite angles (here +58° for the probe and -58° for the pump) the position of the sub-Doppler resonances depends on the frequency of the pump, suggesting the possibility of velocity selection in the opal experiments.

Additionally, we noticed that the intense pump beam induced some light induced atomic desorption (LIAD). Atomic desorption from porous media and organic coatings attracts attention in atomic spectroscopy with cells or photonic bandgap fibres [26] and finds applications in alkali atom dispensing.

## Optical modelling and characterisation of the opals

During the course of the opal project, we also tried to understand the optical properties of the opals and the behaviour of the electric field inside the empty interstitial regions. For this purpose we performed a great deal of characterisation measurements initially using our excitation lasers (894nm, 852nm and 455nm) and subsequently a super-continuum laser source with a compact Ocean Optics spectrometer optimised for visible and near infra-red wavelengths. Eventually, characterisation measurements were compared with two different theoretical models that were developed in the course of the opal project. We also characterised the opals using optical and scanning electron microscopy, which revealed that our opals have numerous defects that accumulate with the increasing number of layers.

### One-dimensional model of stratified index

The first model was developed in the group and considers the opal as a stratified one-dimensional medium consisting of thin layers whose effective index $n_{eff}(z)$ depends on the filling factor $f(z)$ (percentage of the layer occupied by the glass spheres) via the simple formula $n_{eff}(z) = \sqrt{f(z)\varepsilon_{glass} + (1 - f(z))}$ . Here z is the axis perpendicular to the deposition window and $\varepsilon_{glass}$ is the dielectric constant of glass. This model goes one step further from the well-know simplification of considering the opal as a homogeneous medium whose reflective index is given by $n_{eff} = \sqrt{f\varepsilon_{glass} + (1 - f)}$ with a filling factor that depends on the crystallographic system (for a fcc or hcp system and glass spheres of $n_{glass}$=1.45 on finds $n_{eff}$=1.35). The reflection and transmission of an electromagnetic wave incident upon such a stratified opal can be calculated numerically by using the transfer matrix method. This method cannot include scattering losses, which we account for by introducing an ad hoc imaginary part of the effective refractive index that can also be wavelength dependent ( in our case a Rayleigh type of scattering $\lambda^{-4}$ was considered).

The force of the above model stems essentially from its simplicity, which allows a rapid analysis of different opals by changing the number of layers, the angle of incidence and the wavelength of the electromagnetic field. Although the model essentially neglects the there-dimensional structure of the opal, and only accounts for Bragg-type peaks originating from crystallographic planes parallel to the deposition window, its predictions are in partial agreement with optical characterisation (reflection and transmission) measurements. This is probably due to that fact that opals have numerous defects and their periodicity survives predominantly in the z-direction, due to the deposition of layers, whereas it remains very poor in the other directions.

In order to check the validity of the numerical model we performed a set of dedicated characterisation measurements using 20-layer opals of 280nm diameter spheres deposited on microscope slides. For our measurements, we used a fibered super continuum source (400nm-1700nm) and a spectrometer operating at the 350nm-950nm range. Typically, we measured the reflection and transmission of the opals for different angles of incidence, with the white laser incident from the glass (laser incident on deposition window) as well as from the opal side (laser incident directly on the opal). The opals investigated here, were never used in the Cs infiltration experiments, but were chosen to have Bragg peaks at the wavelength range of our detection. The comparison of the theoretical and experimental reflectivity of the opals for both illuminations (opal side and glass side) is shown in Fig. 3.4. The only adjustable parameter in these curves is the imaginary index of refraction that accounts for the scattering losses, chosen by fitting the opal transmission curve as a function of wavelength. In addition, the refractive index of the spheres is considered to be $n_{sphere}$=1.4, a value that is smaller than the index of $SiO_2$ and most types of glass. It is justified by the fact that spheres are not compact but are probably an agglomerate of smaller spheres. Thus $n_{sphere}$=1.4 corresponds to a type of effective index of the sphere. The experimental curves exhibit a big Bragg-type peak at a wavelength that depends on the angle of incidence θ as well as oscillations due to interferences between the front and the back side of the opal (interferences of the Fabry-Perot type, consistent with the overall thickness of the opal). The basic characteristics are also reproduced by the theoretical model, which gives an accurate prediction of the positions of the Bragg peaks and their widths. The amplitude of the reflectivity is almost always overestimated by the theoretical model.

In Fig.3.5 we show the reflectivity of a 10-layer opal of 1μm diameter spheres as a function of incidence angle, irradiated with a 894 nm laser. Our measurements are performed for a laser beam incident from the opal as well as the glass side and for both polarisations. When the laser is incident from the glass side the reflectivity is comparable to the glass/air interface exhibiting also a Brewster-type of angle for TM polarisation. On the contrary, reflectivity from the opal side is almost negligible for incidences smaller than ~50°. This behaviour, that 'inspired' us to measure the reflectivity spectrum of caesium infiltrated opals, finds an explanation in the context of the stratified opal model. Fig.3.5 shows that the effective index of the window/opal/air system exhibits a discontinuity, or a gap, on the glass/opal interface (the index jumps from $n_{glass}$ to $n_{air}$), which is not present in the opal/air interface (a smooth increase of the index). For wavelengths comparable to the sphere diameter, the reflectivity from the glass side is sensitive to the gap between glass and opal, taking values close to 4% (glass/air reflectivity). For longer wavelengths the opal behaves more like a homogeneous medium of index $n_{eff}$=1.35 thus exhibiting small reflectivity for a glass side irradiation and a large reflectivity for an opal side irradiation. The phenomenon was examined in more detailed in Fig.3 of reference [27].

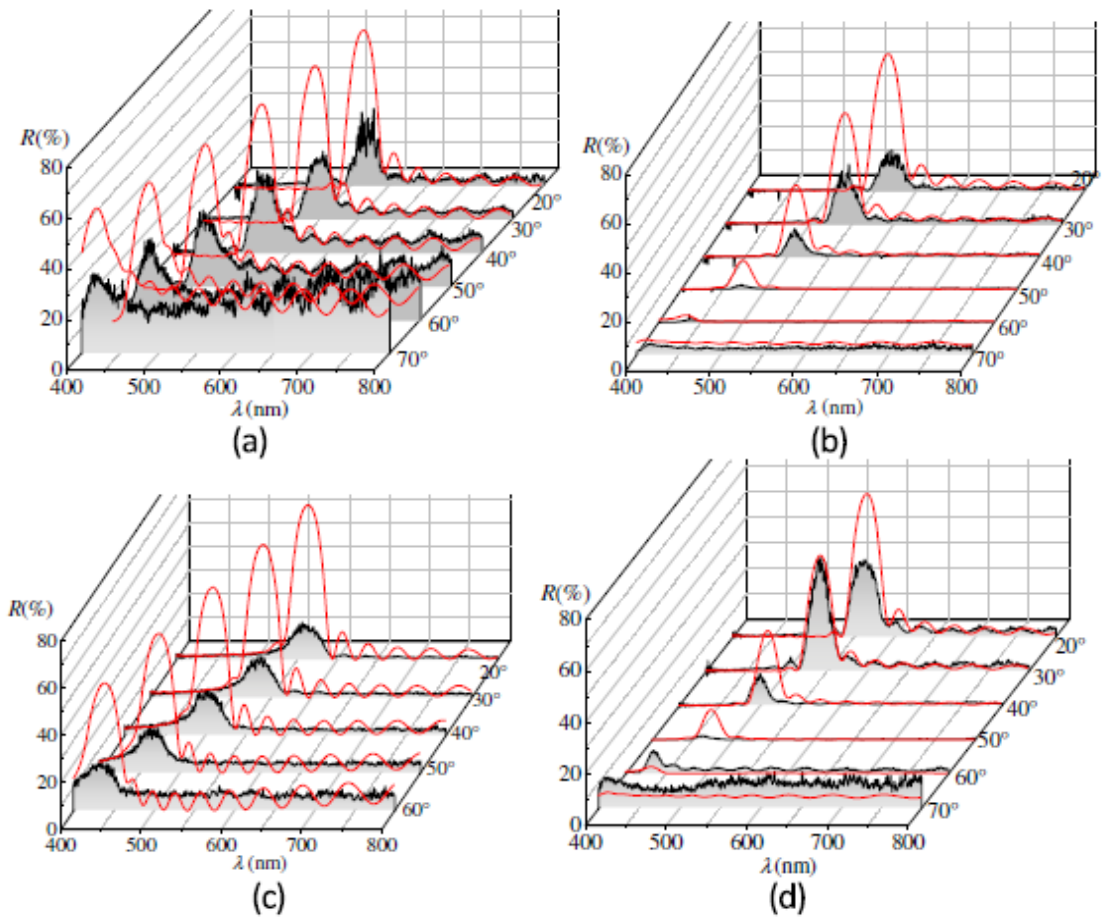

**Fig. 3.4** (from ref [23]) Reflectivity measurements (solid black lines) and the predictions of the stratified index numerical model (solid red lines) as a function of wavelength λ and angle of incidence θ. (a) and (b) correspond to the reflectivity measured from the opal side whereas (c) and (d) from the glass side. The polarization is TE for (a) and (c) and TM for (b) and (d).

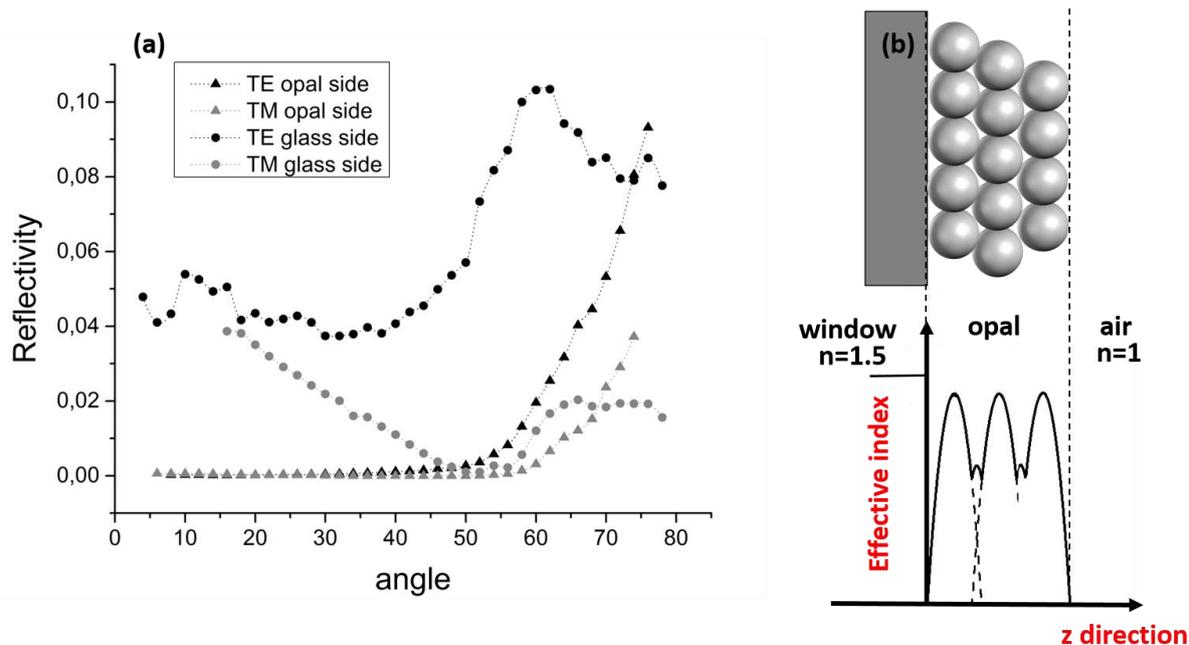

**Fig. 3.5** a) Reflectivity measurements at 894nm for a 10-layer opal made with 1µm diameter spheres. The opal is not infiltrated with Cs vapour. The reflectivity form the glass (window) side has similar characteristics as the glass/air reflectivity for small angles of incidence. The reflectivity from the opal side is negligible for incidences smaller than 50°. A maximum of reflectivity observed at ~55° is probably a second order Bragg peak, mainly visible when the laser is incident from the substrate side. This is probably because the organisation of the first layers (as deposited on the substrate) is much better than the organisation of the last. It should also be noted that the interpretation of optical measurements at very oblique incidences can be difficult. b) Schematic of an opal deposited on a glass window and a qualitative plot of the effective index as a function of distance from the substrate. The window thickness (usually a few mm) and the sphere diameter (smaller than 1µm) are not in scale.

## Three-dimensional model

So far, we have discussed the periodicity of the opals in the direction vertical to the window. Although the organisation of the nanospheres on the deposition plane was undoubtedly very poor, SEM images on monolayer opals revealed a mono-domain crystalline organisation (hexagonal structure) for small areas (in most samples smaller than ~20x20 µm²). We also observed first order diffraction of laser light (various wavelengths were used: 633nm He-Ne laser, a 455nm laser and 894 nm laser) both in transmission and in reflection. Optical characterisation of the diffracted beams and measurement of the diffraction angle θ allowed us to directly evaluate the sphere diameter in most of our samples with an accuracy of ~5%, even after cell fabrication. This proved to be critical, as very often the samples were mislabelled. It should be mentioned that probing a mono-domain organisation for SiO₂ spheres was extremely difficult and required extreme focusing of the laser beams as well as finding the proper spot on the sample surface. Most often a poly-domain area was probed resulting in a multiple diffraction spots or even diffraction ring (Fig. 3.6) rather than diffraction beams [28]. Opals made from polystyrene spheres are much better organised, but cannot be used for Cs infiltration experiments.

The only way to model three-dimensional photonic crystals is by using numerical methods. For this purpose we developed a Finite Element based numerical model in collaboration with Ilya Zabkow and V. Klimov of the Lebedev Institute in Moscow, experts in electromagnetic simulations. The model is very cumbersome, time consuming, and provides more information than a scientist can handle. It gives details on experimental observables, such as the reflection and transmission of an incident laser depending on its polarisation, wavelength λ, angle of incidence θ, as well as the angle of orientation φ of the opal relative to the plane of incidence (Fig. 3.6). Moreover, it gives the exact vectorial distribution of the electric and magnetic fields inside the interstitial regions, as well as inside the spheres. The model assumes a periodic opal with spheres that are almost but not exactly in contact for convergence purposes. All calculations were performed in Moscow by Ilya Zabkov and were limited to opals of up to two or three deposited layers, since calculations became extremely time consuming for thicker opals.

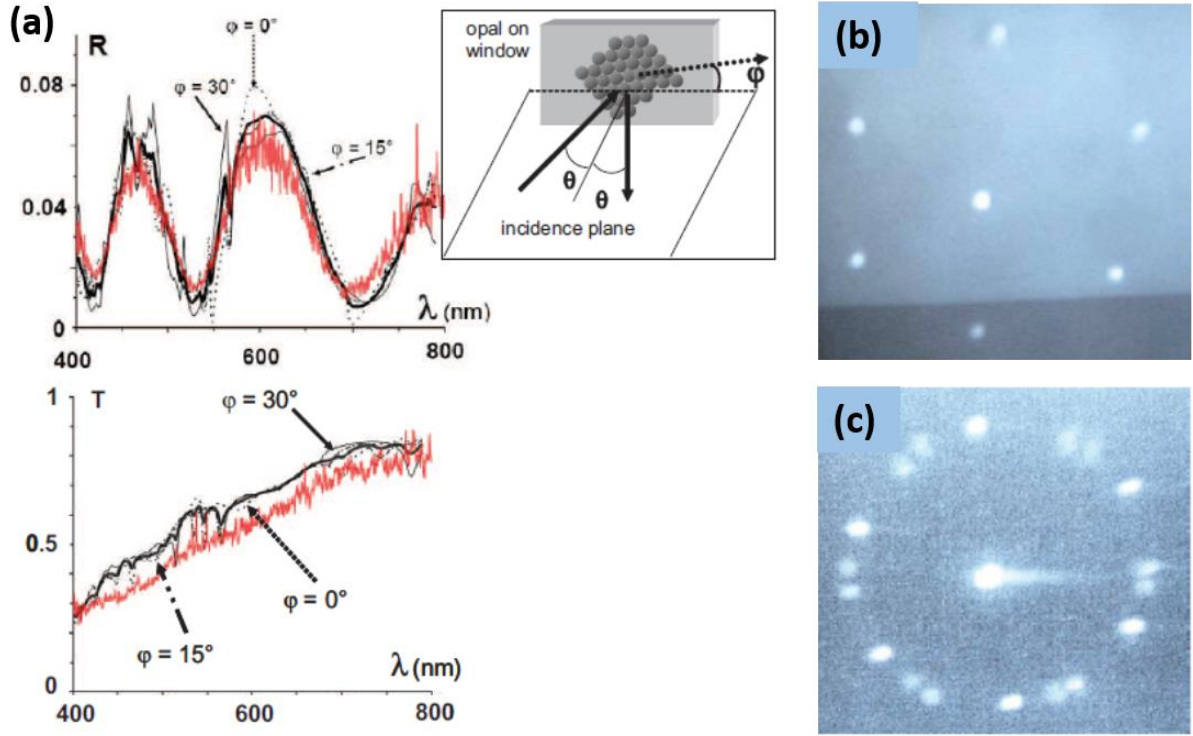

**Fig. 3.6** (a) (from ref. [23]) Reflectivity and Transmission of a monolayer of 735nm diameter spheres deposited on a glass substrate as a function of wavelength. The inset shows the orientation angle φ of the opal with respect to the plane of incidence of the laser beam. The red curves represent the experimental data and the black curves show the theoretical predictions for different values of φ. The thick solid black line represents a randomised opal orientation. (b) and (c) (from ref. [28]) typical diffraction transmission patterns of a tightly focused laser beam one a monodomain (b) and polydomain (c) part of the opal surface. Here the measurements are made with a 633nm laser incident on a monolayer of 800nm diameter spheres.

Again, the model was put to the test by performing a big set of experimental measurements. Most characterisation experiments were done with the white super-continuum source (as previously mentioned) but also with dedicated lasers. Measuring the reflectivity and transmission for a single-layer opal with a monodomain periodicity was extremely challenging and the idea was quickly abandoned. Instead, we made measurements using polarised beams of ~1mm waist, in order to average the effects of opal orientation. Typical results are shown in Fig. 3.6 for a single-layered opal of 735nm nanospheres, with theory and experiment being in almost excellent agreement. It should be noted that assuming the opal as a homogenous dielectric with an effective refractive index ($n_{eff}$~1.35) or even as a one-dimensional multi-layered system (stratified index approach) cannot reproduce the period and amplitude of the oscillations observed in the reflectivity of the opal, despite some qualitative resemblance of the curves.

## Conclusions and final thoughts on the opal experiments

Reflection experiments with Cs infiltrated opals have proved inconclusive in determining the origin and the physical mechanism of the sub-Doppler contribution and the theoretical analysis has proved to be difficult. Nevertheless, opals are so far one of the very few truly three-dimensional miniaturized systems that can provide high-resolution spectroscopy of alkali atoms. The opal system is easy to fabricate, it can be infiltrated with Cs at the appropriate temperatures and it can be probed by a very simple reflection spectroscopy experiment. Here we have not pursued a miniaturisation of the opal cell nor the experimental set-up, as it was out of the scope of our work, but standard microfabrication and miniaturisation technologies could in principle be used towards this purpose [1].

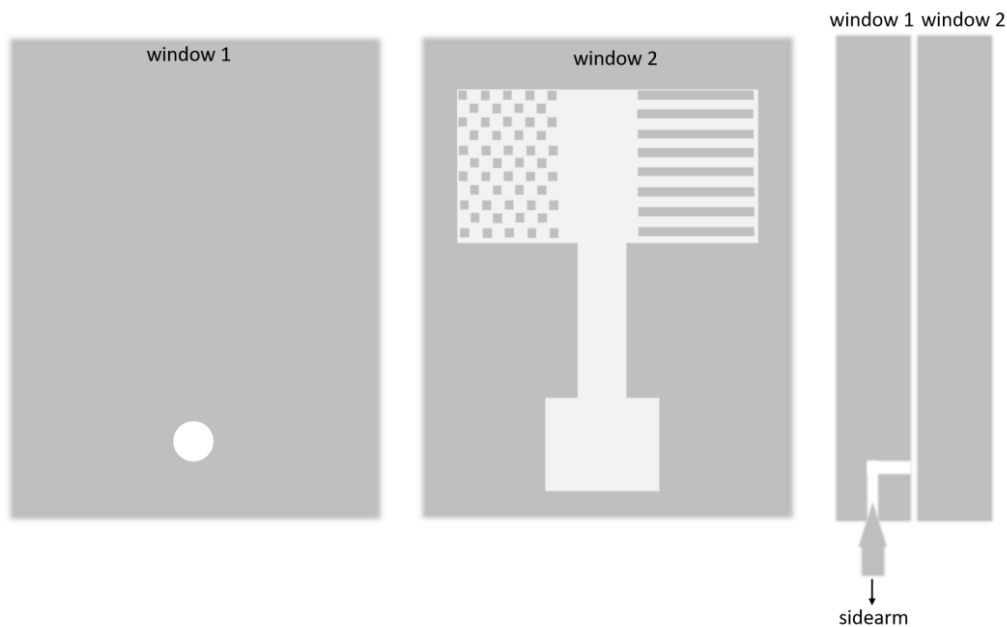

**Fig 3.7** A schematic diagram of the design for a new thin cell containing cubic or long rectangular compartments where atoms are 3D or 2D confined respectively. The design shows a front and side view of the two cell windows. The cell fabrication begins by plasma etching an area (light grey) on one of the cell windows leaving cubic and rectangular gaps with opening as to allow for atom circulation. The depth and the width of the features can be tuned easily between a few hundred nanometres up to a couple of microns. The actual arrangement of the cubes or rectangles can be periodic or random. The second window contains a drilled hole that allows us to funnel atoms into the cell using a Cs sidearm. The two windows are put together either by optical contact or by anodic bonding using process similar to those reported in [29].

Future experiments should focus more specifically on identifying the spectroscopic behaviour of atoms under three-dimensional or bi-dimensional confinement. In principle, the probing method should not be crucial in detecting a sub-Doppler contribution originating from a truly three-dimensional confinement. As such, it would be extremely favourable to deploy the arsenal of microfabrication methods, in order to achieve a complete control of the size and shape of the interstitial regions of confinement. One proposal would be to fabricate the equivalent of a thin nanocell including cubic compartments (with gaps to allow vapour circulation) that can be arranged periodically (or not) in the plane of the cell (Fig.3.7). The atoms confined in these compartments could be probed by a simple transmission measurement with a signal that should not dramatically depend on the angle of incidence (as the compartments are not spheres some dependence in directionality, also in terms of field scattering should be expected). Similarly long rectangles (tube-like) can also be etched, in order to test bi-dimensional atomic confinement. In this case, laser illumination and therefore velocity selection should be perpendicular to the tube axis, a scheme that should be more favourable for observing a possible sub-Doppler contribution.

Confining atoms in an interstice of controlled geometry also simplifies the theoretical analysis of the system. In this case, the calculation of the transient behaviour of atoms should be easier to perform as the atomic trajectories can be more easily identified. Although the actual details of the electric field are probably secondary in identifying a possible sub-Doppler signal, a complete numerical electromagnetic calculation of the electric field in a cubical gap of air embedded in glass, should be much easier to perform, yielding results that can be more easily interpreted and used in a subsequent modelling of the atomic response.

The possibility of fabricating such systems is currently under consideration in our group and is roughly depicted in Fig. 3.7. The fabrication of such a cell can move forward our understanding of confined atom spectroscopy. It could also be considered for exploring different physics, such as the Rydberg blockade phenomenon [12], [13] (probing Rydberg atoms in vapour confined in a radius similar to the blockade radius) or Dicke supperadiance [11].

## Publications resulting from this work

The experimental work on infiltrating opals with Cs vapour and the observation of the sub-Doppler contributions has been largely a collective effort of the SAI members (Daniel Bloch, Isabelle Maurin, myself and the students Philippe Ballin and Elias Moufarej). The stratified index model was mostly developed by Isabelle Maurin and Daniel Bloch, although I personally participated and contributed in some discussions. The three dimensional model was developed by Ilya Zabkov and Vasily Klimov (theoreticians) interfacing with myself and Daniel Bloch (experimentalists). Experimental characterisation of the opals was mostly driven by Elias Moufarej, myself and Daniel Bloch.

There are three peer-reviewed journal publications describing the work done on the opal project. The first [P. Ballin, E. Moufarej, I. Maurin, A. Laliotis, D. Bloch, 'Three-dimensional confinement of vapor in nanostructures for sub-Doppler optical resolution' *Appl. Phys. Lett.* (2012)] is predominantly the experimental work done during the thesis of Philippe Ballin. The second [E. Moufarej, I. Maurin, I. Zabkov, A. Laliotis, P. Ballin, V. Klimov, D. Bloch, 'Infiltrating a thin or single-layer opal with an atomic vapour: Sub-Doppler signals and crystal optics' *EPL*, **108**, 17008 (2015)] is mostly a review of the project focusing on the experimental work done during the thesis of Elias Moufarej and the numerical modelling of opals. Finally the third publication [I. Maurin, E. Moufarej, A. Laliotis, D. Bloch 'Optics of an opal modeled with a stratified effective index and the effect of the interface', *JOSA B*, **32**, 1761 (2015)] is a detailed description of the stratified index model compared to dedicated experimental measurements. The pump-probe experiments on opals are described in a conference proceedings paper [P. Ballin, E. Moufarej, I. Maurin, A. Laliotis, D. Bloch , 'Sub-Doppler optical resolution by confining a vapour in a nanostructure', Proceedings of SPIE, **8770**, 87700J (2013)].

## Atom confinement inside a random porous medium

In parallel with the opal experiments that took place in the University of Paris, the group of atomic spectroscopy in Montevideo was also exploring the physics of atomic confinement, albeit in a radically different system and with different probing methods. This was not a simple coincidence but a product of the long collaboration between the two groups dating back to the 1990's.

Fabricating a sealed cell filled with Rb containing a random porous medium was a very important part of the Montevideo experiment. The fabrication process begins by grinding glass into little pieces whose size can vary from ~1-100µm very approximately. The fragments are then mechanically filtered, allowing separation of pieces according to size. The process can be further refined by sedimentation filtering. Following this step, the glass fragments of a specific size (mostly 10µm, 50µm and 100µm) are placed in a glass tube (made of exactly the same glass as the spheres) and are heated at temperatures about 700°C for several hours. This creates a random porous medium with particles that are bonded together as well as to the cell walls. A picture of the cell is shown in Fig. 3.8 (a). The cell has a small sidearm that can be separately heated at slightly lower temperatures than the rest of cell (and the porous medium in particular) to avoid rubidium condensation inside the medium. At room temperature, the porous medium is slightly coloured probably due to rubidium aggregates that quickly disappear at normal operating temperatures ~100°C.

Here, atoms confined inside the porous medium were probed by fluorescence excitation spectroscopy. Conventionally, fluorescence detection relies on the principle that spontaneous emission of absorbed photons randomly redistributes their direction in space. By detecting scattered photons away from the axis of the propagating laser beam, resonant fluorescence is background free which makes it attractive for sub-Doppler spectroscopy in thin cells [30].  This principle does not apply to a scattering porous medium, which randomises the direction of *all* incident photons, thus making spontaneous emission indistinguishable from the background. The first fluorescence detection measurements performed at low rubidium densities at Montevideo showed that the number of detected photons did not change when the incident laser frequency was scanned around the rubidium resonance line.

In October/November 2011, Arturo Lezama visited Paris13 on a research trip funded by ECOS-Sud.

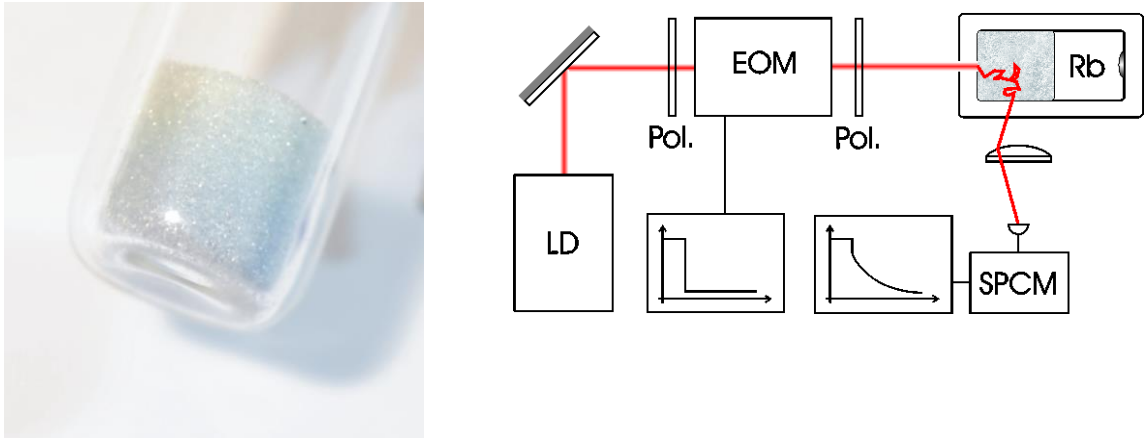

**Fig 3.8** (a) Picture of a porous medium inside a cell containing Rb vapour at room temperature. The slight blue coloration is probably due to Rb aggregates. The porous medium has a size of ~1cm³ and looks very similar to a sugar cube. (b) (from ref. [32]) A schematic diagram of the experimental set-up used for time-resolved experiments. Light pulses are created by use of fast EOM (electro-optic modulator) and a polariser and are shone upon the porous medium containing Rb vapour. The scattered light is detected using a lens system and a sensitive photon counting module.

After discussions on the preliminary experiments performed at Montevideo we decided that time resolved experiments would allow us to separate spontaneous emission from a non-resonant background [31], [32]. The basic principles of time resolved experiments are the following: A laser beam, tuneable in frequency around the D1 transition of Rb at 794nm, illuminates the porous medium infiltrated with Rb, which scatters all photons (resonant and non-resonant) in random directions. A lens system collects the scattered light coming from a remote point of the porous medium (with respect to the point of illumination) onto a very sensitive fibered photon counting avalanche photodiode. The laser beam passes through an electro-optic modulator that rapidly changes the polarisation state of the light, which translates into a rapid change of intensity by use of a polariser. We thus create light pulses having a contrast of about 50% and a rise time around 10nsec. The photon counting photodiode is extremely fast (better than 1ns) but has a dead time of about 30ns which obliges us to work at low powers and average the experiment over many cycles using a train of light pulses of a high repetition rate ~MHz. The laser frequency is slowly scanned around the Rb transitions and monitored by an auxiliary saturated absorption experiment. This process allows us to make time resolved fluorescence measurements as a function of laser frequency. A schematic diagram of experimental set-up is shown in Fig. 3.8 (b).

Typical measurements of scattered light as a function of time for various temperatures of the rubidium reservoir are shown in Fig.3.9 (a). Here, the vapour density *inside* the pores is correlated to the reservoir temperature. The coloured curves are taken with the laser resonant to the rubidium transition whereas the black one shows the off-resonant variation of the scattered light intensity. To better understand the curves one has to keep in mind that the steady state signal should be given by $I = I_o(1 - \alpha + \beta)$, where the term α is equivalent to an *absorption* representing photons that are missing when the laser is resonant with a rubidium transition, whereas β is the *emission* that represents the photons added do the signal due to atomic fluorescence. The transient behaviour between two steady state values $I_1$ and $I_2$ (which is the intensity pulse of our laser) is given by $I(t > 0) = I_2(1 - \alpha - \beta) + (I_1 - I_2)\beta e^{-\frac{t}{\tau}}$, where τ is a constant related to the finite lifetime of the atomic state but also to the time required by the photons to exit the medium that is dependent on radiation trapping and non-radiative de-excitation due to wall collisions. We stress that we only take into account the transient behaviour of emitted photons, while transient effects in the absorption term are ignored. From the black off-resonant curve of Fig. 3.9 (a) we see that due the first 20ns after switching the light intensity are contaminated by the electronics and are therefore discarded in our analysis.

Using the above reasoning, we can extract from our experimental spectra the absorption α and the emission β as a function of laser frequency and rubidium vapour pressure, as well as the decay time τ as a function of Rb pressure. The basic findings are summarised as follows: For low frequencies, the emitted photons are balanced by the absorbed photons, whereas when the density increases the number of spontaneously emitted photon reduces resulting to an observed net absorption in the spectra. This is clearly illustrated in Fig. 3.9 (b). The decay time τ increases as a function of Rb pressure due to radiation trapping, but quickly reaches an upper limit as can be seen in Fig. 3.9 (c). Both these observations are a direct consequence of fluoresce quenching (non-radiative de-excitation) due to collisions of excited atoms with the walls of the porous medium. If one assumes that the probability for a radiative decay after one photon absorption is b (and consequently of non-radiative decay 1-b), then multiple absorption events, due to radiation trapping, significantly reduce the probability that a photon exits the porous medium (for n absorption events the probability of radiative fluorescence decreases to $b^n$). Also as a direct consequence of the non-zero probability of non-radiative de-excitation, the decay time of resonant scattered light in a porous medium is bound to a value $\tau_c = \tau_o \frac{b}{1-b}$, where $\tau_o$ is the lifetime of the Rb D1 state. A detailed derivation of this formula is presented in [28] and in the supplementary materials.

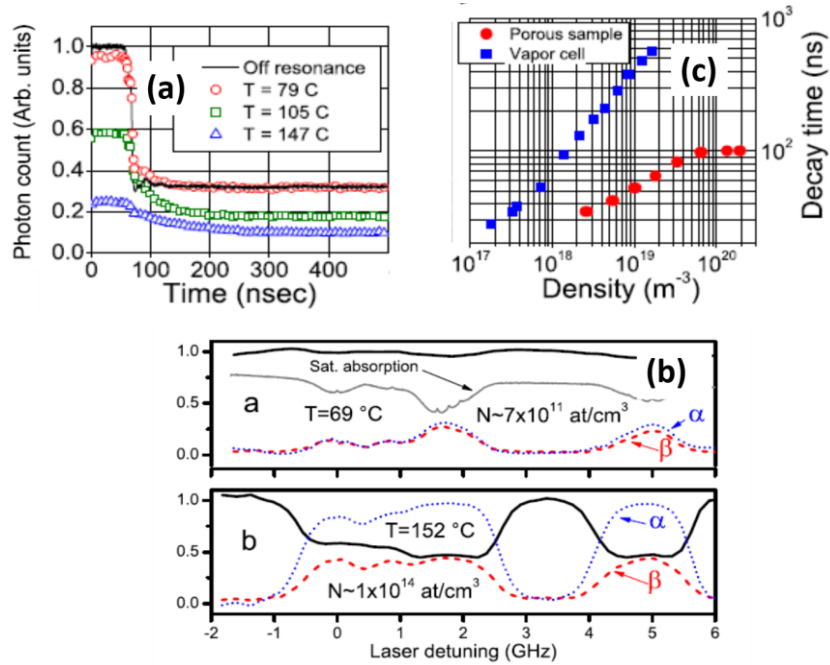

**Fig 3.9** (from ref. [32]) (a) Time dependence of the fluorescence signal after a laser pulse. The black line shows the time dependence of the detected light when the laser is out of resonance while the coloured points (red, green and blue) show the on resonance time dependence of the signal for different Rb densities. (b) Steady state fluorescence spectrum of the D1 line of Rb confined inside the porous medium for two different Rb densities (solid black lines) with a saturated absorption reference (grey line). The relative contributions of absorption (blue) and emission (red), reconstructed from time resolved measurements, are also shown. (c) Lifetime τ of photons inside a conventional cell (blue squares) and inside the porous medium (red circles). The lifetime of photons inside a porous medium (confined atoms) is bound at ~100nm, contrary to the photon lifetime in a conventional cell (free atoms).

The results of [31] represent a detailed analysis of fluorescence measurements inside porous media taking into account spontaneous emission, which was more or less ignored in previous works of similar nature [16]–[18]. Additionally, an important result of our experiments is the measurement of b (probability of non-radiative decay) which is directly linked to the size of the pores of the random medium, by using simple consideration of kinetic gas theory. Here, we concluded that the size of the interstices is ~50μm in consistence with the initial size of the glass particles used to fabricate the porous medium. This is a spectroscopic method for assessing the pore size of random media that is mostly adapted for large pore sizes. It is possible that for nanometric pore sizes, spontaneously emitted photons will not be able to exit the cavity due to large the non-radiative losses, thus compromising the resolution of the measurements. A spectroscopic method for measuring the size of pores was also proposed by T. Svensson and colleagues [18], relying on the measurement of the spectral linewidth of molecular transitions, which for small pore sizes (smaller than ~100nm) can be dominated by wall-to-wall collisions. Porous media find extensive applications in in catalysis, pumping and gas storage and porosimetry is usually performed by methods such as mercury intrusion or by evaporation of propanol.

It is also worth mentioning in passing that fluorescence quenching was also observed in normal vapour cells filled with dense atomic vapours. It was observed that when irradiated at normal incidence, with respect to the cell window, fluorescence quenching is velocity selective (mostly efficient for atoms that arrive towards the surface) thus resulting in a sub-Doppler contribution in the backward emitted fluorescence signals [32].

## Sub-Doppler spectroscopy inside a porous medium

In parallel with the experiments on time-resolved fluorescence, the Montevideo group also performed a detailed study of the backscattered light from a random porous medium. The original idea was to measure the atomic signal carried on the coherent backscattering from the medium. The group believed at the time that weak localisation effects that favour self-intersecting paths of scattered photons would give rise to a sub-Doppler component to the atomic signal, equivalent to a saturated absorption signal. Although the idea has an inherent charm, in retrospect it might lack solid basis as weak localisation effects are based on a single photon interference and can probably not give signals equivalent to saturated absorption that is essentially a pump-probe effect.

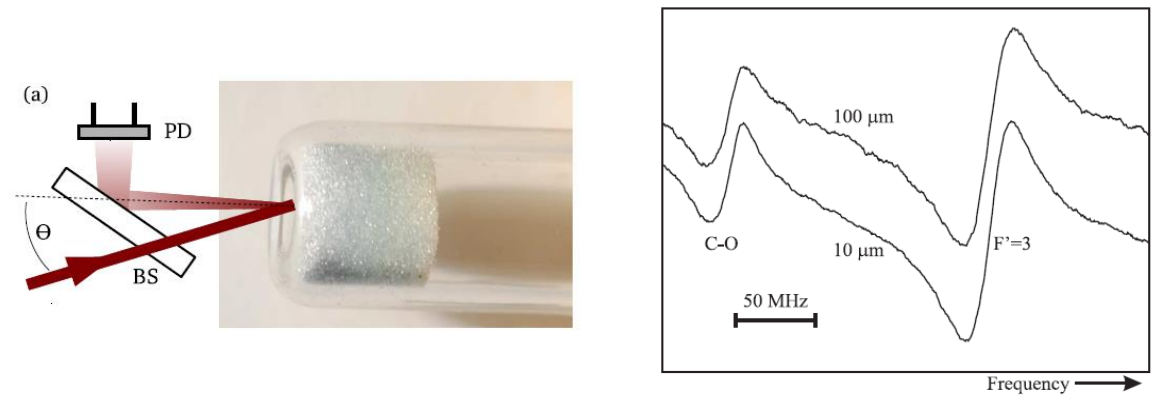

**Fig. 3.10** (a) Schematic of the experimental set-up and (b) Scattered light spectrum, after FM demodulation, of the $^{85}$Rb F=3→F'=3 transition and the nearby crossover resonance. We show spectra for 10μm and 100μm pore sizes with a detection angle of θ~0°. Both figures are taken from ref. [33].

The first experimental measurements revealed that saturated absorption is indeed present in the backscattered light but very quickly, it was discovered that the signal does not display the same angular sensitivity, as coherent backscattering, who has a sharp angular spread around the exact backward direction. Rather, the saturated absorption signal was present for a wide variety of angles (very wide angular spread) displaying a residual Doppler broadening similar to the one observed in a typical saturated absorption set-up, inside a vapour cell when an angle is introduced between pump and probe beams. This overwhelming experimental observation has lead the Montevideo group to attribute the observed sub-Doppler resonances to a more conventional saturated absorption spectroscopy that takes place near the surface of the porous medium (depth comparable to the scattering mean free path which can be several tens of microns depending on the size of the pores). For small penetration depths, the incident laser beam retains its directionality (scattering is not yet very efficient) and saturates (*pumps*) specific velocity class of the atomic medium, which is consequently *probed* by the backscattered light. Although the physical origin of this phenomenon (backscattering saturated absorption) is much more mundane than the original concept of *coherent* backscattering saturated absorption, this represents a simple, very robust and eventually scalable system providing reliable frequency references.

The experimental set-up that was used for these experiments is shown in Fig. 3.10(a). A laser beam resonant on the D1 Rb transition is incident on the porous medium and the backscattered light on a solid angle of about ~$10^{-5}$ sr is collected on a photodiode. The angle θ between the detected scattered light and the incident laser beam is changed by simply moving the position of the photodetector. For small angles θ~0 the scattered light is collected by use of a beam splitter. The actual orientation of the porous medium (or the cell containing the porous medium) plays absolutely no role in our measurements. The experiments were performed with two different porous media of 10µm and 100µm pore size. In order to increase the relative weight of the sub-Doppler contributions, a frequency modulation (FM) is applied on the incident laser. A typical FM spectrum of the backscattered light at θ=0 is shown in Fig. 3.10 (b). A more detailed description of the experiment is given in [33]. The linewidth of the single and cross-over resonances as a function of the detection angle θ for a random medium with 10µm and 100µm pore sizes is shown in Fig.3.11. For small angles the saturated absorption lines observed in the porous media (10µm and 100µm size pores) are ~25MHz wide, slightly broader than the saturated absorption linewidths of ~15MHz observed in a vapour cell, probably limited by parameters such laser linewidth, FM excursion or power broadening. The larger linewidth observed for porous media is probably due to scattering of the incident laser light that slightly randomises the incident k-vectors even at small penetration depths. For porous media the cross-over resonances are narrower than the single transition resonances, an effect qualitatively analysed in [33].

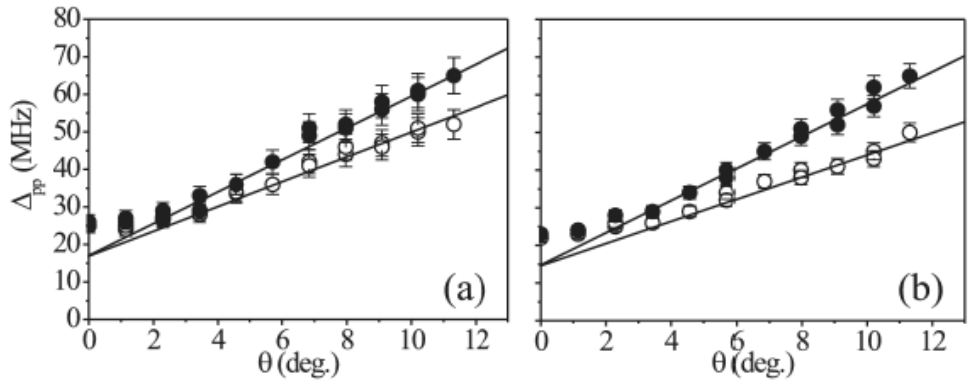

**Fig. 3.11** (from ref. [33]) (a), (b) Linewidth as a function of detection angle for a porous medium with a mean pore size of 100µm (a) and 10µm (b). The solid circles represent the linewidth of single transition resonances whereas open circles represent the linewidth of the crossover resonances

## Final thoughts on the experiments with porous media

Since the publication of the above experiments [31], [33] the group in Montevideo has explored the possibility of performing pump-probe experiments inside random porous media. In the first experiments, performed without my participation, both pump and probe beams are resonant with the D1 line of rubidium and the experimental spectra are compared to a simple model solving the rate equations and assuming a randomisation of the k-vectors of both pump and probe beams [34]. The ultimate goal of pump-probe experiments with atoms that are three dimensionally confined in the pores of a random medium, is to access high lying energy states of alkali atoms and in particular Rydberg states. It has been shown that, due to extremely high dipole interactions, a single Rydberg atom induces an energy level shift to its neighbours strong enough to suppress their interaction with a resonant laser beam. This phenomenon quoted as the 'Rydberg blockade' allows the observation of collective behaviour of atomic ensembles. Although the phenomenon has been mainly observed with cold atom samples, recent experimental studies have demonstrated the possibility of probing Rydberg atoms using thermal vapours confined in thin cells [12] or hollow core fibres [13], paving the way to the observation of collective phenomena using simple, room temperature experiments with vapour cells. The porous medium fabricated and studied in Montevideo is an attractive system for performing Rydberg spectroscopy as it confines atoms in interstices whose effective size ranges from a few to several tens of microns, comparable to the Rydberg blockade radius.

Towards this end, the group first attempted to probe the high lying state $5D_{3/2}$ of rubidium via a two-step excitation ($5S_{1/2} \rightarrow 5P_{3/2}$ at 780 nm and $5P_{3/2} \rightarrow 5D_{3/2}$ at 776nm). In this experiment the $5D_{3/2}$ population was detected via the blue (almost violet) fluorescence at 420nm emitted form rubidium atoms decaying through the $6P_{1/2} \rightarrow 5S_{1/2}$ de-excitation channel. Subsequently, the group probed Rydberg atoms inside random media with 10µm and 100µm pore sizes. Rydberg spectroscopy was done using a 794nm probe laser, resonant with the D1 line of Rb and a 476nm pump laser that can excite nS and $nD_{3/2}$ levels with principal quantum numbers n ranging from 30 to 40. The spectroscopic set-up is similar to the ones reported in [12], using two superposed lasers focused on the porous medium. In contrast to previously reported experiments, detection was performed on the backscattered light of the probe laser. Preliminary measurements were presented in [35] for $33D_{3/2}$, $32D_{3/2}$ and $31D_{3/2}$ Rydberg levels for both 10µm and 100µm pore sizes. For 10µm pores spectral broadening is dominated by Casimir-Polder interactions, (Doppler broadening ~1GHz is prominent only for 100µm pores). These experiments also provide a stepping-stone for studying Casimir-Polder interactions with Rydberg atoms in complex geometries.

## Publications resulting from this work

So far, my collaboration with the Montevideo group has given two peer reviewed publications. The first one describes the time resolved fluorescence experiments [S. Villalba, H. Failache, A. Laliotis, L. Lenci, S. Barreiro, A. Lezama, Opt. Lett. **38** (2013)]. I had an important contribution in the conception of the experiment, which started from discussions between Arturo Lezama and myself during his visit in Paris. I also participated in the experiments, data analysis and all the discussions that led to a better understanding of the physical mechanism of the fluorescence quenching by wall collisions and its manifestation in our experimental data. The mathematical model that describes part of the data was developed mainly by Arturo Lezama. My participation in the second experiment demonstrating sub-Doppler resonances in the backscattered light was less significant [S. Villalba, A. Laliotis, L. Lenci, D. Bloch, A. Lezama, H. Failache Phys. Rev. A. **89**, 023422 (2014)]. The basic idea of exploring the backscattered or coherently backscattered light came from the Montevideo group and the first experiments were performed without my participation. Nevertheless, I did participate significantly in the detailed investigation of the saturated absorption resonances as a function of angle and in the discussions leading to their interpretation. The fabrication of rubidium cells with porous media was made possible due to the ingenuity of H. Failache with crucial help from S. Villalba.

The Franco-Uruguayan collaboration has been  supported by various projects but the main contributor is the ECOS-SUD program that sponsors regular visits of researchers and students between the groups. Personally, I have been a participant in an ECOS program (2008-2011), I currently direct another ECOS project (2015-2018) with Daniel Bloch and Horacio Failache.  I have been an invited research fellow in Montevideo for a period of 2 months and I have also obtained other small travelling grants to support the collaboration. In total, I have made many academic visits to Montevideo and I have strongly participated in the development of the experiments on porous media.

# Chapter 4: Casimir-Polder interaction at finite temperature: the theory

Interactions between neutral atoms or molecules and macroscopic surfaces have been a subject of study for the past century or so. The simplest model of this interaction, that of an electric dipole attracted towards its mirror induced image, was laid out by Lennard-Jones in 1932 [1] for the purposes of describing particle adsorption or desorption to dielectric (metallic) walls. If we denote the atom (molecule)-surface distances as z, the dipole image approach gives an interaction energy proportional to $-C_3/z^3$, where $C_3$ is the van der Waals coefficient. The validity of this approach is restricted to distances much greater than size of the electronic cloud. For argument's sake we will here set the limit to one nanometre (z>1nm). Closer to the wall the distortion of the electronic cloud becomes important, leading to an additional repulsive interaction and a potential trapping well with a minimum positioned a few angstroms away from the surface.

In 1948, H. B. G. Casimir and D. Polder looked at the atom-surface interaction problem from quite a different point of view [2]. In their work, the atom interacts with a fluctuating vacuum field, which is modified by the presence of the reflective surface (boundary). Using this approach for ground state atoms, they predicted that the $-z^{-3}$ law would break down at distances comparable to the reduced wavelength of atomic transitions (typically for z>100nm). At large separations, they calculated that the interaction energy is proportional to $-z^{-4}$. This phenomenon can be ascribed to the influence of retardation, suggesting the limitations of the static model of instantaneous interaction between dipoles. It is however worth mentioning that one cannot reproduce the results of the QED approach [2] simply by accounting for retardation in the semi-classical picture of the dipole that interacts with its image. For these reasons we will henceforth use the term Casimir-Polder interaction as the generic term to describe the long-range interactions (z>1nm) between atoms or molecules and surfaces. Eventually, the Casimir-Polder retardation was experimentally demonstrated 45 years after H. B. G. Casimir's and D. Polder's seminal work using a beam of ground state sodium atoms flying through a metallic cavity [3].

The separation between excited and ground state atoms is a subject of experimental relevance. Spectroscopic experiments rely on the measurement of energy differences between energy states [4], [5] and beams of highly excited, long-lived Rydberg atoms have been used for Casimir-Polder measurements [6]. These techniques are predominantly sensitive to the Casimir-Polder potential of excited states. In contrast, most techniques deploying cold atoms measure the Casimir-Polder energy for atoms in their ground state [7], [8]. Excited state atoms have a strikingly different behaviour in the far–field, where it displays an oscillating behaviour analogous to that of a classical dipole antenna [9-11]. These QED oscillations have been demonstrated using ions trapped several centimetres away from a surface [12].

A central question of Casimir-Polder interactions that has occupied physicists for a very long time, concerns the influence of thermal fluctuations of the electromagnetic field. It was recognised since Lifsitz (see for example ref. [13] and a discussion in [14]) that at distances comparable to what is now called the thermal wavelength $\lambda_T = \frac{\hbar c}{k_B T}$ the Casimir-Polder interaction is proportional to temperature, $T$, and switches back to the inverse cube law $(-T/z^3)$. It was later predicted that at the special case when the surface is out of equilibrium with respect to the rest of the environment an increase of thermal effects is expected and the interaction is proportional to $\sim \frac{(T_S^2 - T_E^2)}{z^2}$, where $T_E$, $T_S$ are the temperatures of the environment and the surface respectively [15]. An out of equilibrium experiment using a Rb BEC at several microns away from a silica surface was the first demonstration of the thermal effects in Casimir-Polder interaction [14].

More recently, a different aspect of temperature effects was brought to light, related to the dielectric properties of materials and in particular evanescent polariton modes (surface modes). It was shown that due to thermal excitation of surface modes, thermal emission of dielectrics can be dramatically different in the near field (nanometric distances ~100nm) [16]. In particular, near field thermal emission can in some cases be almost monochromatic compared to the well-known broadband far-field thermal emission (blackbody radiation). Intense evanescent vacuum, or thermal fields have a strong influence on the Casimir-Polder interaction [17, 18], allowing for an efficient tuning of the atom-surface interaction if the atomic dipole couplings and the surface resonances of the dielectric are carefully chosen.

The theoretical results that will be presented here follow mostly on the footsteps of the perturbation theory approach presented in the work of Wylie and Sipe [10, 11] published in 1984 and 1985. Their work is of importance in particular for spectroscopic experiments as it uses perturbation theory to calculate the lifetimes and energy shifts of atoms at any state (excited or ground) as a function of distance. Wylie and Sipe also provide a framework to study temperature effects although at the time, they did not consider temperature to be relevant parameter for experiments. It is approximately around the turn of the century that a demonstration of temperature effects appeared experimentally feasible [17, 19] (see also discussions in [18]). An extension of the formalism of [10, 11], calculating Casimir-Polder shifts at non-zero temperatures was done in our group [20]. Other theoretical works were also presented on the same subject [21, 22].

Finally, before going into the details of the Casimir-Polder interaction it is worth stressing its close relationship with the Casimir effect [23] between two macroscopic surfaces. The interest in the Casimir force sparked in the end of 1990's after demonstration of precision measurements between a plane surface and a sphere using a torsion pendulum [24] and an atomic force microscope (AFM) [25] and shortly after using micromechanical systems (torsional device) [26,27]. Most Casimir measurements use metallised objects to avoid systematic errors due to accumulation of charges. The comparison of experimental measurements with theoretical predictions requires taking into account, amongst other parameters, finite temperature corrections (thermal Casimir force) and the effects of the metallic dielectric constant. The two effects are intricately connected as the choice between plasma or Drude model for the dielectric constant (close to zero frequency where no experimental data of the dielectric constant of metals are available) changes significantly the thermal corrections in the Casimir force [28-31]. The interpretation of Casimir measurements has been a subject of intense scientific debate.

## Distance dependent shift and linewidth at finite temperatures

In what follows, we will not discuss the details of the perturbation theory approach that leads to the calculation of Casimir-Polder shift and the distance dependent atomic lifetime in front of a dielectric surface at finite temperatures. According to [20] the (free) energy shift, $\delta F_a$, of a given atomic state $|a\rangle$ is given as:

$$\delta F_a = \sum_b \left\{ n(\omega_{ab}, T) \mu_\alpha^{ab} \mu_b^{ba} Re[G_{\alpha\beta}(z, |\omega_{ab}|)] - 2 \frac{k_B T}{\hbar} \sum_{p=0}^{\infty} {}' \mu_\alpha^{ab} \mu_b^{ba} G_{\alpha\beta}(z, i\xi_p) \frac{\omega_{ab}}{\omega_{ab}^2 + \xi_p^2} \right\} \quad (4.1)$$

The first summation is on all the allowed dipole couplings $|a\rangle \rightarrow |b\rangle$ and the term inside the curly brackets is the contribution $\delta F_{a \rightarrow b}$ of each individual dipole coupling to the overall energy shift.

$$\delta F_{a \rightarrow b} = n(\omega_{ab}, T) \mu_\alpha^{ab} \mu_b^{ba} Re[G_{\alpha\beta}(z, |\omega_{ab}|)] - 2 \frac{k_B T}{\hbar} \sum_{p=0}^{\infty} {}' \mu_\alpha^{ab} \mu_b^{ba} G_{\alpha\beta}(z, i\xi_p) \frac{\omega_{ab}}{\omega_{ab}^2 + \xi_p^2} \quad (4.2)$$

We use the Einstein notation, implying a summation over the index variables $\alpha$ and $\beta$ that denote the Cartesian coordinate components. The prime symbol signifies that the first term of the sum should be multiplied by 1/2. The transition frequency $\omega_{ab} = \omega_b - \omega_a$ depends on the energy difference between the two levels. It takes a positive sign for an upward coupling (absorption) and a negative sign for a downward coupling (emission). Also $\xi_p = 2\pi \frac{k_B T}{\hbar} p$ are the Matsubara frequencies and $\mu_\alpha^{ab}$ are the dipole moment matrix elements. $G_{\alpha\beta}(z, \omega)$, are the elements of the linear susceptibility matrix, $\vec{\vec{G}}$, as defined in [10, 11]. We briefly remind that the linear susceptibility matrix gives the reflected displacement field, at a point $\vec{r}$, of a dipole $\vec{\mu}(\omega)$ oscillating at frequency $\omega$, positioned at $\vec{r'}$, via the relation $\overrightarrow{D}(\vec{r},\vec{r}',\omega) = \overleftrightarrow{G}(\vec{r},\vec{r}',\omega)\,\vec{\mu}(\omega)$. In our case $\overleftrightarrow{G}$ is evaluated at $\vec{r} = \overrightarrow{r'}$ because we are interested in the interaction of the dipole with its own reflected field. In the case where cylindrical symmetry applies (flat surface), the linear susceptibility matrix $\overleftrightarrow{G}(z,\omega)$ is simply a function of distance $z$ and frequency $\omega$. Finally $n(\omega_{ab},T) = 1/\left(e^{\frac{\hbar\omega}{k_B T}} - 1\right)$ is the occupation factor for Bose-Einstein statistics.

The first term of Eq. (4.2) will be herein called resonant term, as it depends on the value of the linear susceptibility at the resonance frequency and the second term will be called non-resonant. The resonant term is due to spontaneous emission (downward couplings) or due to stimulated emission/absorption from thermal fields and therefore, reminiscent of the interaction of a classical dipole antenna with its own reflected field [9]. Due to retardation (propagation) in the far field ($z \gg \lambda/4\pi$) it decays slowly, oscillating between attraction and repulsion with a period of $\lambda/2$, ($\sim z^{-1}\cos\frac{4\pi}{\lambda}z$). The non-resonant term includes the influence of vacuum fluctuations integrated on all frequencies. It is that term, first derived in [2], that in the far field decays as $z^{-4}$ (at zero temperature), and eventually as $-T/z^3$, at distances $z \gg \lambda_T$, in the case of finite temperature. In the near field ($z \ll \lambda/4\pi$) the linear susceptibility matrix reduces to the following analytical expression:

$$G_{xx}(z,\omega) = G_{yy}(z,\omega) = \frac{1}{2}G_{zz}(z,\omega) = \frac{1}{(2z)^3}\frac{\varepsilon(\omega)-1}{\varepsilon(\omega)+1} \quad (4.3)$$

Replacing the linear susceptibility equations Eq. (4.3) into the energy shift of Eq. (4.1) we get the following expression:

$$\delta F_a = -\frac{1}{z^3}\sum_b C_3^{a\to b}\left\{-2n(\omega_{ab},T)Re\left[\frac{\varepsilon(|\omega_{ab}|)-1}{\varepsilon(|\omega_{ab}|)+1}\right] + 4\frac{k_B T}{\hbar}\sum_{p=0}^{\infty}{}'\frac{\varepsilon(i\xi_p)-1}{\varepsilon(i\xi_p)+1}\frac{\omega_{ab}}{\omega_{ab}^2+\xi_p^2}\right\} \quad (4.4)$$

The coefficients $C_3^{a\to b} \sim \frac{A_{ab}|\lambda_{ab}|^3}{256\pi^4}$ are proportional to the transition probability of the $|a\rangle \to |b\rangle$ transition and the cube of transition wavelength $\lambda_{ab}$. More details on the calculation of the van der Waals coefficient (essentially the term inside the sum of Eq. (4.4)) are given in [32]. For convenience, we call the quantity inside the curly brackets the image coefficient $r(\omega_{ab},T)$. For a perfect reflector $r(\omega_{ab},T) = 1$ and the energy shift simply becomes $\delta F_a = -\frac{1}{z^3}\sum_b C_3^{a\to b}$.

The lifetime of the atomic level $|a\rangle$ also depends on the distance from the surface, due to the change of the local density of states of the electromagnetic field. This is due to the existence of frustrated modes (modes that are evanescent in the dielectric but propagating in vacuum), surface-polariton modes (evanescent at both sides) or by modification of propagating modes upon reflection with the surface [33]. The linewidth of level $|a\rangle$ (the inverse of the lifetime) can be expressed as $\gamma_a = \gamma_a^o + \delta\gamma_a(z)$ where $\gamma_a^o$ represents the linewidth far away from the surface (in the volume), including the natural linewidth and additional broadening effects such as collisional broadening, and $\delta\gamma_a(z)$ includes all the modifications induced by the surface.

$$\delta\gamma_a(z) = \sum_b 2n(\omega_{ab},T)\mu_a^{ab}\mu_b^{ba}Im[G_{\alpha\beta}(z,|\omega_{ab}|) \quad (4.5)$$

In the far field, $z \gg \lambda/4\pi$, the role of evanescent modes diminishes and the linewidth decays as $\delta\gamma_a(z) \sim z^{-1}\sin\frac{4\pi}{\lambda}z$ due to the reflection of propagating modes. In the near field, $z \ll \lambda/4\pi$, it is accustomed to use Eq. (4.3) for the linear susceptibility to make the approximation:

$$\delta\gamma_a(z) = -\frac{2}{z^3}\sum_b C_3^{a\to b}\left\{2n(\omega_{ab},T)Im\left[\frac{\varepsilon(|\omega_{ab}|)-1}{\varepsilon(|\omega_{ab}|)+1}\right]\right\} \quad (4.6)$$

Nevertheless, it should be stressed that in Eq. (4.6) the influence of propagating and frustrated modes is ignored. In the case of a lossless dielectric, $Im[\varepsilon(\omega)] = 0$ and Eq. (4.6) leads to the erroneous conclusion that $\delta\gamma_\alpha(z) = 0$. For a correct description of the distance dependent linewidth one has to fall back to Eq. (4.5) correctly accounting for retardation (propagation) effects.

## Thermal excitation of surface modes

We will now focus at the temperature dependence of the Casimir-Polder energy shift and the distance dependent atomic linewidth in the near field of the interaction. A quick inspection of Eq. (4.4) reveals that the resonant part of the van der Waals coefficient essentially depends on the real part of the quantity $S(\omega) = \frac{\varepsilon(\omega)-1}{\varepsilon(\omega)+1}$, which we will call surface response. The non-resonant term also depends on the surface response evaluated at the imaginary Matsubara frequencies $i\xi_p$, whose physical meaning can be rather obscure. Furthermore, the quantities $S(i\xi_p)$ cannot directly inferred from an experimental measurement, which suggests that correctly evaluating the non-resonant term requires essentially an analytical function of the dielectric constant and therefore the knowledge of the dielectric constant at all frequencies. In this chapter, we will only try to provide a clear physical interpretation of the near field temperature dependence and as such, we will limit ourselves to a simple model that describes a dielectric with a single resonance, whose dielectric constant is given by the following expression:

$$\varepsilon(\omega) = \varepsilon_\infty + \frac{(\varepsilon_o - \varepsilon_\infty)\omega_V{}^2}{\omega_V{}^2 - \omega^2 - i\Gamma\omega} \quad (4.7)$$

The same model has been used extensively in various theoretical works in the field [16, 20] and to fit experimental data of the dielectric constant [34, 35], although the latter usually requires accounting for additional resonances. Here we will call $\omega_V{}^2$ volume resonance, $\Gamma$ a dissipation constant and $\varepsilon_\infty$, $\varepsilon_o$ give the dielectric constant at the two extremes of the spectrum. Using Eq. (4.7) one can find a simple analytical expression for the surface response.

$$S(\omega) = S_\infty + \frac{(S_o - S_\infty)\omega_S{}^2}{\omega_S{}^2 - \omega^2 - i\omega\Gamma} \quad (4.8)$$

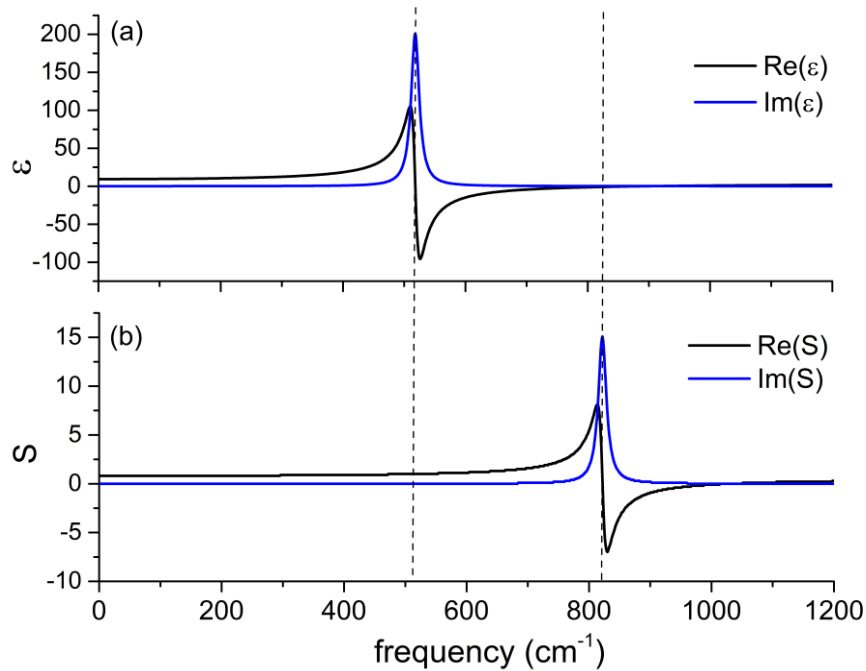

**Fig. 4.1** Real and imaginary parts of the dielectric constant (a) and the surface response (b). Here we use Eqs. (4.7), (4.8) with parameters $\varepsilon_\infty = 3.1$, $\varepsilon_o = 9.32$, $\omega_V = 2\pi\,518\;cm^{-1}$ and $\Gamma = 0.02\;\omega_V$.

The above expression is very similar to Eq. (4.7), with a notable difference of the resonance frequency, which is given by the equation $\omega_S = \sqrt{\frac{\varepsilon_o+1}{\varepsilon_\infty+1}}\,\omega_V$. Naturally, the constants $S_o, S_\infty$ give the surface response at the extrema of the spectrum. In Fig. 4.1(a) and Fig. 4.1(b) we show a graphic representation of Eq. (4.7) and Eq. (4.8) respectively for a set of parameters that roughly correspond to the surface response of sapphire (in reality sapphire has a much more complicated dielectric constant with four resonances).

Using the analytical expression of the dielectric constant of Eq. (4.7) we can get exact analytical expressions for the image coefficient $r(\omega_{ab}, T)$ [36]. In the case of $\Gamma \ll \omega_V$ we can write the following approximation:

$$r(\omega_{ab}, T) \cong Re[S(\omega_{ab})] + \frac{2\omega_{ab}\omega_S(S_o - S_\infty)[\Gamma^2 + 2(\omega_{ab}^2 - \omega_S^2)]}{[\Gamma^2 + 2(\omega_{ab}^2 - \omega_S^2)]^2 + 4\Gamma^2\omega_S^2}\coth\left(\frac{\hbar\omega_S}{2k_BT}\right) \quad (4.9)$$

Or when $\Gamma = 0$ (approximation valid if $\omega_{ab}$ is far away from $\omega_S$) we get the very simple expression:

$$r(\omega_{ab}, T) = Re[S(\omega_{ab})] + \frac{\omega_{ab}\omega_S(S_o - S_\infty)}{(\omega_{ab}^2 - \omega_S^2)}\coth\left(\frac{\hbar\omega_S}{2k_BT}\right) \quad (4.10)$$

In Fig. 4.2 we show the image coefficient as a function of atomic transition frequency, $\omega_{ab}$, calculated for a temperature of $T$=800 K. We show the exact solution using the same parameters as before (Fig. 4.1), as well as the approximations of Eqs. (4.9) and (4.10).

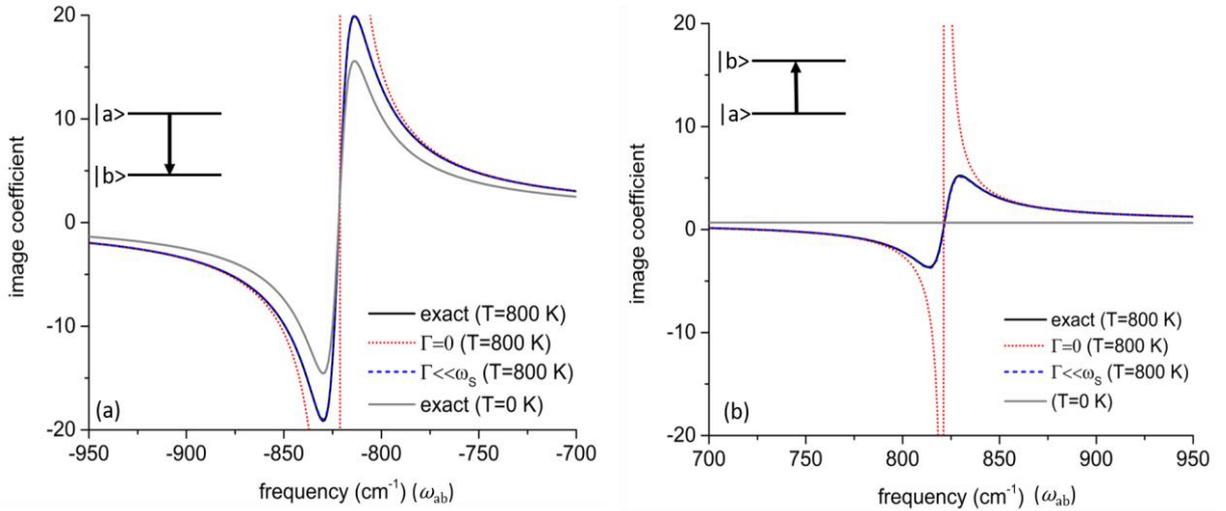

**Fig. 4.2** Plot of the image coefficient as a function of the transition frequency $\omega_{ab}$ for negative (a) and positive (b) frequencies for a temperature of $T$=800 K. We assume that the dielectric constant is given by Eq. (4.7) and use the same parameters as those in Fig. 4.1. The black line is the exact numerical solution, the blue dashed line makes the approximation of Eq. (4.9) ($\Gamma \ll \omega_S$) and the red dotted line assumes that $\Gamma = 0$. The blue and black lines are practically indistinguishable. We also show the exact calculation for T=0 K. For downward couplings the resonant behaviour is still presented at T=0 K due to spontaneous emission.

It is useful to write the temperature dependent part of the image coefficient ignoring any terms that survive at zero temperature:

$$\Delta r(\omega_{ab}, T) \cong \frac{4\omega_{ab}\omega_S(S_o - S_\infty)[\Gamma^2 + 2(\omega_{ab}^2 - \omega_S^2)]}{[\Gamma^2 + 2(\omega_{ab}^2 - \omega_S^2)]^2 + 4\Gamma^2\omega_S^2}\,n(\omega_S, T) \quad (4.11)$$

Which in the case of $\Gamma \to 0$ gives the simple expression:

$$\Delta r(\omega_{ab}, T) = \frac{2\omega_{ab}\omega_S(S_o - S_\infty)}{(\omega_{ab}^2 - \omega_S^2)} n(\omega_S, T) \quad (4.12)$$

The most intriguing aspect of the above equations is that they only involve the number of photons in the polariton frequency $n(\omega_S, T)$ whereas the number of photons at the transition frequency is irrelevant. Additionally, the temperature dependence is sensitive to the 'difference' between the transition and the polariton frequencies. Stronger effects are generally to be expected for resonant conditions ($\omega_{ab} \cong \omega_S$, but not exactly at $\omega_{ab} = \omega_S$) with a sign that flips when the polariton frequency is blue or red detuned with respect to the transition frequency (see Fig 4.2). From Eqs. (4.11) and (4.12) we can see that the Casimir-Polder interaction of atoms or molecules against metallic surfaces should also be insensitive to temperature as metals have plasmonic resonances on the UV side of the spectrum [37].

It is also useful to relate the above results to the effects of surface thermal emission on the atomic properties. Generally the black body radiation spectrum $I_{BB}$ is given by $I_{BB} = \hbar \omega n(\omega, T) \rho(\omega)$, where $\rho(\omega)$ is the density of states of the electromagnetic field. In the near field, the thermal spectrum of dielectrics is 'coloured' by surface modes. In this case, the density of states depends on position (local density of evanescent states) which according to [15] is given by:

$$\rho_{ev}(\omega, z) = \frac{1}{8\pi^2 \omega z^3} Im[S(\omega)] \quad (4.13)$$

Using Eq. (4.13) one can calculate the a.c. Stark shift induced on the atomic two level system, $|a\rangle \to |b\rangle$ by the thermal emission of the surface. In the case of $\Gamma \to 0$ the calculations can easily be performed analytically and they reproduce the energy shift predicted by Eq. (4.12). Calculations of the (far-field) black body radiation induced Stark shift for many energy levels of alkali atoms [38] was performed by Farley and Wing [39] in a similar fashion.

According to the above considerations, the near field temperature dependence of the Casimir-Polder interaction finds a simple transparent explanation. It is the a.c. Stark shift induced by the near field thermal emission of the surface. When the polariton frequency $\omega_S$ (centre frequency of the evanescent thermal field) is red detuned with respect to the atomic transition, $\omega_{ab} > \omega_S$, the atom is more strongly attracted towards the surface (increase of the $C_3$ coefficient with temperature). When the polariton frequency is blue detuned the attraction is reduced and can even turn to repulsion (decrease of the $C_3$ coefficient with temperature). For a downward dipole coupling (negative frequencies), the effects are reversed as can be seen in Fig. 4.2.

## Publications resulting from this work

The results and the reasoning presented in this chapter conveys my point of view on the subject of the near field Casimir-Polder temperature dependence, which is due to thermal excitation of the surface polariton modes of the surface. It is a result of many discussions that took place, over many years, in the extended group SAI (including many ex-members or outside visitors). The theoretical description of the Casimir-Polder shift as a function of temperature was published in 2006 [M-P. Gorza and M. Ducloy, EPJD, 2006] but the main results were probably derived much earlier than that. The analytical expressions of the Casimir-Polder shift in the case of a single polariton resonance was given in a later publication [A. Laliotis and M. Ducloy, PRA, 2015].

# Chapter 5: Casimir-Polder interaction at finite temperature: the experiments

The group SAI uses selective reflection to measure the Casimir-Polder interaction of excited state atoms. Selective reflection (SR) spectroscopy measures the reflection of a near resonant light (laser) beam on the interface between an atomic (molecular) vapour (gas) and a dielectric surface (window of a vapour cell), in principle transparent at the laser frequency. It is a spectroscopic technique that is essentially sensitive to dispersion, i.e the changes of the refractive index of the atomic medium around a resonance frequency. It is generally accepted in selective reflection theory that atomic collisions interrupt the interaction of the atoms with the electromagnetic field, making it necessary to account for a transient regime of interaction with the laser light [1, 2]. Collisions with the surface lead to an enhancement of the contribution of slow atoms that spend most of their lifetime (typical time of alignment of the atomic dipole to the electric field) close to the surface at distances where the field phase does not vary significantly ($z<\lambda/2\pi$). As such, slow atoms add a logarithmic contribution on the reflectivity, which can be greatly pronounced by taking the derivative of the signal, in practice translating to a frequency modulated (FM) detection [3]. Therefore, selective reflection is a high-resolution spectroscopic technique, linear with laser power that probes atoms that are typically at nanometric distances away from the reflecting surface. This combination (high resolution and nanometric probing) makes selective reflection an ideal technique for measuring atom-surface interactions, with the notable advantage: it essentially bypasses the extremely complicated problem of deterministically placing atoms close to the surface, which is, in one way or the other, encountered in most experimental measurements of the Casimir-Polder interaction. The drawback is that one has no control of the probing depth apart from changing the detection wavelength.

The basic theory of selective reflection including atom-surface interaction was first described in [4]. It was shown that atom-surface interactions can have a profound impact on the selective reflection spectra. When plotted as a function of normalised frequency $\Delta = \frac{2(\omega - \omega_o)}{\gamma}$, the FM selective reflection lineshape depends exclusively on a dimensionless parameter $A = \frac{2C_3}{\gamma z_{SR}^3}$, where $\omega$ is the laser frequency, $\omega_o$ the transition frequency, $\gamma$ the transition linewidth, $z_{SR} = \frac{1}{k} = \frac{\lambda}{2\pi}$ is the characteristic selective reflection probing depth and $C_3$ is the van der Waals coefficient. The parameter A represents the ratio between the van der Waals shift at the characteristic probing depth divided by the spectral resolution of the experiment defined by the transition linewidth. The above considerations are valid under certain assumptions. Firstly, the Doppler shift $ku_p$ (where $k$ is the wave vector and $u_p$ the most probable velocity) is considered much larger than the transition linewidth $\gamma \ll ku_p$. The effects of Doppler broadening on the FM selective reflection spectra can be subsequently added as a *correction*, which is however, calculated for the A = 0 case. Secondly, the atom-surface interaction is considered to be in a pure van der Waals regime $\delta F = -\frac{C_3}{z^3}$, where $\delta F$ is the *difference* between the energy shifts of the upper and lower probed atomic states.

To extract the van der Waals coefficient from experiments we fit the measured SR spectra to a library of theoretical curves calculated for different values of the parameter A. The experimental spectrum is compared with a theoretical curve, of a given A, by dilating the frequency axis, to obtain $\gamma$. We also allow for an overall offset of the spectrum and adjustment of the global amplitude of the theoretical curve to reflect experimental instabilities. We then change the dimensionless A value and repeat the process until the best fit is identified. This provides a measurement of the $C_3$ coefficient and the transition linewidth $\gamma$. In most cases we probe multiple (partially resolved) hyperfine components of an atomic transition. Linearity of SR spectroscopy imposes the relative amplitudes and the frequency spacing of the hyperfine components of a given manifold to their theoretical values, allowing us to reconstruct seeming complicated lineshapes. Finally, we also allow for a global pressure shift (the same for all hyperfine components).

The first SR experiments were done on the D2 line of cesium [5] and subsequently on the second cesium resonance $6S_{1/2} \rightarrow 7P_{3/2}$ at 455 nm [6, 7]. At first, the theoretical curves were simply compared, rather than fitted to the experimental spectra. The effects of the atom-surface interaction are so striking that a simple comparison usually suffices. The automated fitting process, developed subsequently adds nevertheless to the accuracy of the $C_3$ measurement and to the prestige of the method. The possibility of coupling atomic resonances to surface polariton was theoretically investigated in [9] and the experiments demonstrating surface repulsion were performed a few years later [10]. Around that time, the possibility of experimentally demonstrating near field thermal effects was also recognized [10, 11]. In what follows, I will describe the experimental efforts of the SAI group towards this direction, which took more than a decade to bear fruit.

## Experiments with fluorine surfaces

Selective reflection measurements are performed on vapour cells that are very typically made of glass (BK7 or silica). Glass reacts with caesium vapour at elevated temperatures, around 200-400 C depending on the type of glass. This makes it a poor candidate to study thermal effects. Sapphire cells on the other hand can usually be heated at elevated temperatures without any sign of degradation or reactivity of the surface with the aggressive alkali metal vapour. Sapphire, however, has an isolated resonance at $\lambda \sim 12.35$ µm, suggesting that temperatures about ~1200 K are required for thermally exciting polaritons on the sapphire surface. Working with sealed vapour cells at high temperatures can be technologically challenging and therefore at the initial stages of the experiment, the group oriented its efforts into finding alternative dielectrics with surface resonances further in the mid infrared, where surface waves can be excited at milder experimental conditions. The surface resonances of a long list of dielectrics is given in [12]. Eventually it was decided that the best candidates for our experiments are $CaF_2$ and $BaF_2$, dielectrics with a large transparency window in the visible and near infra-red that have resonances at 24 µm and 35 µm respectively, in partial coincidence (resonance) with the $8P_{3/2} \rightarrow 7D_{5/2}$ dipole coupling of caesium at 36 µm. The basic idea of the experiment was to probe by selective reflection spectroscopy the third resonance of caesium $6S_{1/2} \rightarrow 8P_{3/2}$ at 387.7nm on a $CaF_2$ of $BaF_2$ window and measure the van der Waals coefficients of the $Cs(8P_{3/2})$-$CaF_2$ interaction as a function of temperature. A schematic of the caesium energy levels involved in this experiment is shown in Fig. 5.1(a). The $C_3$ coefficient is a sum of all the allowed dipole contributions from the $8P_{3/2}$ level, but, as can be seen in Table 5.1, it is largely dominated by the $8P_{3/2} \rightarrow 7D_{5/2}$ coupling, which is resonant with the $CaF_2$ and $BaF_2$ polaritons. Therefore, a strong thermal dependence of the van der Waals coefficient is expected even for modest temperatures.

Theoretical predictions of the van der Waals coefficient requires knowledge of the dielectric constant of bulk material, at a wide range of frequencies. Additionally, for experimental measurements performed at elevated temperatures one has to take into account the effects of temperature on the dielectric constant itself, notwithstanding the QED effects described in Chapter 4. For this purpose, the group performed dedicated measurements of $CaF_2$ and $BaF_2$ dielectrics in collaboration with the group of Patrick Echegut and Domingos Meneses in the University of Orleans (CEMHTI-UPR3079). The experiments rely on a reflectivity measurement on the dielectric-vacuum interface using a broadband source and Fourier Transform Spectroscopy. The measured frequency range is typically between 100cm$^{-1}$ to 2000cm$^{-1}$ with the resolution of a few cm$^{-1}$ (1cm$^{-1}$ ~ 30GHz). The windows are heated on a hotplate at temperatures up to 500 C. The reflectivity data is then fit to the following semi-quantum model of the dielectric constant:

$$\varepsilon(\omega) = \varepsilon_\infty + \sum_j \frac{f_j \Omega_j^2}{\Omega_j^2 - \omega^2 - i\Gamma_j(\omega)\omega} \quad (5.1)$$

It is similar to the classic model used in Eq. (4.7) with a damping constant that is a complicated function of frequency, due to multiphonon processes in the crystal. Eq. (5.1) has been extensively used in the group of Orleans and reproduces the experimental reflectivity data [13] more faithfully than the classical model, albeit with a significant increase of the fitting parameters. In the case of $CaF_2$ and $BaF_2$, both materials with a single resonance, the damping constant is a sum of 6 and 4 'extended-Gaussian' functions (see [13] for more details) requiring the use of 18 and 12 fitting constants respectively.

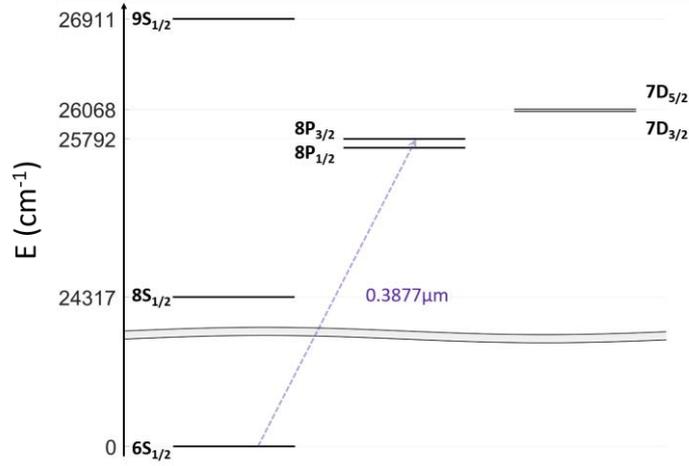

**Fig. 5.1** Relevant caesium energy levels for the experiments with fluorine surfaces. The transition probed with selective reflection spectroscopy is the third resonance of caesium ($6S_{1/2}\rightarrow 8P_{3/2}$) at 387.7nm.

(a) CaF₂

| Cs(8P$_{3/2}$) | $\lambda$ (µm) | $C_3$ (perfect reflector) | $C_3$ ($T=0$) | $C_3$ ($T=200$) | $C_3$ ($T=400$) | $C_3$ ($T=600$) | $C_3$ ($T=800$) | $C_3$ ($T=1000$) |
|---|---|---|---|---|---|---|---|---|
| 8S$_{1/2}$ | −6.78 | 12.07 | 2.17 | 1.99 | 1.27 | 0.38 | −0.56 | −1.52 |
| 7D$_{3/2}$ | 39.05 | 5.32 | 3.11 | 2.92 | 1.93 | 0.69 | −0.63 | −1.99 |
| 7D$_{5/2}$ | 36.09 | 37.79 | 21.8 | 19.98 | 11.19 | 0.19 | −11.5 | −23.48 |
| 9S$_{1/2}$ | 8.94 | 11.63 | 5.19 | 5.44 | 6.41 | 7.61 | 8.88 | 10.19 |
| 8D$_{5/2}$ | 4.92 | 3.7 | 1.51 | 1.55 | 1.71 | 1.9 | 2.09 | 2.3 |
| Total | | 73.71 | 34.94 | 33.04 | 23.73 | 12.05 | −0.37 | −13.09 |

(b) BaF₂

| Cs(8P$_{3/2}$) | $\lambda$ (µm) | $C_3$ (perfect reflector) | $C_3$ ($T=0$) | $C_3$ ($T=200$) | $C_3$ ($T=400$) | $C_3$ ($T=600$) | $C_3$ ($T=800$) | $C_3$ ($T=1000$) |
|---|---|---|---|---|---|---|---|---|
| 8S$_{1/2}$ | −6.78 | 12.07 | 3.14 | 2.86 | 2.09 | 1.23 | 0.34 | −0.56 |
| 7D$_{3/2}$ | 39.05 | 5.32 | 3.01 | 0.71 | −5.84 | −13.19 | −20.77 | −28.43 |
| 7D$_{5/2}$ | 36.09 | 37.79 | 21.07 | −11.44 | −105.46 | −211.34 | −320.54 | −431.09 |
| 9S$_{1/2}$ | 8.94 | 11.63 | 5.13 | 5.49 | 6.49 | 7.62 | 8.79 | 9.97 |
| 8D$_{5/2}$ | 4.92 | 3.7 | 1.52 | 1.58 | 1.75 | 1.94 | 2.13 | 2.33 |
| Total | | 73.71 | 35.06 | 0.42 | −99.69 | −212.4 | −328.64 | −446.33 |

**Table. 5.1** (from ref [11]).Contribution of each dipole coupling to the $C_3$ van der Waals coefficient between Cs(8P$_{3/2}$) and (a) CaF2, (b) BaF2 at different temperatures. $C_3$ is measured in kHz µm³ and the temperature in Kelvin. The negative sign denotes a downward transition. The $C_3$ value, given by the sum of each individual contribution, is also shown at the end of each table. The contributions to $C_3$ were calculated using measurements of the dielectric constant [13] at room temperature.

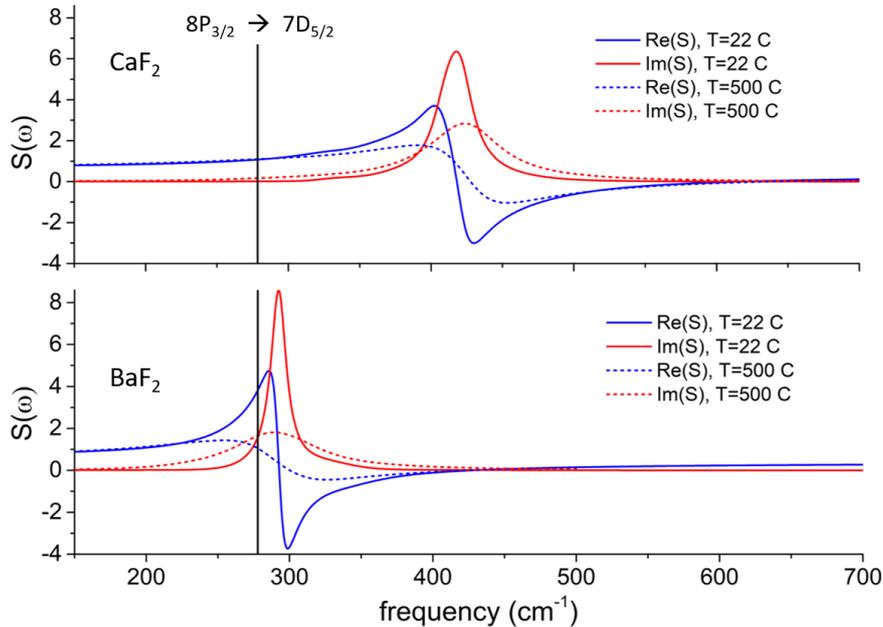

**Fig. 5.2** Real (blue) and imaginary (red) rats of the surface response S at room temperature (solid lines) and at T=500 C (dashed lines) for CaF₂ and BaF₂. The position of the 8P$_{3/2}$→7D$_{5/2}$ dipole coupling at 277 cm⁻¹ (~36 µm) is shown with a solid black vertical line.

In Fig.5.2 we show the real and imaginary parts of the surface response $S(\omega)$ for CaF$_2$ and BaF$_2$ dielectrics as extracted from the experiments. For clarity, we show the position of the dominant $8P_{3/2} \rightarrow 7D_{5/2}$ coupling with a vertical dashed line, which lies on the wings of the CaF$_2$ and in the heart of the BaF$_2$ resonance. Table 5.1 shows the exact temperature dependence of the most important dipole couplings for both dielectrics. The total temperature dependence of the C$_3$ coefficient of the Cs($8P_{3/2}$)-CaF$_2$ as well as the Cs($8P_{3/2}$)-BaF$_2$ interaction is shown in Fig. 5.3. A substantial decrease of the C$_3$ coefficient is predicted in the CaF$_2$ case, whereas for BaF$_2$, a spectacular dive of the coefficient is expected leading to strong repulsion even at room temperatures.

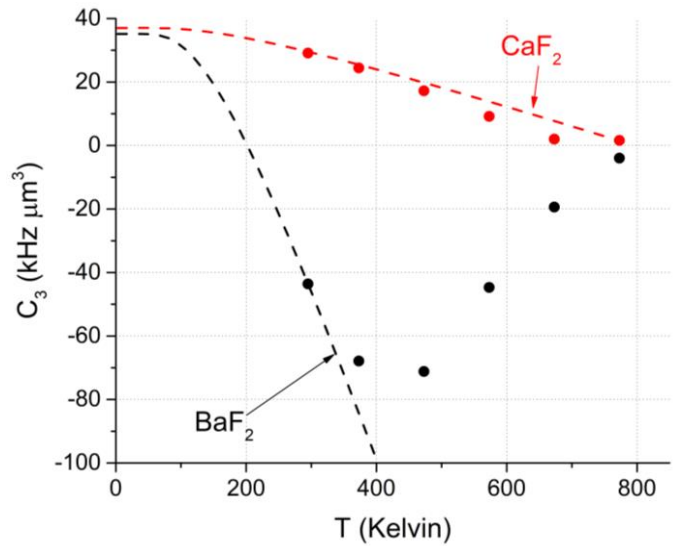

**Fig. 5.3** Theoretical predictions for the Van der Waals, C$_3$, coefficient as function of temperature for CaF$_2$ (red) and BaF$_2$ (black). The dashed lines show the predictions ignoring the temperature dependence of the dielectric constant itself and accounting solely for QED effects. The points show the theoretical predictions using the experimental measurements of the dielectric constants given in ref. [13].

## Selective reflection measurements on fluorine surfaces

Although fluorine surfaces appear to be favourable for measuring thermal effects on the Casimir-Polder interaction their use in vapour cell fabrication is limited. Sealed vapour cells are most commonly made out of glass which can be moulded at reasonable temperatures (~500 C). Using other dielectrics usually requires gluing (joint) between the dielectric in question and glass, which can then be melted to seal the cell. All-sapphire cells have been made in the group of D. Sarkisyan in Armenia bypassing altogether the use of glass but sapphire windows need to be glued to a sapphire (alumina) body. Gluing fluorine windows to sapphire or glass at elevated temperatures results in cracking due to the difference in thermal expansion coefficients. To overcome this problem, the two groups (in Paris13 and Ashtarak, Armenia) conceived a highly unconventional vapour cell design shown in Fig. 5.4.

The cell is T-shaped with a core made out of an alumina body with a sapphire and YAG window glued at each side of the main cylindrical core. A hole is drilled in the main body of the cell, onto which the sidearm is attached. The other side of the sidearm is glued (joint) to a glass tube that eventually seals the cell. Inside the cell and in close contact with the YAG window there is a CaF$_2$ tube (approximately 5 cm long). Selective reflection experiments are performed on the CaF$_2$-vapour interface. To avoid any parasitic signals from caesium vapour trapped in the gap between CaF$_2$ tube and YAG window (about 100 μm wide), this part of the cell is maintained in low temperatures using a Peltier element. Attempts to fabricate a cell with a BaF$_2$ tube were unsuccessful due to chemical attack of the aggressive caesium vapour, resulting in a rapid degradation of the dielectric's transparency. The cell is heated by three separate and independent ovens: One surrounding the sapphire window, one surrounding the CaF$_2$-vapour interface and finally, one that controls the sidearm temperature and therefore the vapour density inside the cell.

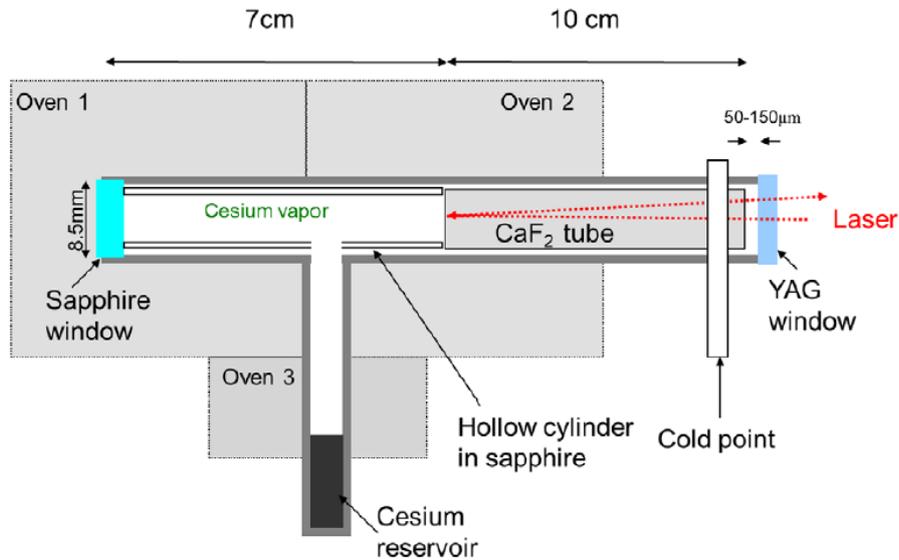

**Fig. 5.4** (from ref. [14]) Schematic of the caesium vapour cell containing a $CaF_2$ tube.

The first $CaF_2$ cell arrived in the University of Paris13 around the same time as myself (September 2005). Our initial characterisation experiments quickly revealed the presence of a significant amount of impurities due to insufficient outgasing that hindered the observation of narrow linewidth signals. A series of characterisation experiments on the D1 line of caesium resulting in a simultaneous measurement of the $Cs(6P_{1/2})$-sapphire and $Cs(6P_{1/2})$-$CaF_2$ atom-surface interaction was published in [15]. However, the main experiment on the third resonance of caesium at 388nm proved much more difficult to tackle in a contaminated cell, due to an extremely poor signal to noise ratio. Eventually the group decided to fabricate a new cell, identical to the first, only this time without the unwanted impurities.

Selective reflection experiments were performed on the $6S_{1/2} \rightarrow 8P_{3/2}$ line (388.7 nm) and eventually on the $6S_{1/2} \rightarrow 8P_{1/2}$ line (388.8 nm) of caesium on both $CaF_2$ and sapphire interfaces. Initially we used an extended cavity laser with a UV diode, whose power was limited and frequency stability was rather dubious. Eventually, this source was replaced by an amplified, frequency-doubled 780 nm laser diode with a final output power of approximately 100 mW at UV wavelengths. The source with the UV diode was frequency modulated by applying a voltage on the piezoelectric element attached to the grating of the extended cavity laser. The frequency-doubled source was modulated by double-passing the beam through an acousto-optic modulator. A saturated absorption was performed in a slightly heated (~80 C) sapphire vapor cell. Additionally, we used a stable Fabry–Perot cavity with a free spectral range of 83 MHz as a frequency marker. These auxiliary experiments allowed us to determine the absolute frequency of the laser throughout our scans with an accuracy of a few MHz. Soon after the start of selective reflection experiments, we noticed that cesium vapor had actually infiltrated the $CaF_2$ tube. This extremely curious effect that strongly suggests serious degradation of the dielectric's chemical composition was not followed by a degradation of the off-vapour resonance transparency (absorption was measured on vapour resonance). This allowed us to complete a long series of selective reflection experiments at various temperatures on the third resonance of cesium. The $C_3$ coefficient of the $Cs(8P_{3/2})$-sapphire interaction was measured to be ~65kHz $\mu m^3$ and practically independent of temperature, in consistence with measurements reported in [16]. The measurements on the 'supposed' $CaF_2$ interface are shown in Fig. 5.5. Although the van der Waals coefficient is in complete disagreement with theoretical predictions, the quality of the fits was satisfactory and the collisional shift and broadening extracted from the measurement was consistent with the ones found on the sapphire interface. We also attempted to fit our spectra using different atom-surface potentials [11]. This resulted to good quality fits, albeit with a collisional shift that was inconsistent with our measurements on the sapphire interface, as well as previously reported values in literature (~60MHz Torr$^{-1}$).

Finally, we mention briefly that saturated absorption measurements on the third cesium resonances revealed previously unobserved resonances that we studied, as a small spin-off project, in the idle moments of waiting for new cells to arrive. We finally attributed these resonances to light-assisted photoassociated cesium dimers at almost room temperatures (T~80 C) [17].

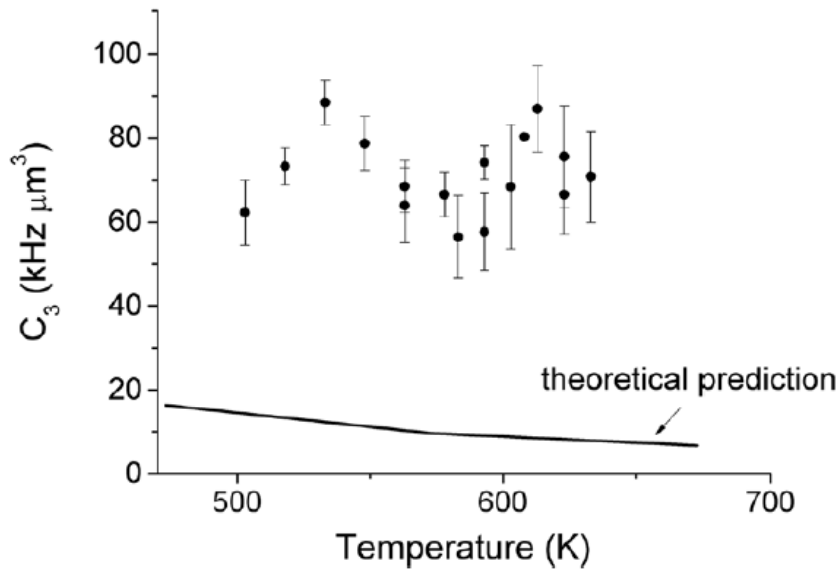

**Fig. 5.5** (from ref. [14]) $C_3$ coefficient of the Cs($8P_{3/2}$)-CaF$_2$ interaction as a function of temperature.

## Experiments in sapphire cells

The failed attempts to measure the thermal Casimir-Polder interaction are a testimony to the difficulty of atom-surface interaction experiments that can critically depend on the state of the surface, which is extremely difficult to characterise, especially if it is in a sealed caesium cell. The group flirted with the idea of breaking the vapour cell to perform rigorous characterisation measurements of the CaF$_2$ tube but we never had the courage to see this vapour cell 'sacrifice' through, as at the time it was the only relatively high temperature cell available to us.

Instead, we oriented our efforts towards obtaining high temperature sapphire cells. This choice was dictated by the knowledge that sapphire is a robust and durable material that cannot be corroded by the chemically aggressive alkali metal vapour even at very elevated temperatures. The technology to fabricate all-sapphire high-temperature cells was developed in Ashtarak Armenia by the group of D. Sarkisyan but the Armenian group was initially reluctant to embark on the fabrication of a new sapphire cell. We thus borrowed an all-sapphire cell, also fabricated in Armenia, from the group of G. Pichler in Zagreb, Croatia. The cell (shown in Fig. 5.6(a)) consists of a 12cm long alumina tube with a sapphire windows glued on each side. The windows had no wedge (making separation of the two reflected beams difficult) and the surface quality was unknown. However, our initial tests inspired a great deal of optimism, sufficient to overcome the Armenian group's reluctance that finally agreed to fabricate a dedicated high temperature sapphire cell specially designed for our experiments. The cell, shown in Fig. 5.6(b), consists of an 8-cm-long cylindrical sapphire tube onto which two sapphire windows are glued. The primary window is 'super-polished' with the c-axis perpendicular to the surface. The average surface roughness was measured by the fabrication company to be 0.3 nm. The window has a small wedge that allows selection of the reflection from the vapour interface. The mineral gluing is capable to resist temperatures up to 1200 K. The secondary window is of a lower surface quality. The 7-cm-long sidearm, parallel to the cell's main body, is glued on a small hole drilled on the secondary window. Both cells were heated with three independent ovens (Fig. 5.6) that allowed us to control independently the caesium vapour pressure and the temperature of the windows.

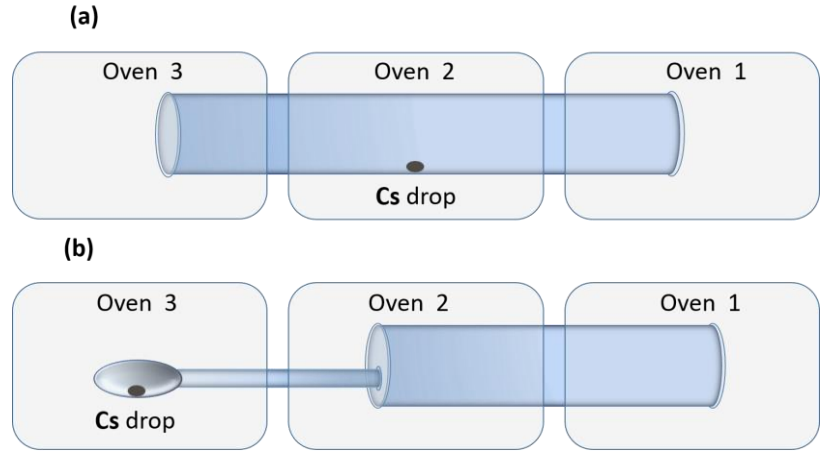

**Fig. 5.6** Schematic of the sapphire high temperature cells used in our experiments.

The Cs(8P$_{3/2}$) level does not have strong dipole couplings resonant with the sapphire polariton at ~12.35 µm (810 cm$^{-1}$) and the C$_3$ coefficient of the Cs(8P$_{3/2}$)-sapphire interaction is independent of temperature [13]. The caesium 7P$_{1/2}$➔6D$_{3/2}$ coupling at 12.15 µm (823 cm$^{-1}$) is strongly resonant with the sapphire [10] but at the time laser sources at 459 nm, necessary to probe the 6S$_{1/2}$➔7P$_{1/2}$ transition, were hard to find and relatively expensive compared to our available funds for the project. For this purpose, we decided to probe the Cs(7D$_{3/2}$) level whose van der Waals coefficient is largely dominated by the 7D$_{3/2}$➔5F$_{5/2}$ dipole coupling at 10.8 µm (924 cm$^{-1}$), that lies at the wings of the sapphire polariton resonance. The 7D$_{3/2}$ level is probed via selective reflection at the 6P$_{1/2}$➔7D$_{3/2}$ transition (672 nm) using a prior pumping step at 6S$_{1/2}$➔6P$_{1/2}$ (894 nm). In Fig. 5.7(a) we show the caesium energy levels relevant to this experiment and in Table 5.2 we show the contributions of the most important dipole couplings to the C$_3$ coefficient of the Cs(7D$_{3/2}$)-sapphire interaction. From Table 5.2 we can see that the 7D$_{3/2}$➔5F$_{5/2}$ dipole coupling carries almost all of the temperature dependence of the van der Waals coefficients. The other couplings have minor contribution, as there are not resonant with the sapphire polariton (Fig. 5.7(b). As the frequency of the surface wave (810 cm$^{-1}$) is red detuned compared to the atomic frequency (924 cm$^{-1}$) the thermally excited surface (evanescent) waves attract caesium atoms towards the surface resulting in an increase of the C$_3$ coefficient as a function of temperature, which can be seen in Fig. 5.6(c). In this experiment, high temperatures are required to thermally populate the sapphire polariton modes and observe significant thermal effects on the near field of the atom-surface interaction.

In Fig.5.7(b) we show the measured surface response $S(\omega)$ of sapphire (ordinary axis) at T=22 C and T=500 C, also measured by dedicated measurements in the University of Orleans [10]. The polariton resonance broadens and slightly shifts with increasing temperature but its wings (where the 7D$_{3/2}$➔5F$_{5/2}$ is positioned) are largely unaffected. We also show the surface response deduced from measurements performed almost 50 years ago by A. S. Barker [18]. Again, differences between the two measurements are only observed close to the polariton central frequency. Consequently, the theoretical predictions of the C$_3$ coefficient show little dependence on the measurement or the model used to describe the dielectric constant of the material (Fig. 5.7(c)). Sapphire birefringence should also be taken into account in the calculations of the van der Waals coefficient [19], but at the time of the experiment, the only available measurements of the extra-ordinary axis of sapphire were performed by A. S. Barker [18]. Nevertheless, taking sapphire birefringence into account has a small effect on our theoretical predictions. Further calculations show that the atom–surface interaction is independent of the actual axis orientation of sapphire to within 1%. Finally it is worth mentioning that a considerable uncertainty in the theoretical predictions stems from the uncertainty of the values of the transition probabilities which are essentially extracted from calculations the can show variations on the order of 5%. This uncertainty is represented by the grey shaded area in the plot of Fig.5.7(c).

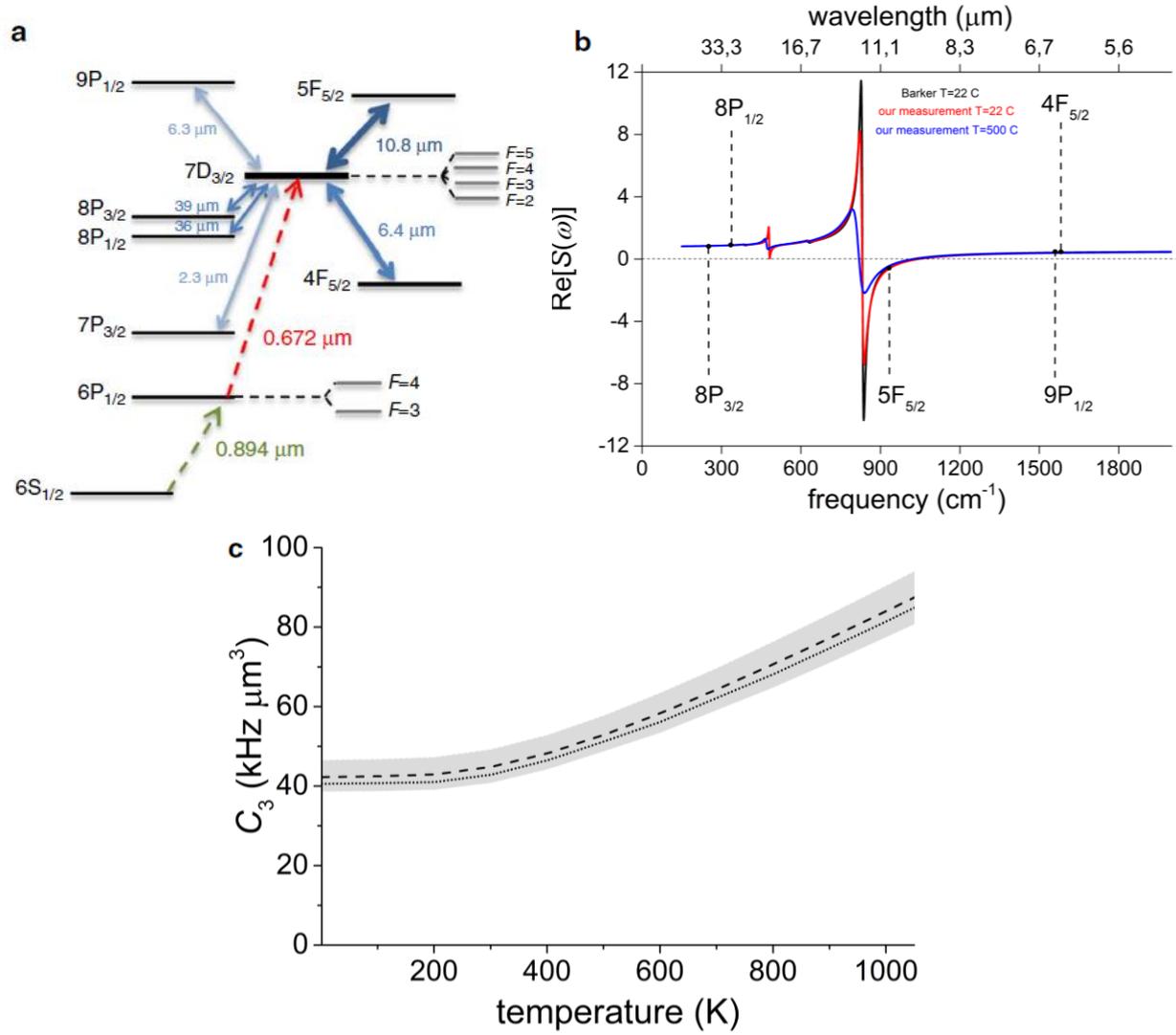

**Fig. 5.7** (a) Schematic diagram of the caesium energy levels relevant for the Cs(7D$_{3/2}$)-sapphire selective reflection experiment. (b) Real part of the image coefficient of sapphire as measured by A. S. Barker [18] at room temperature (black mine) and by our group in collaboration with the University of Orleans [13] at room temperature (red line) and at 500 C (blue line). (c) Theoretical predictions of the C$_3$ coefficient of the Cs(7D$_{3/2}$)-sapphire interaction using the dielectric constant of Barker (dashed line) and our measurements (dotted line). The grey shaded area represents a 5% error due to the uncertainty on the values of the transition probabilities.

**Table 1 | Contributions of individual atomic transitions.**

| Cs (7D$_{3/2}$) | $\omega$ (cm$^{-1}$) | $\lambda$ (µm) | C$_3$ (ideal conductor) | C$_3$ (T = 0 K) | C$_3$ (T = 200 K) | C$_3$ (T = 400 K) | C$_3$ (T = 600 K) | C$_3$ (T = 800 K) | C$_3$ (T = 1,000 K) |
|---|---|---|---|---|---|---|---|---|---|
| 7P$_{1/2}$ | −4,282.2 | −2.33 | 0.89 | 0.39 | 0.39 | 0.38 | 0.37 | 0.35 | 0.34 |
| 4F$_{5/2}$ | −1,575.4 | −6.35 | 3.37 | 0.61 | 0.6 | 0.52 | 0.4 | 0.21 | 0 |
| 8P$_{1/2}$ | −338.7 | −29.52 | 11.49 | 12.09 | 12.09 | 12.42 | 12.89 | 13.53 | 14.17 |
| 8P$_{3/2}$ | −256.1 | −39.05 | 5.32 | 5.07 | 5.06 | 5.14 | 5.27 | 5.44 | 5.61 |
| 5F$_{5/2}$ | 923.7 | 10.83 | 36 | 23.2 | 23.52 | 28.54 | 37.64 | 48.91 | 61.5 |
| 9P$_{1/2}$ | 1,589.4 | 6.29 | 1.41 | 0.86 | 0.86 | 0.89 | 0.96 | 1.04 | 1.13 |
| Total | | | 59.6 | 42.8 | 43.1 | 48.5 | 58.2 | 70.2 | 83.4 |

The values of the global C$_3$ coefficient, in kHz µm$^3$, as well as the individual contributions of the most important dipole couplings for different temperatures. The C$_3$ values for an ideal conductor are also given. Here we use the values of the dielectric constant as measured at T = 300 K, ignoring the effects of temperature on the dielectric properties of sapphire. These are analysed in detail in ref. 39, but in our case they are negligible.

**Table 5.2** (from ref. [20]) Contributions of individual dipole coupling to the van der Waals coefficient.

# Selective reflection experiment on the $6P_{1/2} \rightarrow 7D_{3/2}$ transition

The experimental set-up used for this experiment is shown in Fig.5.7. The selective reflection probe is a 672 nm laser tuned on the $6P_{1/2} \rightarrow 7D_{3/2}$ transition. Atoms are pumped on the $6P_{1/2}$ level using a strong pump laser resonant with the $6S_{1/2} \rightarrow 6P_{1/2}$ transition at 894 nm. The pump is a distributed Bragg reflector (DBR) laser, locked on a saturated absorption slope of the $6S_{1/2} \rightarrow 6P_{1/2}$ transition after passing through an acousto-optic modulator that allows us to switch it on and off at 10kHz (AM modulation). Its total power before reaching the selective reflection cell is about 20mW, which is focused on a beam waist size of about 800 µm. The pump is usually absorbed within a few tens of microns within the cell depending on the exact intensity and caesium vapour density. The selective reflection laser is an extended cavity laser. It was made using the box of an old Toptica DL100 and replacing the grating and the laser diode. Its total power output at 672 nm was 10-15mW. Its frequency was modulated at 1kHz with an excursion of about 5MHz, using the piezoelectric element attached on the grating. The laser stability was rather poor and the frequency drifted significantly in the course of a few seconds, making it difficult to observe undistorted narrow signals. Eventually, it was locked on the slope of a Doppler (linear) absorption profile (after FM demodulation) and its frequency was scanned inside the Doppler width by adding an offset voltage to the error signal. This technique limited the available scanning range but did wonders for the frequency stability of the laser and the overall quality of the scans. An auxiliary saturated absorption set-up and a 60 cm long Fabry-Perot cavity (with a bad finesse that resulted in a sinusoidal variation of the FP transmission with frequency) was used to provide absolute and relative frequency references that allowed us to reconstruct the frequency scale of our scans. Finally, the selective reflection beam of about 500 µm waist was superposed with the pump on the selective reflection cell. The beam power was low (~50µW) to avoid saturation of the atoms. The reflection was measured on a silicon photodiode and the signal was demodulated on the on the AM frequency, to clean it from any parasitic signals that did not originate from the atoms, and then on the FM frequency, to obtain the derivative of the selective reflection signal.

In our experiments we benefit from collisions to create an almost thermal population on the intermediate $6P_{1/2}$ level [10, 21, 22]. While our pump laser stays locked on resonance with one hyperfine component of the upper $6P_{1/2}$ state (e.g on the $6S_{1/2}(F=4) \rightarrow 6P_{1/2}(F'=3)$ transition), collisions redistribute the the $6P_{1/2}$ excitation to all velocities of both hyperfine components. This allows us to probe the $6P_{1/2}$ (F=4)$\rightarrow 7D_{3/2}(F'=3,4,5)$ transitions with a $6P_{1/2}$ population that has a quasi-thermal velocity distribution. This is very important for interpreting the selective reflection signals since the theory developed in [4] was derived under the assumption of a broad atomic velocity distribution. The same technique was also used in [8].

It is also worth mentioning that at one point we used two 672 nm lasers to perform our experiments. The first laser was scanned, as previously described, on the $6P_{1/2} \rightarrow 7D_{3/2}$ transition, while the second (newly bought DL100 from Toptica) was locked on a saturated absorption set-up and then frequency shifted by an acousto-optic modulator. A beat signal on a fast photodiode provided us with the absolute frequency of the scanned (selective reflection) laser. Unfortunately, weeks after the time-consuming installation of the experiment was completed, the shiny new Toptica laser broke down, leaving us high and dry and forced to work with our old homemade laser system. The few measurements we took during these weeks indicated that our frequency scale reconstruction from SA and FP references was satisfactory.

Our experimental protocol is the following: We perform selective reflection measurements at different window temperatures from 500 K to about 1000 K. At every window temperature, we change the caesium vapour density and therefore the linewidth of the transition due to the influence of atomic collisions. This changes the selective reflection spectra but not the van der Waals coefficient, which is independent of the atomic density and velocity. We also change the hyperfine manifold, either $6P_{1/2}$ (F=4)$\rightarrow 7D_{3/2}$ (F'=3,4,5) or $6P_{1/2}$ (F=3)$\rightarrow 7D_{3/2}$ (F'=2,3,4) with pumping on the $6S_{1/2}(F=4) \rightarrow 6P_{1/2}(F'=3)$ and $6S_{1/2}(F=4) \rightarrow 6P_{1/2}(F'=4)$ transitions respectively. Again, this changes the spectra (different relative amplitudes of the hyperfine components) but not the van der Waals coefficients. The experimental selective reflection spectra with their corresponding fits at the two extreme window temperatures are shown in Fig. 5.8. The quality of the fits is quite remarkable. After considering all our different measurements, the error bar in the measurement of the $C_3$ coefficient is about 10% [20].

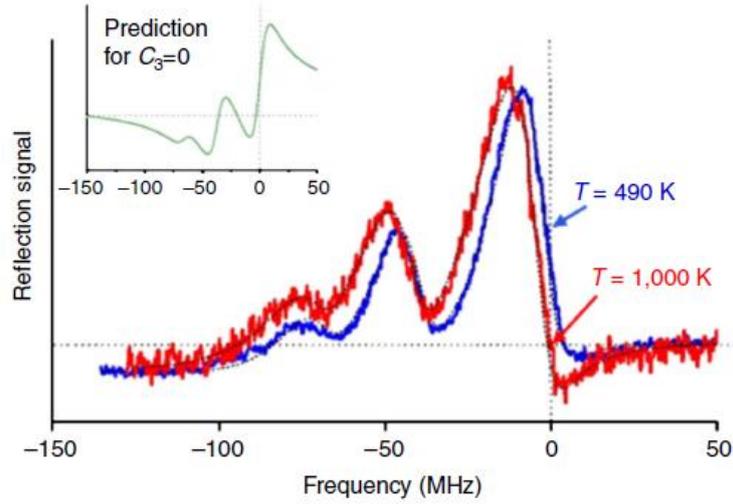

**Fig. 5.8** (from ref. [20]) Normalized frequency modulated selective reflection spectra on the $6P_{1/2}$ (F=4) - $7D_{3/2}$ (F'=3, 4, 5) transition for a window temperature at T=490 K (blue) and at T=1,000 K (red). The frequency axis is referenced to the hyperfine component of the F=4→F'=5 transition. In the inset, we show the predicted spectrum for $C_3$=0. The observed lineshape distortion of the experimental curves is evidence of the atom–surface interaction. The dashed black lines are the respective fits, demonstrating the same linewidth Γ=19MHz and an increase of $C_3$ with temperature from 55 to 86 kHz μm³.

In Fig. 5.9 we show the $C_3$ coefficient as a function of temperature as measured in both sapphire cells. Both measurements are in agreement with the theoretical predictions. A slight systematic difference between the two measurements could be an effect of the surface quality.

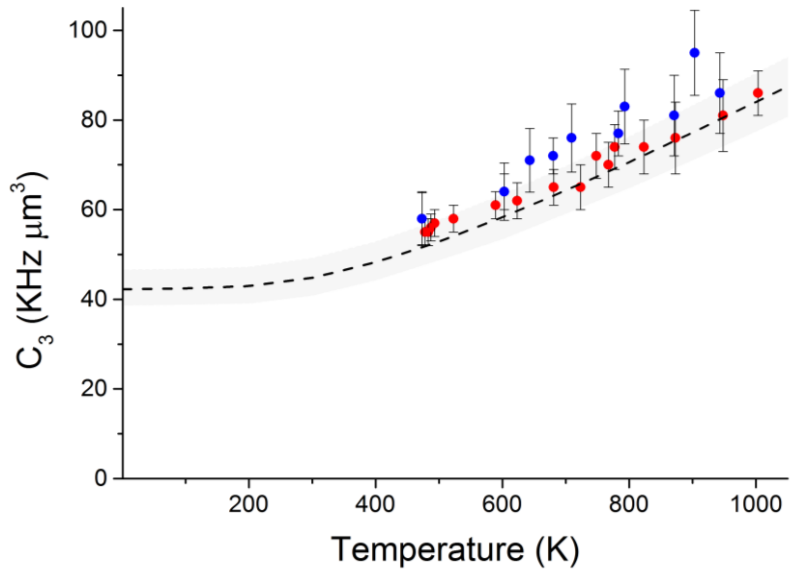

**Fig. 5.9** van der Waals coefficient of the Cs($7D_{3/2}$) sapphire interaction as a function of temperature. Red points represent measurements on a super polished window of the vapour cell in Fig.5.5(b) whereas blue points represent the measurements made on the old cell of Fig. 5.5(b) with windows of unknown quality. The dashed line represents the theoretical predictions using with the dielectric constant of ref. [18] and the grey shaded area of 5% plotted around the theoretical curve represents the uncertainty assigned to the transition probabilities.

## Strongly resonant atom-polariton coupling

The Cs($7D_{3/2}$) experiment, demonstrating near field thermal effects in the Casimir-Polder interaction, started towards the end of 2008 and the results were published in July 2014. During these five long years, we managed to acquire two laser modules emitting at 459 nm and 455 nm for probing the $6S_{1/2}$→$7P_{1/2}$ and $6S_{1/2}$→$7P_{3/2}$ transitions of the second caesium resonance. The van der Waals coefficient of these two levels is dominated by the $7P_{1/2}$→$6D_{3/2}$ dipole coupling at 12.15 μm (823 cm⁻¹) and by the $7P_{3/2}$→$6D_{5/2}$ coupling at 14.6 μm (685 cm⁻¹) respectively (Fig. 5.10(a)). The sapphire polariton frequency at 810 cm⁻¹ (12.35μm) is strongly resonant and slightly red detuned with respect to $7P_{1/2} \rightarrow 6D_{3/2}$ whereas it is relatively far blue detuned compared to $7P_{3/2} \rightarrow 6D_{5/2}$. This suggests a strong increase of the van der Waals coefficient with temperature for the Cs($7P_{1/2}$) level and a small decrease for the Cs($7P_{3/2}$) level. This radically different behaviour also demonstrates that thermal fields can tune the atom-surface interaction down to a complete suppression or even repulsion.

The resonant coupling between the $7P_{1/2} \rightarrow 6D_{3/2}$ transition and the sapphire polariton can also strongly modify the linewidth (lifetime) of the Cs($7P_{1/2}$) level due a direct absorption of energy from a thermally excited polariton, with a subsequent transfer to the Cs($6D_{3/2}$) level. This is a quantum analogue to near field heat transfer experiments, performed up to now solely with classical objects. The distance dependent linewidth, $\delta\gamma(z)$, given by Eq. (4.6) is essentially proportional to the imaginary part of the surface response divided by $z^3$. Since both shift and linewidth have an $z^{-3}$ distance dependence we can define the parameter $\Gamma_3 = \frac{\delta\gamma(z)}{2} z^3$ in analogy to the van der Waals coefficient $C_3$. $\Gamma_3$ can also be considered as an imaginary or dissipative part of the van der Waals coefficient [23]. The distance dependent linewidth also requires in principle a summation on all the allowed dipole couplings, however, it is only the very resonant $7P_{1/2} \rightarrow 6D_{3/2}$ coupling that carries weight in this calculation.

Accurate predictions of the $C_3$ and $\Gamma_3$ should take into account sapphire birefringence, especially for the Cs($7P_{1/2}$)-sapphire interaction. The first series of experiments in collaboration with the University of Orleans [13] only measured the dielectric constant of the ordinary sapphire axis. Additionally, the windows used for the measurement were not of similar quality to the super-polished window of our high temperature vapour cell. For this purpose, a new series of measurements was performed, again in collaboration with the University of Orleans, with superpolished windows at various temperatures up to ~1000 K, from which the dielectric constant of both ordinary, $\varepsilon_{ord}$, and extraordinary, $\varepsilon_{ext}$, axes was extracted. In Fig. 5.11 we plot the image response for various temperatures as extracted from our experiments, showing also the results of A. S. Barker [18] for comparison. In our calculations we use the geometric mean $\sqrt{\varepsilon_{ext}\varepsilon_{ord}}$ which is the parameter relevant in atom-surface interaction experiments when the extraordinary axis is perpendicular to the window surface [9, 19]. The sapphire resonance appears to be sensitive to window temperature and possibly to the exact sample fabrication conditions (impurities, humidity…), surface roughness as well as the theoretical model used for the dielectric constant. Fig.5.12(a) shows the theoretically predicted $C_3$ coefficient of the Cs($7P_{1/2}$) and Cs($7P_{3/2}$) interaction with sapphire. The temperature effects predicted using Barker's measurements are more pronounced, although one has to keep in mind that they were exclusively performed at room temperature. In Fig.5.12 (b) we show the $\Gamma_3$ coefficient as a function of temperature only for the Cs($7P_{1/2}$) level, as the effects predicted for Cs($7P_{3/2}$) are negligible.

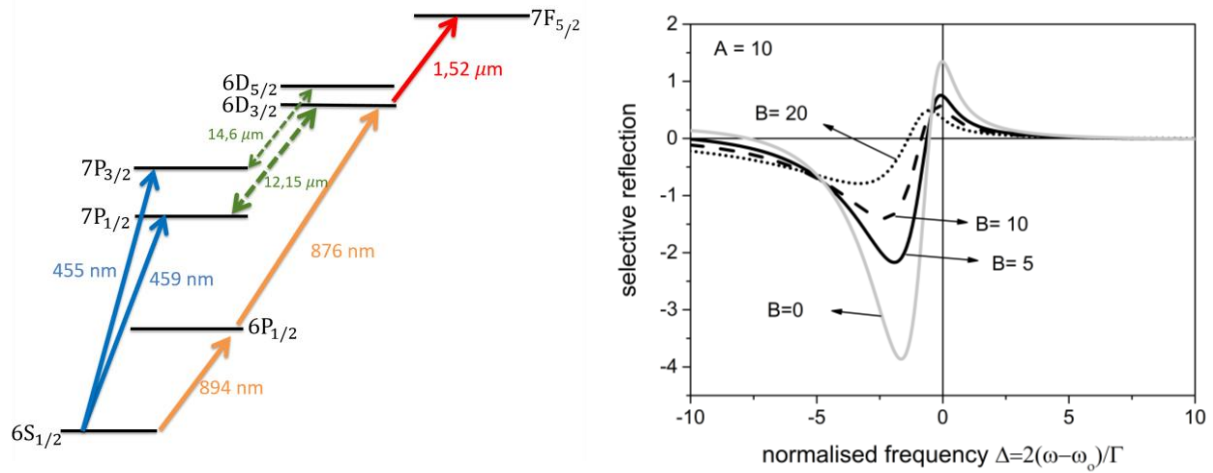

**Fig. 5.10** (a) Caesium energy level relevant for the Cs($7P_{3/2}$) and Cs($7P_{1/2}$) experiment (b) Theoretical curves for A=10 and different values of B.

The influence of the distance dependent linewidth on selective reflection spectra is taken into account by an additional dimensionless parameter $B = \frac{2\Gamma_3 k^3}{\gamma}$ (the imaginary part of A). When $A$ is positive, increasing the parameter B, for a given A, decreases the amplitude of the curve and increases its apparent linewidth but does not alter significantly the fundamental symmetry of the curve (Fig. 5.10(b)). This suggests that, contrary to A, the parameter B cannot be uniquely defined by the fit process and additional constraints on the amplitude and/or linewidth are required to extract the value of B. For this purpose, we decided to perform simultaneous measurements (identical atomic vapour density and window temperature) on the $6S_{1/2} \rightarrow 7P_{1/2}$ and $6S_{1/2} \rightarrow 7P_{3/2}$ transitions. Consequently, analysis of the $6S_{1/2} \rightarrow 7P_{3/2}$ spectra, where $B$ can safely be ignored, can constrain the amplitude used to fit the $6S_{1/2} \rightarrow 7P_{1/2}$ spectra.

The selective reflection experiments started in September 2014 and completed approximately two years later. We used two Toptica laser diodes of the DL100 type emitting about 10mW of power, FM modulated with an excursion of ~1-2MHz at frequencies of ~1kHz. As previously a Fabry-Perot cavity and an auxiliary saturated absorption reference is used to calibrate the frequency scale. Selective reflection measurements were performed at different temperatures, different vapour pressures and different hyperfine level manifolds for both $6S_{1/2} \rightarrow 7P_{1/2}$ and $6S_{1/2} \rightarrow 7P_{3/2}$ transitions. The measured $C_3$ coefficient for Cs($7P_{1/2}$) and Cs($7P_{3/2}$) is shown in Fig.5.13. The fit quality was similar to the one achieved in the Cs($7D_{3/2}$) experiment. By assuming that B=0, the amplitude ratio of the two transitions was found to be equal to the theoretically predicted value and the collisional broadening was found to be very similar for both transitions, corroborating independent measurements of collisional shift and broadening performed in a thin cell by saturated absorption spectroscopy.

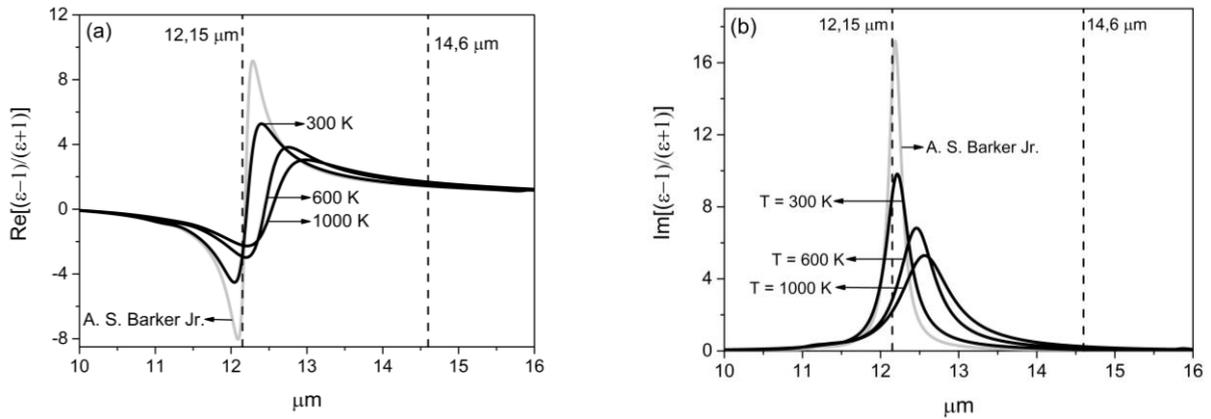

**Fig. 5.11** (from thesis of Joao Carlos de Aquino Carvalho) real (a) and imaginary part of the image response as measured by A.S. Barker [18] (grey lines) at room temperature and by the University of Orleans (black lines) at various temperatures. The frequency of the dipole couplings $7P_{1/2} \rightarrow 6D_{3/2}$ and $7P_{3/2} \rightarrow 6D_{3/2}$ are shown for comparison. Here ε is the geometric mean of the ordinary and extraordinary axes of sapphire.

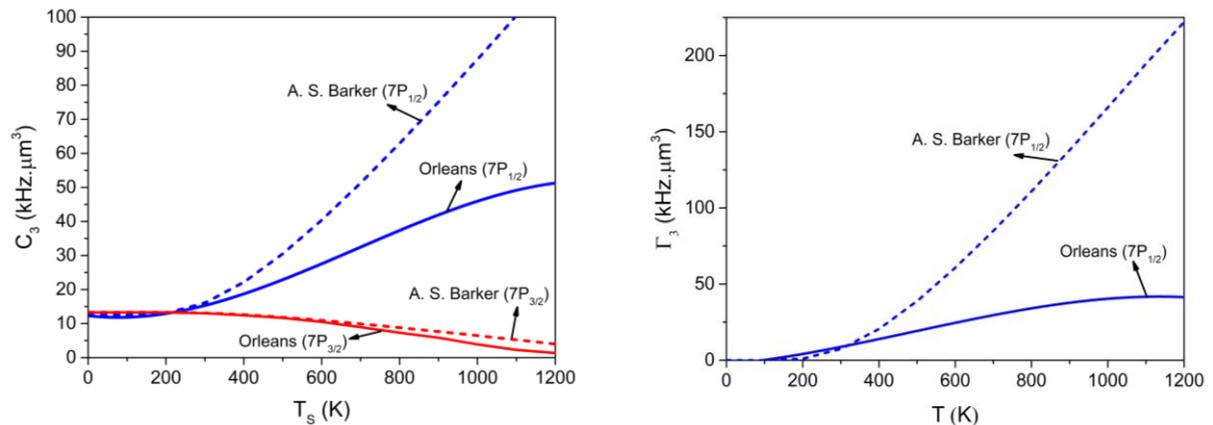

**Fig. 5.12** $C_3$ (a) and $\Gamma_3$ (b) coefficients for the Cs($7P_{1/2}$) (blue) and Cs($7P_{3/2}$) interaction with sapphire. The coefficients are calculated using the dielectric constant measured by A. S. Barker [18] and by the University of Orleans.

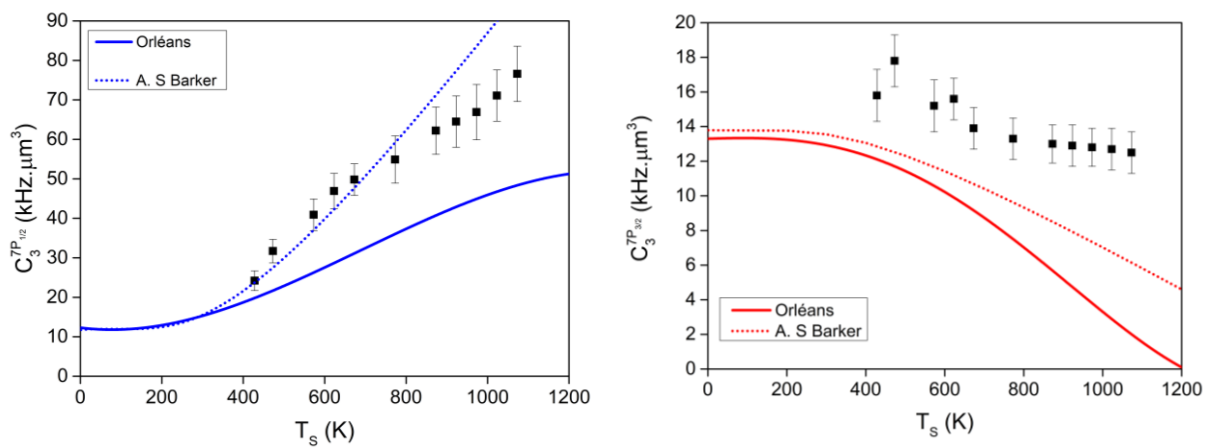

**Fig. 5.13** $C_3$ coefficients for the Cs(7P$_{1/2}$) (a) and Cs(7P$_{3/2}$) (b). Experimental measurements (points) are compared to the theoretical predictions (solid and dashed lines).

## Final thoughts and future of the near field energy transfer experiments

The surprising results of our experiments on the second caesium resonance suggest either that the values of $\Gamma_3$ are much smaller than the theoretical predictions or that some of our underlying assumptions in the interpretation of selective reflection spectra are here not valid. The first hypothesis essentially means the imaginary part of the surface response $Im[S]$ (at 823 cm$^{-1}$) is smaller than measured by broadband measurements of the reflectivity. This would be unexpected since the real part of the image response, $Re[S]$, seems to be in relative agreement with the theoretical predictions (van Waals coefficient shown in Fig. 5.13). Additionally, measurements on the caesium 6P$_{1/2}$→6D$_{3/2}$ line [10] were also 'compatible' with theoretical predictions, suggesting if anything that $Re[S]$ is larger than measured by reflectivity measurements (Barker and Orleans). The above considerations indicate that broadband measurements of the surface reflectivity are probably a limited experimental tool for obtaining information on the surface response function close to polariton resonances. Direct measurements of the near field thermal emission (imaginary part of the surface response) have been performed [24] but no data is available for sapphire. In this respect, our experiments can be seen as an alternative method for probing the real and imaginary part of the surface response or at least as a strenuous experimental test of existing measurements and experimental methods.

Our selective reflection experiments on the second caesium resonance do not require an intermediate pumping step. This suggests that in principle their interpretation should be straightforward as most of the main assumptions made in selective reflection calculations [4] (two-level system, uniform lower state population with distance …) appear to be valid. A question mark remains on the subject of the velocity distribution, which is considered constant (flat. The assumption is justified by the fact that mostly slow atoms contribute to the FM selective reflection signal and the contribution of atoms that travel large distances (much greater than $\lambda/2\pi$) within their lifetime essentially averages to zero. In the case of our experiments, however, the distance dependent lifetime decreases as $z^{-3}$ and it might be possible that even fast atoms spend most of their lifetime at nanometric distances away from the surface. Thus, solving the selective reflection integrals with a Maxwell-Boltzmann distribution might reveal interesting information related to our experiments. An additional benefit of this calculation is the possibility to estimate the effects of quantum friction (in both vertical and horizontal direction) that can be exalted in the presence of surface polaritons [25].

It is worth mentioning that the energy transfer from Cs(6D$_{3/2}$) atoms to sapphire polaritons (via 6D$_{3/2}$→7P$_{1/2}$ *emission* in a polariton mode) was demonstrated experimentally by direct detection of the Cs(7P$_{1/2}$) population with a selective reflection measurement on the 7P$_{1/2}$→10D$_{3/2}$ at 1.298μm [25]. For this purpose we have decided to implement a similar scheme to detect the near field population of Cs(6D$_{3/2}$) transferred by *absorption* of a thermally excited sapphire polariton. The basic principle of this pump-probe experiment is summarised in Fig. 5.10(a). The 459 nm laser excites atoms to the $7P_{1/2}$ and near field absorption of a polariton excitation further *pumps* the atoms to the $6D_{3/2}$ level. The near field $Cs(6D_{3/2})$ population is then *probed* by a selective reflection laser beam, superposed with the blue laser beam, tuned on the $6D_{3/2} \rightarrow 7F_{5/2}$ line at 1524nm. Naturally, population transfer from $7P_{1/2}$ to $6D_{3/2}$ can also happen via collisions. Reference experiments on the $6D_{5/2} \rightarrow 7F_{5/2}$ line (1.534µm) as well as experiments with a first excitation step on the $7P_{3/2}$ level can help us decouple the two mechanisms. The group has obtained an extended cavity diode laser covering wavelengths from 1500nm-1600nm but so far, we have not managed to see a reproducible and reliable signal on the $6D_{3/2} \rightarrow 7F_{5/2}$ line. Experiments are still underway.

## Personal contribution and publications resulting from this work

The experiments on the near field temperature dependence of the Casimir-Polder interaction started in 2005 around the time I arrived in the group as a post-doc. A first paper [A. Laliotis *et al.* Appl. Phys. B (2008)] describes the tests performed on the $CaF_2$ cell and the measurements of the $C_3$ coefficient of the $Cs(6P_{1/2})$ level against a sapphire and $CaF_2$ surface. The selective reflection experiments on the third resonance of caesium were part of the thesis of Thierry Passerat de Silans (2006-2009). A part of the experiments was performed while I was a post-doc at Imperial College (2006-2008). The measurements on a $CaF_2$ surface did not produce the expected results but the main findings were published several years later in [T. Passerat de Silans *et al.* Laser Phys. (2014)]. The additional resonances observed by saturated absorption on the third caesium resonance ($6S_{1/2} \rightarrow 8P_{3/2}$) were reported in [T. Passerat de Silans *et al.* Phys. Rev. A (2010)]. The first demonstration of near field temperature effects on the van der Waals coefficient of the $Cs(7D_{3/2})$-sapphire interaction published in [A. Laliotis *et al.* Nat. Commun (2014)] came with a lot of personal effort, as no PhD student was working on the experiment and the group was simultaneously putting a lot of efforts on the opal experiment. Joao Carlos de Aquino Carvalho started his thesis in September 2014 and he has performed the series of experiments mainly on the second resonance of caesium. Joao's efforts have been herculean and although the main goal of his thesis, the demonstration of near field energy transfer to atoms, remains elusive we believe that good publications will result from his work.

# Chapter 6: Beyond the van der Waals approximation: retardation effects and metasurfaces

Selective reflection measurements explore the near field regime of the Casimir-Polder interaction where the inverse cube approximation is considered to be valid. The validity of the van der Waals approximation is partly guaranteed by the fact that our experiments normally probe high lying excited states whose $C_3$ coefficients are dominated by low energy dipole couplings (mid infrared), whereas selective reflection probing is performed at visible or near infrared wavelengths (probing depth of $\lambda/2\pi$). In our work published in [1], we performed explicit calculations of the retarded Casimir-Polder potential of the Cs($7D_{3/2}$) state (see Fig.5 of [1]). Within the limits of our experimental accuracy, the van der Waals approximation is valid for distances as large as half a micron away from the surface, whereas the typical probing depth of the experiment is ~100nm.

For lower lying atomic states, things can be less straightforward. On one hand, the dipole couplings of lower lying states tend to be more energetic. On the other hand the $C_3$ coefficients are in general smaller making the contribution of the ground state $C_3$ coefficient (spectroscopy is sensitive to the difference between upper and lower state coefficients) more dominant. Ground state Casimir-Polder potentials almost exclusively depend on the non-resonant component (a QED term) of the atom-surface interaction, which plummets quickly to a $z^{-4}$ dependence, as discussed in Chapter 4. In contrast, excited state potentials are also sensitive to the resonant term (classical interaction of a dipole with its image) that displays a slowly decaying oscillatory behaviour ($\sim z^{-1} \cos \frac{4\pi}{\lambda} z$). An exact calculation of the difference of Casimir-Polder potentials between the Cs($6P_{1/2}$) and Cs($6S_{1/2}$) states (probed by a selective reflection experiments on the D1 line of caesium) was presented in [2]. Although the main point of [2] was to discuss thermal effects, the calculation revealed that the potential did not follow a $z^{-3}$ dependence. In Fig. 6.1 we show the effective $C_3^{eff}$ van der Waals coefficients for the Cs($6P_{1/2}$) and Cs($6S_{1/2}$) levels against a sapphire surface as well as their difference (inset of the figure) that represents the spectroscopically relevant quantity. Here, $C_3^{eff} = -\delta F \times z^{-3}$, where $\delta F$ is the Casimir-Polder potential energy [2-5] . We can see that at a probing depth of ~100nm, $C_3^{eff}$ is about 1.5kHz $\mu m^3$ instead of its theoretical value 1.1 kHz $\mu m^3$. Additionally, experimental selective reflection measurements (~1.4kHz $\mu m^3$), although performed in a vapour cell containing impurities, seemed to agree more with the former than the latter value. At the time, the result was unexpected and slightly counterintuitive since it suggested that retardation effects resulted in a slower rather than a faster decay of the spectroscopic Casimir-Polder potential as a function of distance, giving a larger rather than smaller spectroscopic van der Waals coefficient.

In October 2014 I spent one month in the Nanyang Technical University in Singapore working with Martial Ducloy and David Wilkowski on a new experiment aiming at performing selective reflection measurements on the interface between a caesium vapour and a nanostructured planar metamaterial (a kind of metasurface). One of the possible directions of this experiment was to probe the $6S_{1/2}$→$5D_{5/2}$ electric quadrupole transition in the hope of demonstrating enhancement of the transition probability next to the metasurface. It is during this period that we also started thinking about the Casimir-Polder interactions of the Cs($5D_{5/2}$) level with a dielectric surface. Eventually calculations demonstrated that the Cs($5D_{5/2}$) level has a small van der Waals coefficient, almost the same as the one of the ground state Cs($6S_{1/2}$). This means that a selective reflection experiment on a simple, flat dielectric (sapphire) surface would be sensitive exclusively to retardation effects of the difference between the Casimir-Polder potentials of the two states. The exact calculations that quantitatively demonstrate the above argument are also shown in Fig. 6.1. Knowing that probing electric quadrupole transitions (so called forbidden transitions) is easier said than done, I have searched for dipole couplings (of caesium and rubidium) that could be just as sensitive to retardation. It seems, however, that the $6S_{1/2}$→$5D_{5/2}$ transition at 685 nm is the most favourable for this type of experiment.

Retardation effects have been measured before [6] with ground state sodium atoms and in experiments with cold atoms that bounced off an evanescent blue detuned wave (atomic mirrors) [7, 8]. Initial experiments performed in the group of A. Aspect [7] were inconclusive, but similar experiments performed some 20 years later by C. Zimmermann, S. Slama and colleagues provided a measurement of the retarded Casimir-Polder potential of the rubidium ground state [8]. These experiments concern exclusively ground state atoms. Additionally, the oscillations of the excited state Casimir-Polder potential where shown using trapped ions at record distances ~20cm from a distant mirror [9]. Retardation effects in selective reflection experiments, examined here, are of somewhat different nature, as they concern the difference between ground and excited state potential at very small (nanometric) distances from the surface.

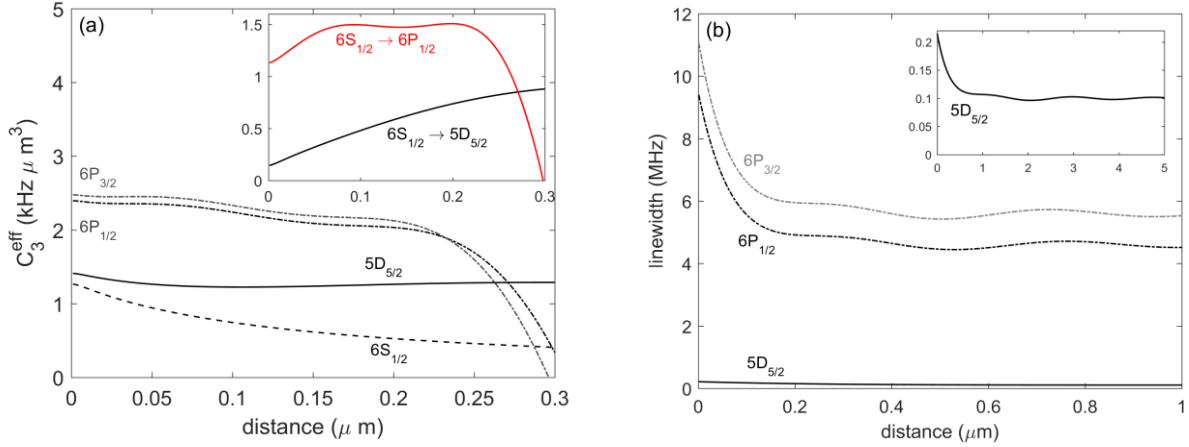

**Fig. 6.1.** (from ref. [3]) (a) Effective van der Waals coefficient for the cesium levels $6S_{1/2}$ (black dashed line), $6P_{1/2}$ and $6P_{3/2}$ (black and gray dash-dotted lines, respectively), as well as $5D_{5/2}$ (black solid line) against a sapphire surface. The inset shows the difference of the effective van der Waals coefficients for the $6P_{1/2} \rightarrow 5D_{5/2}$ (black solid line) and the $6S_{1/2} \rightarrow 6P_{1/2}$ (red solid line) transitions. (b) Distance dependent linewidth for three principal transitions $6S_{1/2} \rightarrow 6P_{1/2}$ (black dash-dotted line), $6S_{1/2} \rightarrow 6P_{3/2}$ (gray dash-dotted line) and $6S_{1/2} \rightarrow 5D_{5/2}$ (black solid line in the main figure and the inset).

Following the publication of [2] we started examining more closely the effects of retardation on selective reflection spectroscopy. The idea was to include the fully retarded Casimir-Polder potential in the calculation of selective reflection spectra. Selective reflection integrals are particularly difficult to tackle. In the simple case of the van der Waals interaction, a lot can be done analytically (some, but not all the analytic tricks of the calculation are reported in [10]). However, the Casimir-Polder interaction has no exact analytical expression and one has to resort to a fully numerical solution of the selective reflection integrals. A lot of this hard and tedious work was done in collaboration with Paolo Pedri. During our works on the problem, we discovered that while in the near field the distance dependence of the linewidth is usually significantly smaller than the atomic energy shift and in most cases can be safely ignored, this is not the case when one considers the complete Casimir Polder potential. In the far field, linewidth ($\delta\gamma$ /2) and shift ($\delta F$) oscillate with the same amplitude and frequency and a phase shift of $\pi/2$ (see Chapter 4). As such, ignoring the distance dependent linewidth in a fully retarded calculation has no realistic physical justification and can lead to erroneous results or even, in some cases, to divergent selective reflection integrals. For this purpose in Fig.6.1(b) we show the distance dependent linewidths of the $6S_{1/2} \rightarrow 6P_{1/2}$ and the $6S_{1/2} \rightarrow 5D_{3/2}$ transitions.

The details of our calculations are given in [3]. Here we will simply outline the main results of our analysis. In Fig. 6.2(a) we show simulated selective reflection spectra on the $6S_{1/2} \rightarrow 5D_{5/2}$ transition at 685 nm and similarly for the $6S_{1/2} \rightarrow 6P_{1/2}$ transition at 894 nm (Fig. 6.3(a)) for different vapour pressures and therefore transition linewidths (natural linewidth plus collisional broadening). The solid black lines represent the selective reflection spectra using the fully retarded Casimir-Polder (retarded SR spectra), the grey lines represent the selective reflection spectra assuming a van der Waals approximation (vdW SR spectra) and the dashed lines represent the best fits of the retarded spectra using an ad hoc van der Waals coefficient (SR fits). One easily notices that retarded SR and vdW SR spectra are significantly different. Surprisingly, however, the fits of the retarded SR spectra with an ad hoc $C_3$ coefficient are in most cases quite satisfactory making difficult to measure unequivocally the effects of retardation. In Fig. 6.2(b) and Fig.6.3(b) we show the evolution of the ad hoc $C_3$ coefficients with transition linewidth. We can clearly see that they are not constant, but as the transition linewidth increases, the ad hoc van der Waals coefficient approaches its theoretical value, whereas for narrow linewidths, SR seems to probe the Casimir-Polder interaction at a finite distance ~100nm. This phenomenon has a rather transparent interpretation: Due to big Casimir-Polder shifts, atoms that are very close to the surface, experience a large detuning parameter $\Delta$ (laser detuning divided by linewidth) that reduces their relative contribution to the SR spectrum. When the transition linewidth increases due to collisional broadening, $\Delta$ decreases, thus enhancing the contribution of atoms that are closer to the surface. Figs 6.2 and 6.3 suggest that retardation effects could manifest experimentally as a linewidth dependent ad hoc (fitted) van der Waals coefficient.

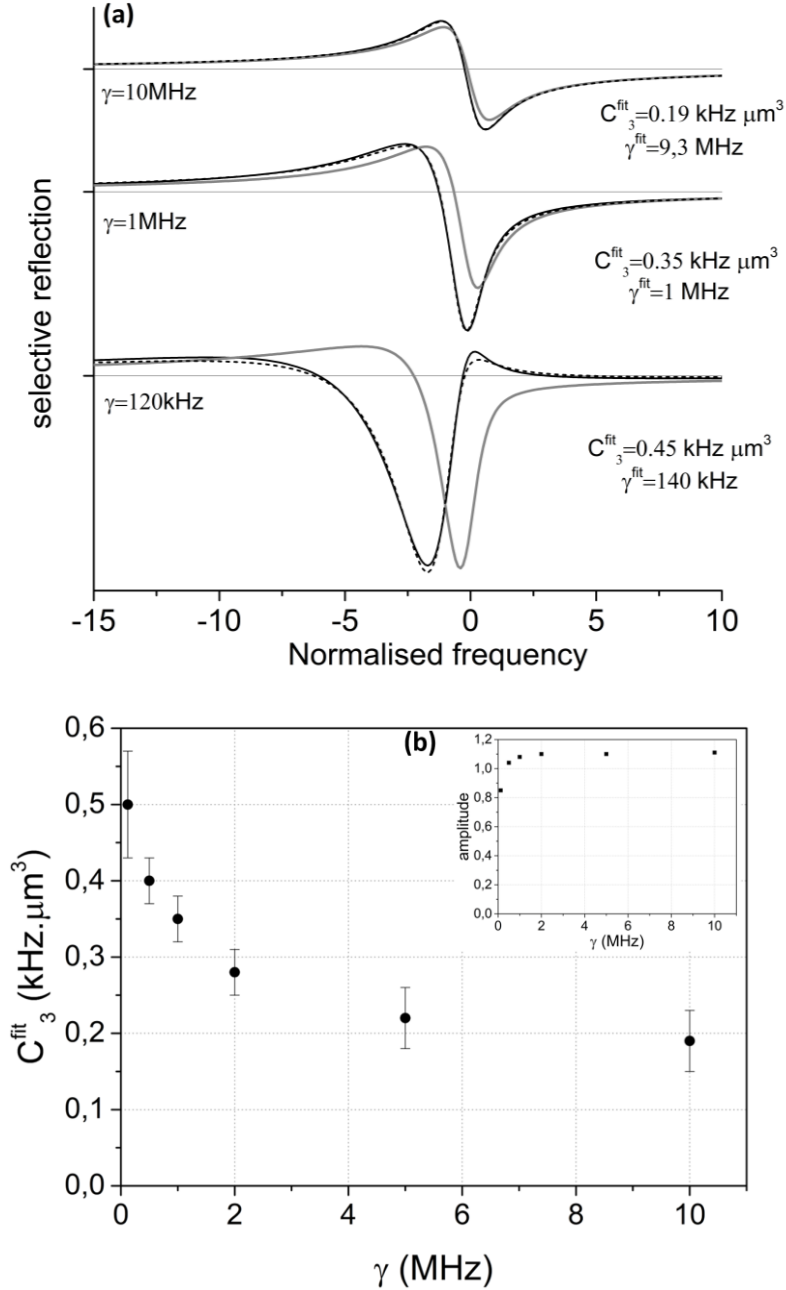

**Fig. 6.2.** (from ref. [3]) (a)The black lines represent the simulated selective reflection spectra on the $6S_{1/2} \rightarrow 5D_{5/2}$ transition, using a fully retarded Casimir-Polder potential with a transition linewidth of $\gamma$=120kHz (natural linewidth) as well as $\gamma$=1MHz and $\gamma$=10MHz. The spectra are given as a function of the detuning parameter (normalized frequency). The grey lines represent the expected SR lineshapes assuming a pure van der Waals atom-surface potential (i.e using the theoretical prediction of $C_3$=0.15kHz $\mu m^3$). The dashed curves are the best fits of the exact SR lineshapes using an ad hoc van der Waals coefficient $C_3^{fit}$. (b) The ad hoc van der Waals coefficient as a function of the transition linewidth. The inset shows the amplitude ratio between the fully retarded SR spectra and the corresponding fits.

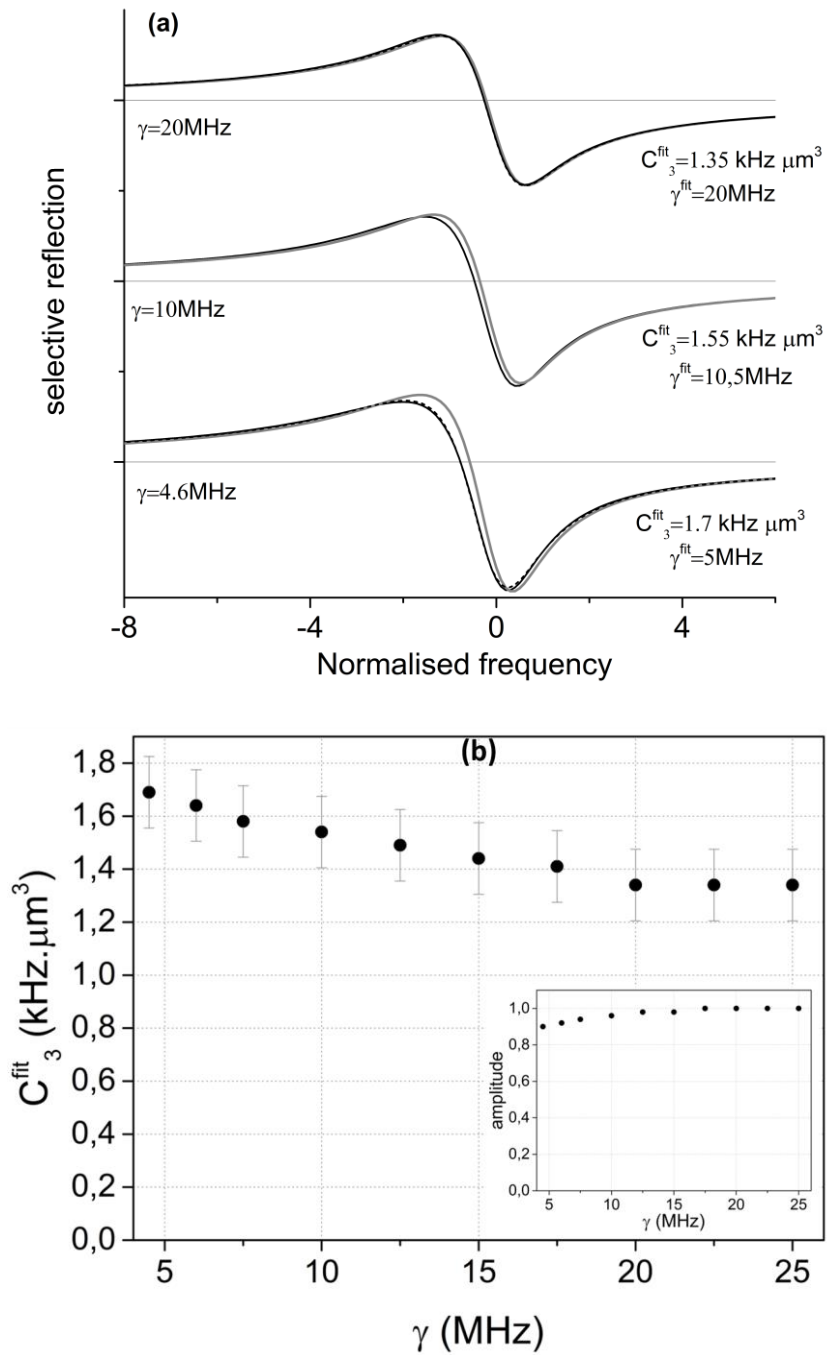

**Fig. 6.3.** (from ref. [3]) The same as before (Fig. 6.2) but for the $6S_{1/2}\rightarrow6P_{3/2}$ transition.

## Selective reflection on a planar metamaterial

When I arrived in Singapore in October 2014, the group had just finished a set of selective reflection experiments on the interface between an atomic caesium vapour and a metallic planar metamaterial. The metamaterials consisted of a thin metallic layer of ~50 nm thickness onto which rectangular slits of 70nm width and length varying from 170-240nm were etched with a period varying from 380-520nm (square unit cell) respectively. The coupling of plasmonic to propagating waves makes the metallic silver film (normally opaque) semi-transparent in certain wavelengths. The planar metamaterials fabricated in NTU had plasmonic resonances (and transmission windows) ranging from red (~600nm) to near infrared wavelengths (~950nm). The basic idea behind the experiment was to couple these plasmonic resonances to the $6S_{1/2}\rightarrow6P_{3/2}$ transition of caesium at 852 nm.

The Singapore group eventually published their experiments in [11], but at the time of my arrival in Singapore we also started thinking how we can use the data in order to make a quantitative measurement of the Casimir-Polder interaction between Cs(6P$_{3/2}$) and the planar metamaterials. Planar metamaterials are in essence three-dimensional structures suggesting that the van der Waals approximation is no longer valid. Treating the complete problem is almost impossible to tackle as it requires a complete calculation of the Casimir-Polder interaction for a structured surface [12] and a calculation of the selective reflection spectra including propagation optics. In our case, we decided to follow a mean-field approach, which can be justified by the fact that atoms fly over in parallel to many unit cells of the metamaterial within their lifetime. This suggests that the metamaterial effects are averaged and it can be treated as a planar layer with an effective dielectric constant. The effective dielectric constant of the metamaterials was extracted by FDFD simulations and broadband (far-field) reflection and transmission measurements of the metamaterials. The traditional fitting methods used in our group in Paris13 had to be further developed to include the effects of propagation optics in the effective metamaterial layer. The fits also included an imaginary part of the van der Waals coefficient.

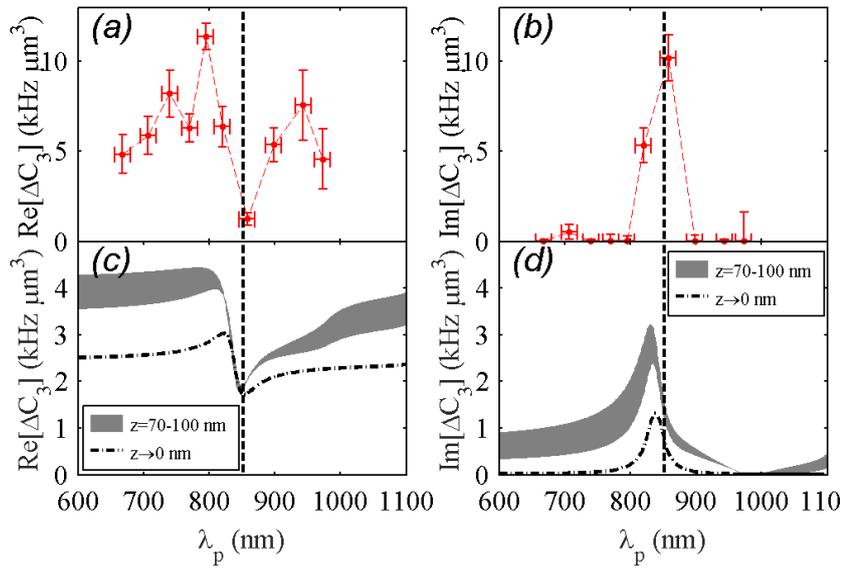

**Fig.6.4** (from ref. [13] )The van derWaals coefficient as a function of the plasmon resonance. Real part (a) and imaginary part (b) extracted from the fits of the SR signals. (c) and (d) are the real and imaginary parts of the theoretically predicted C$_3$ coefficients. The dot-dashed curve corresponds to the non-retarded case (z→0). The retarded contribution is taken into account by considering an effective distance ranging from 70 to 100 nm. It corresponds to the shaded gray surface. The vertical dashed lines indicate the position of the atomic resonance.

The experimental results for the real (a) and imaginary (b) parts of the spectroscopic C$_3$ coefficient (difference between excited and ground state) are shown in Fig. 6.4. The real part of the van der Waals coefficient $Re[C_3] = \delta F(z) \, z^3$ displays a resonant dispersive behaviour as the plasmonic resonance of the metamaterial is scanned around the atomic resonance (D2 line at 852 nm). When the plasmonic resonance is red detuned with respect to the atomic resonance, an increase of the C$_3$ coefficient (10kHz µm$^3$) is observed with respect to its non-resonant value (5kHz µm$^3$), whereas when the plasmonic resonance is blue detuned the spectroscopic C$_3$ coefficient goes down almost to zero (1kHz µm$^3$). The imaginary (dissipative) part $Im[C_3] = \frac{\delta\gamma(z)\, z^3}{2}$ represents a decrease in the atomic lifetime associated with the increased density of modes around the plasmonic resonance.

Theoretical predictions of the van der Waals coefficient are shown in Fig 6.4 (c) (real part) and (d) (imaginary part). The dashed line represent the non-retarded prediction assuming a pure van der Waals approximation. The theoretical and experimental curves follow similar trends but with strikingly different amplitudes. In fact, the measured effects are surprisingly much stronger than the predicted ones. The grey shaded areas show the effective C$_3$ coefficient at a depth between 70-100 nm, corresponding approximately to the probing depth of our experiment, including a fully retarded calculation of the Casimir-Polder potential (and considering a semi-infinite metamaterial). It appears that retardation effects are significant since the plasmonic and the probing wavelengths are almost the same. The fully retarded predictions seem to reproduce well the off-resonant values of the $C_3$ coefficient. The predicted amplitude of the resonance is also larger but is still far from its experimentally measured value. More details on the experiments and the calculations are given in [13].

## Personal contribution and publications resulting from this work

The possible influence of Casimir-Polder retardation effects was recognised in [A. Laliotis and M. Ducloy Phys. Rev. A (2015)]. The observation was a fortunate by-product, as the paper treated predominantly thermal effects of the Casimir-Polder interaction. Following this, I started working on the calculation of selective reflection integrals using the fully retarded Casimir-Polder potential [J. C de Aquino Carvalho et al. Phys. Rev. A (2018)]. Soon I was forced to ask the help of a trained theoretician (Paolo Pedri) which made the task much easier and less lonely. Martial Ducloy was also involved in these efforts providing invaluable intuition and help with the details of the original selective reflection calculation [10]. After the foundations of the calculation were laid, Joao Carlos de Aquinho Carvalho got also involved in the project. Joao essentially performed all the fits but also participated in the discussions and interpretations. I should acknowledge the help of Marie-Pascale Gorza with the calculations of the retarded Casimir-Polder potential.

The Singapore experiment was launched by M. Ducloy and D. Wilkowski in collaboration with N. Zheludev who is a metamaterials expert and director of the CDPT in NTU. I had no personal contribution in the experiment and data collection and I am not an author of the first paper [S. A. Aljunid et al. NanoLett. (2016)]. The second paper [E. A. Chan et al. Sci. Adv. (2018)] essentially treats the same data (same experiment) from a Casimir-Polder perspective. In this publication, my contribution was significant. I participated in the interpretations and discussions of the measurements and I did the theoretical predictions in particular including the effects of retardation. The final published version of the paper was mainly written by David Wilkowski, M. Ducloy and me.

# Chapter 7: Beyond atoms: selective reflection on a molecular gas

Most spectroscopic experimental studies of vapours close to surfaces, or under confinement, have been limited to alkali atoms. This stems from the availability of laser sources at these wavelengths and from the fact that alkali transitions are generally very strong, providing clean spectroscopic signals with particularly good signal to noise ratio. Molecules, in gas form are extremely more difficult to investigate under confinement, because most transitions (rovibrational but also electronic) are generally extremely weak (in terms of absorption strength). In fact, molecular rovibrational spectroscopy usually requires long interaction lengths achieved either by fabrication of extremely long cells (a molecular gas cell of several meters exists here in the LPL) or by more elaborate multi-passage techniques. Yet, there is a lot to gain from working with molecules.

One major advantage of molecular rovibrational spectroscopy is that it can provide frequency references based on molecular transitions that span from near to mid or far-infrared wavelengths. This includes the telecommunication window essentially covered by a series of rovibrational transitions of $C_2H_2$ and HCN shown in Fig.7.1 (from 'Wavelength Standards for Optical Communications' [1]). In most cases, however, working with somewhat compact systems requires the use of high gas pressures, which in turn induces collisional broadening that exceeds even the Doppler width and eventually limits the available frequency resolution. The transmission measurements performed in Fig.7.1 for example are performed in cells whose size is ~10cm and the gas pressure around ~100Torr (collisional broadenings ~15-20MHz/Torr while natural linewidth is negligible). Hollow core fibres were first filled with molecular gazes (acetylene in particular) in 2005 [2]. The advantage of fibre-based systems is that they are compact allowing at the same time the possibility of long interaction lengths. In [1] for example 1m long fiber-cells were fabricated using pressures as low as 6Torr. At this pressure, the Doppler width (~450MHz at FWHM) is the dominant limitation of the available resolution. An alternative method for fabricating compact systems of molecular spectroscopy is the confinement of molecular gases in random porous media [3]. Light scattering inside the porous medium increases the optical path and the absorption from the molecular gas. One limiting factor in terms of resolution in these systems can be transient time broadening due to collisions with the interstitial walls. Although the above systems are compact, they remain essentially macroscopic.

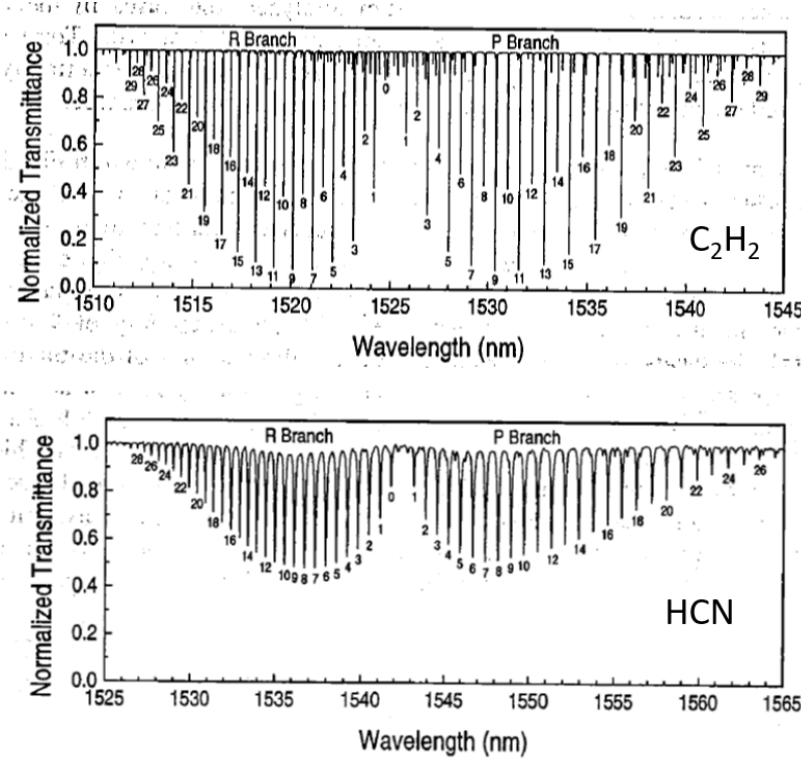

**Fig. 7.1** (from 'Wavelength Standards for Optical Communications' [1]) Rovibrational transmission spectrum of acetylene ($C_2H_2$) in a 5cm long cell at a 50Torr pressure and hydrogen cyanide (HCN) in a 22.5cm long cell at 100Torr pressure.

An alternative path is the use of selective reflection spectroscopy, a technique that provides sub-Doppler linear signals. Additionally, selective reflection probes a thin layer of gas (close to the cell window) whose thickness, $\lambda/2\pi$, is micrometric or nanometric depending on the optical wavelength $\lambda$. Performing selective reflection on a molecular gas opens therefore the path towards a complete miniaturisation of high-resolution molecular spectroscopy. Thin cell spectroscopy, also high-resolution due to the Dicke narrowing effect for a cell thickness below $\lambda/2$ [4, 5] is a step further towards that direction as it represents a compact system that can be integrated and miniaturised using microfabrication techniques. Finally, one can also envisage confining molecular gazes in opals [6], like the ones discussed in Chapter 2 of this document. Towards this end, molecules offer the advantage of working at room temperatures and controlling the gas pressure (density) with a vacuum valve rather than heating up a caesium container. Furthermore, most molecules available in gas phase are generally less chemically aggressive than alkali vapours.

An additional advantage of selective reflection or thin cell spectroscopy is that it allows us to envisage the first spectroscopic measurements of the Casimir-Polder interaction between a surface and a molecule in a specific quantum state. Molecule-surface interactions were first measured by deflection of molecular beams by A. Shih, D. Ruskin and P. Kusch but the accuracy of these experiments performed in the 70's is rather limited [7-9]. More recently, the molecule surface interaction has been recognised to play an important role in macromolecule interferometry, performed primarily in the group of M. Arndt in Vienna, and measurements of the van der Waals coefficient between molecules and dielectric surfaces was performed by analysis of the interferometric fringes of macromolecule diffraction in nanogratings [9, 10]. Measurements of the molecule-surface force have also been performed by attaching molecules on the tip of an AFM (atomic force microscope) cantilever [12]. Nevertheless, these measurements are limited to certain big molecules and they are not quantum state selective.

The Casimir-Polder molecule-surface interaction also attracts a lot of attention from a theoretical point of view, with an increasing number of theoretical studies published on the subject. It's worth mentioning studies that propose a chiral component in the Casimir-Polder interaction when both the surface and the molecule have chiral properties [13] and studies that investigate the effects of anisotropy (molecular orientation with respect to the surface) on the molecule-surface interaction and how this can lead to preferential orientation of molecules next to surfaces [14]. It was also suggested that the anisotropy of molecule-surface interactions could be used to characterise the planarity of surfaces [15]. Finally, it should also be mentioned that the distance dependent lifetime of molecular rovibrational states was calculated in order to estimate the trapping lifetimes of molecules next to the surface of a molecule-chip [16] (the molecular analogue of an atom-chip).

## Selective reflection experiments on a molecular gas

The first selective reflection measurements on a molecular gas started in 2015 in collaboration with Sean Tokunaga and Benoit Darquié, both LPL colleagues from the group MMTF (Metrologie Molecules et Tests Fondamentaux). The group MMTF has a long experience in working with molecules, in particular rovibrational spectroscopy of molecules with $CO_2$ lasers around the 10.6μm window. The molecules of choice were $SF_6$ and $NH_3$, extensively studied and well-known in the group MMTF. Our choice was primarily dictated by the fact that $SF_6$ and $NH_3$ have very strong rovibrational transitions at relatively large wavelengths (large probing depth and small Doppler broadening) making them ideal candidates for observing the first selective reflection signal with a molecular gas.

The first measurements were performed using the $CO_2$ laser set-up of the group MMTF. In these experiments we used two lasers. The first was locked on a narrow $OsO_4$ transition (linewidth < 20kHz) while the second was scanned, by tuning the cavity length, at a range of about 50-60MHz, essentially limited by the gain of the $CO_2$ medium. The frequency modulation (FM) of the second laser was achieved by modulating the cavity length. This imposed limitation on the maximum available excursion (~500kHz) and modulation frequency (5kHz). A frequency beat between the two lasers provided the absolute frequency of the scanned laser, whose linewidth was well-below the kHz range (~10Hz). The accuracy and precision of frequency scale reconstruction of our scans was unprecedented for selective reflection experiments. However, the laser amplitude noise was significant and was added to a huge parasitic signal due to the change of laser power during the scan attributed to the narrow width of the laser gain medium. Additionally, the system was very cumbersome and difficult to use. Nevertheless, the first selective reflection measurements on $SF_6$ inspired sufficient optimism, precipitating the decision to build a dedicated set-up for selective reflection measurements on a molecular gas.

Our laser of choice was an Alpes Lasers quantum cascade laser (QCL) emitting in the 945-950cm⁻¹ (150GHz) window with an output power of about 15mW, depending on the operating temperature (from -30 C to 30 C). The frequency scan and the frequency modulation of the QCL source are achieved simply by changing the laser current (much like a DBR laser). The FM excursion is typically from 0.5-1MHz at a frequency of about 5-10kHz. The frequency window of the QCL was specifically chosen to probe the saP(1) transition of $NH_3$ (J=1→J'=0) and a multitude of $SF_6$ rovibrational transitions (see Fig. 7.2). Additionally, the QCL can probe rovibrational transitions of ethylene ($C_2H_4$), methanol ($CH_3OH$) and other heavier molecules such as trioxane ($C_3H_6O_3$).

The gas cells used in our experiments are made out of standard stainless steel vacuum components (see Fig. 7.3). The cell windows are made out of ZnSe, a dielectric transparent from the visible to mid-infrared whose refractive index is about 2.4 depending on the wavelength. The selective reflection window has a wedge and is AR coated at the outer surface to minimize as much as possible unwanted interferences. The cells are attached to two vacuum valves and a pressure gauge (see Fig. 7.3). The first valve leads to the pumping station whereas the second leads to a gas bottle containing either $NH_3$ or $SF_6$. The above system allows us to pump and fill the cell with the desired pressure of molecular gas.

The experimental set-up is similar to the ones used for atomic selective reflection experiments. The reflection from the ZnSe window (selective reflection cell) is measured with a HgCdTe photodiode and demodulated at the FM frequency with a lock-in amplifier. An auxiliary saturated absorption set-up is used in order to provide an absolute frequency reference in a saturated absorption cell. The QCL source exhibits an important frequency drift that hinders the observation of narrow signals. For this purpose we frequency stabilize the laser in a Doppler absorption profile (linear absorption cell) and we scan the laser by adding a DC offset in the error signal.

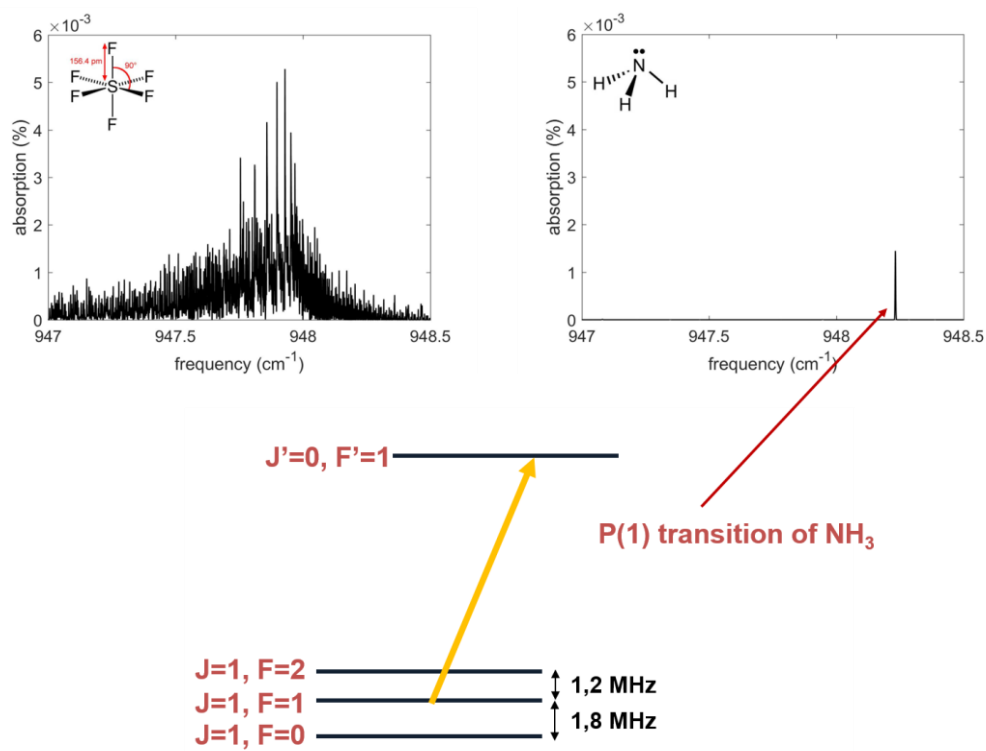

**Fig. 7.2** Linear (Doppler) absorption (%) profile for $SF_6$ and $NH_3$ molecules at a pressure of 1μTorr and a length of L=10cm, reproduced using HITRAN data. For comparison, the D1 absorption at these conditions is about 50%. The hyperfine structure of the saP(1) line of ammonia is also shown.

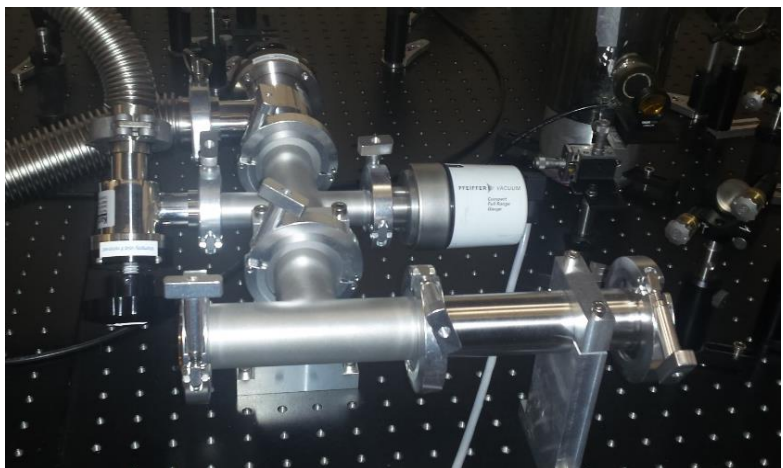

**Fig. 7.3** Photograph of the selective reflection cell used in our experiments. We can see a pressure gauge and two vacuum valves that lead to the pumping station and the $NH_3$ or $SF_6$ gas bottle.

The absorption spectra of $SF_6$ and $NH_3$ for a pressure of P=1μTorr and a length of L=10cm are shown in Fig.7.2. $SF_6$ presents a huge number of transitions that are unresolved with a Doppler width limited resolution ($ku_p$=17 MHz, FWHM width =85 MHz) but normally resolved or partially resolved with sub-Doppler resolution spectroscopy. This makes the $SF_6$ spectrum difficult to interpret. $NH_3$ on the other hand presents a very strong isolated line with a Doppler width ($ku_p$=51MHz, FWHM width =29 MHz) sufficiently large that allows us to comfortably lock and scan our laser. For these reasons our experiments with the QCL laser have so far focused on the P(1) line of $NH_3$. The hyperfine structure of the P(1) line, due to electric quadrupole of the nitrogen nucleus, was measured in [17] and is also shown in Fig. 7.2.

The resolution of our set-up is in fact limited by the linewidth of the QCL source, depending also on the FM excursion. Therefore, in our case the saturated absorption component (including the crossover transitions) are only partially resolved (see Fig. 7.4). The separation of the main peaks is however sufficient to allow us to calibrate the frequency scale of the laser scans. The frequency calibration is also independently corroborated using the linear absorption profile of ammonia. In Fig. 6.4 we show selective reflection spectra on the ammonia saP(1) transition for pressures ranging from 1Torr down to 15mTorr. Each spectrum is usually the average of 4-5 individual scans each lasting approximately 5-10mins. The derivative of the relative variation of reflectivity (FM demodulated signal divided by its DC value) around the $NH_3$ transition does not exceed 2ppm ($2x10^{-6}$) for a frequency excursion of about ~0.5MHz. For pressures below ~ 50mTorr, the hyperfine structure of ammonia is clearly resolved. We also show theoretical curves that satisfactorily reproduce the experimental scans. The curves are produced by using standard selective reflection theory [18] disregarding van der Waals interactions, including the distortions induced by the FM modulation and the effects of the laser linewidth, which is considered to have a Gaussian distribution with a FWHM of about 0.6MHz. The amplitude of the hyperfine components is fixed to its theoretical value (1, 5, 3 for F=0→F'=1, F=1→F'=1, F=2→F'=1 respectively). The theory curves are adjusted for an overall amplitude and an offset in order to fit the experimental data. From Fig.6.4 we can infer that the collisional broadening of the ammonia line is about 25MHz/Torr consistent with other measurements reported in literature for rovibrational transitions of ammonia (but not necessarily the same one). It is evident that the resolution of our scans is limited essentially by laser linewidth, which in turns limits our sensitivity of a $C_3$ (van der Waals coefficient) measurement. Furthermore, the $C_3$ sensibility is compromised by a fluctuating baseline (fluctuations that can be an important fraction of the signal ~$10^{-6}$). It is probably due to baseline fluctuations that theory experiment do not match very well on the wings of the spectra.

We should mention that reflection spectroscopy was performed on alkali dimers [19] but signals were broad (on the GHz scale) and inconclusive. Thin cell spectroscopy of $CO_2$ was also performed at high pressures with a broadband source and Fourier Transform Spectroscopy to study the effect of molecule-surface collisions on the transition linewidth [20]. Here we perform high-resolution (sub-MHz) selective reflection spectroscopy on a micrometric layer of molecular gas.

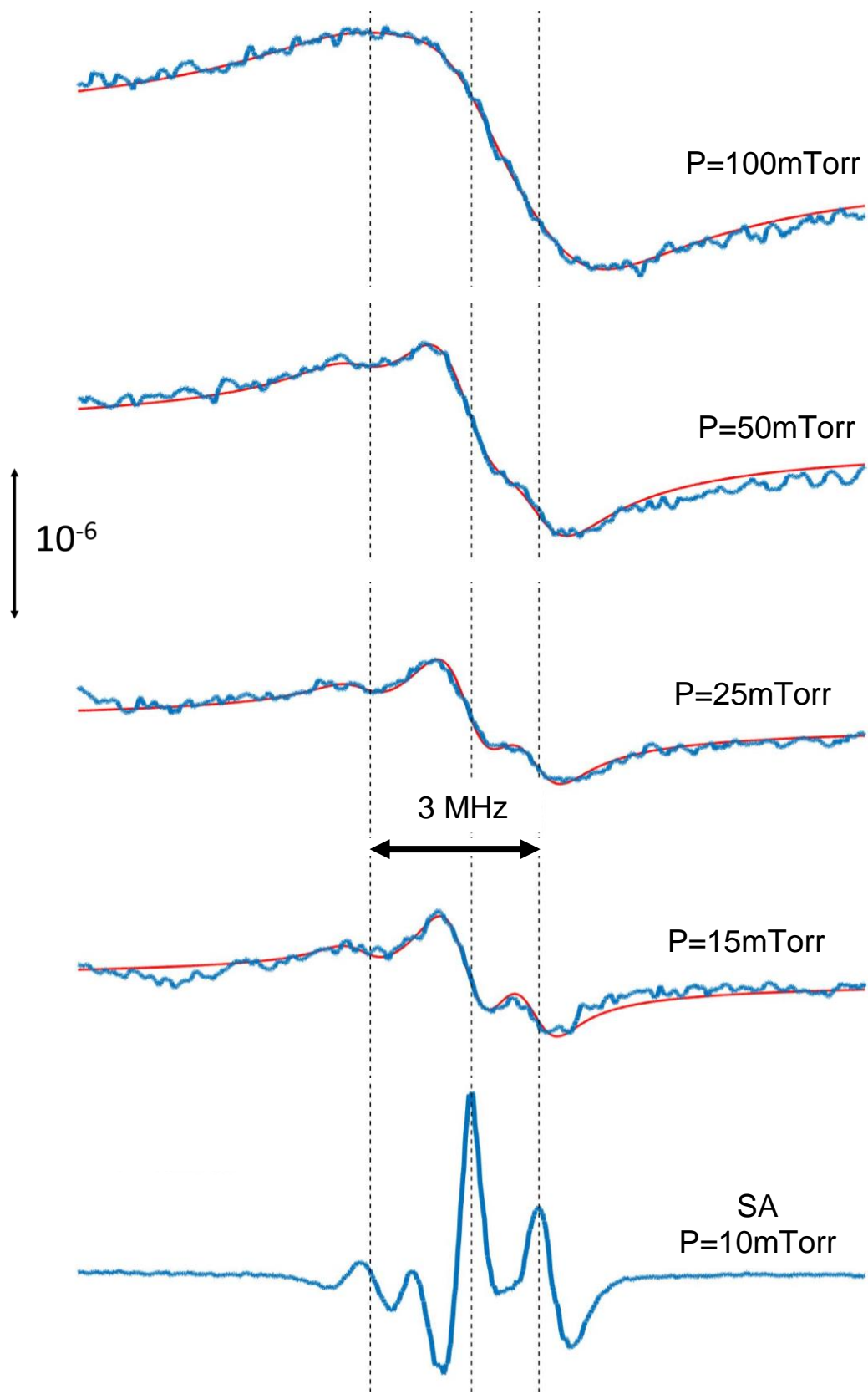

**Fig. 7.4** Selective reflection spectra of the NH₃ saP(1) transition (blue lines) at different pressures (FM demodulated signal divided by the DC signal) . The FM excursion is 0.5MHz. The last spectrum is a typical saturated absorption spectrum (second harmonic) used to calibrate the frequency scale of the scans. The red lines represent the theoretical predictions accounting for the FM distortion, finite Gaussian laser linewidth (FWHM 0.6MHz) and a pressure broadened linewidth of 25MH/Torr.

# Final thoughts about the molecule experiment

The theoretical analysis of selective reflection signals suggests that so far our measurements are not sensitive to the Casimir-Polder interaction between molecules and surfaces. Precise theoretical predictions of the van der Waals coefficients of molecule-surface interactions are indeed very hard to make as they require knowledge of the transition probabilities (and frequencies) of all electronic and rovibrational transitions of the molecules. This has been achieved by means of quantum chemistry calculations for some specific molecules [20, 10]. The molecules that we investigate here are similar in terms of polarizability to $CO_2$ and it is reasonable to assume that the $C_3$ coefficients for the ground molecular state should be about 0.5-1kHz $\mu m^3$. However, as our measurements are sensitive to the difference between $C_3$ coefficients between the two probed quantum molecular states (here between ground and first vibrational level), the expected spectroscopic $C_3$ coefficient is smaller. Selective reflection measurements should therefore be sensitive to the anisotropy between the rovibrational states probed and additionally to a pure rovibrational component (vibrational transitions) of the van der Waals coefficient. In our case for example we start from state (v=0, J=1, M=0,±1) to (v=0, J=0, M=0) of $NH_3$, where the quantum number M signifies the projection of the angular momentum (and therefore molecular orientation) on the quantisation axis, here considered perpendicular to the surface [15]. Light polarisation only allows M±1 to M=0 transitions and therefore the spectroscopic van der Waals coefficient should be sensitive to the anisotropy of the Casimir-Polder interaction between different orientations of the $NH_3$ molecule (tetrahedron) with respect to the surface. Exact calculations would depend on the molecular geometry (essentially the anisotropy of the molecular polarizability) but an order of magnitude estimate would bring the spectroscopic $C_3$ coefficient down to the 0.01-0.1 kHz $\mu m^3$ range. Polar, highly anisotropic molecules are more favourable, whereas for $SF_6$, no anisotropy is expected. For comparison, we mention that the spectroscopic $C_3$ coefficient for the caesium D1 transition is ~1 kHz $\mu m^3$. Probing molecular electronic transitions (visible or UV wavelengths) is more favourable for a measurement of a spectroscopic $C_3$ coefficient, however, electronic transitions are less well known than rovibrational ones.

In addition to the previous considerations, one should account for the probing depth of selective reflection in the mid-infrared range. Here $\lambda$=10.6$\mu m$ and the probing depth should be approximately $\lambda/2\pi$~1.5$\mu m$ more than an order of magnitude larger with respect to atomic transitions that are in the near-infrared range. Assuming that the near-field Casimir-Polder interaction is proportional to $z^{-3}$ (at this distance, retardation will be non-negligible but for argument's sake we will here ignore it) suggests a reduction of the $C_3$ sensitivity by 3-4 orders of magnitude. It is instructive to reason in terms of the parameter $A = \frac{16\pi^3 C_3}{\lambda^3 \Gamma}$ that describes the effects of the Casimir-Polder interaction on the selective reflection spectra. Considering the signal to noise ratio of our measurements, it is sensible to assume that the minimum measurable value of A should not be much smaller than A=0.01 and that the maximum resolution (limited here by laser linewidth) is $\Gamma$~0.5MHz. Using these numbers one finds a minimum measureable $C_3$ value of about 12 kHz $\mu m^3$ well above the previous rough estimates.

One possible solution that could increase the sensitivity of our measurements to the $C_3$ coefficient of the molecule-surface interaction would be to increase the frequency resolution of our measurements to ~1kHz. Reducing the QCL linewidth down to these levels is possible by locking to a saturated absorption line [22] or even further by locking the QCL to a $CO_2$ laser (linewidths ~10Hz) [23] or even further by locking to an infrared ultra-stable laser referenced to frequency primary standards (linewidths ~0.1Hz) [24]. One should nevertheless keep in mind that an increase in resolution is accompanied by a proportional decrease in signal amplitude (since the FM excursion should also be reduced). This price is too high to pay in our experiments were our available signal is ~$10^{-6}$ (1ppm of the reflectivity).

One way around this predicament is the use of thin cell spectroscopy. In the case of thin cells, the characteristic probing depth is no longer defined by the wavelength but by the thickness, L, of the cell (one can assume a probing depth of ~L/2). Reducing cell thickness results to a linear loss of signal amplitude but is followed by $L^3$ increase of the $C_3$ sensitivity. For a cell thickness of ~300nm one should expect a 5-fold reduction of the available signal with respect to selective reflection measurements and a minimum measureable $C_3$ coefficient of ~0.01 kHz $\mu m^3$. These numbers are within experimental reach and could lead to a measurement of the Casimir-Polder interaction with molecules.

Thin cells have the additional advantage of being a truly compact 'quasi' miniaturised system (true miniaturisation is a still a step further requiring microfabrication processes [25]). The system we propose is shown in Fig.7.5(a). Two ZnSe windows transparent from visible to mid-infrared wavelengths separated by a ring-shaped spacer of the appropriate thickness (depending on the experiment from 100nm to 5μm) are put together by means of mechanical pressure. The spacer also acts as a joint and seals the cell. A hole is drilled in one of the windows and a glass tube leading to a KF flange allows pumping and filling of the cell with molecular gases. The system can eventually be sealed with a given gas pressure and form a compact system ideal for high resolution rovibrational spectroscopy (molecular frequency references) of essentially any molecule provided that the cell thickness is smaller than $\lambda/2$ (Dicke narrowing) and that its transitions are strong enough to measure in a single passage transmission spectroscopy. In this design the outside facets of the cells are AR coated whereas the inside faces are left uncoated. This is done in the hope of a 'clean' Casimir-Polder interaction measurement.

Molecular references are most valuable in the telecommunications window and it is very important to probe $C_2H_2$ and HCN transitions by high-resolution spectroscopy. Nevertheless, rovibrational transitions of both these molecules are much weaker than the ones of $NH_3$ or $SF_6$ at 10.6μm. For this purpose, we have designed a multicell with many compartments comprising an additional multipass system in order to increase the signal to noise ratio while at the same time maintaining the compactness of the system. The system is shown in Fig.7.5(b). Here all windows are made out of glass and they are AR coated. As previously, the system can be pumped and filled with molecules via the drilled hole on one of the external windows.

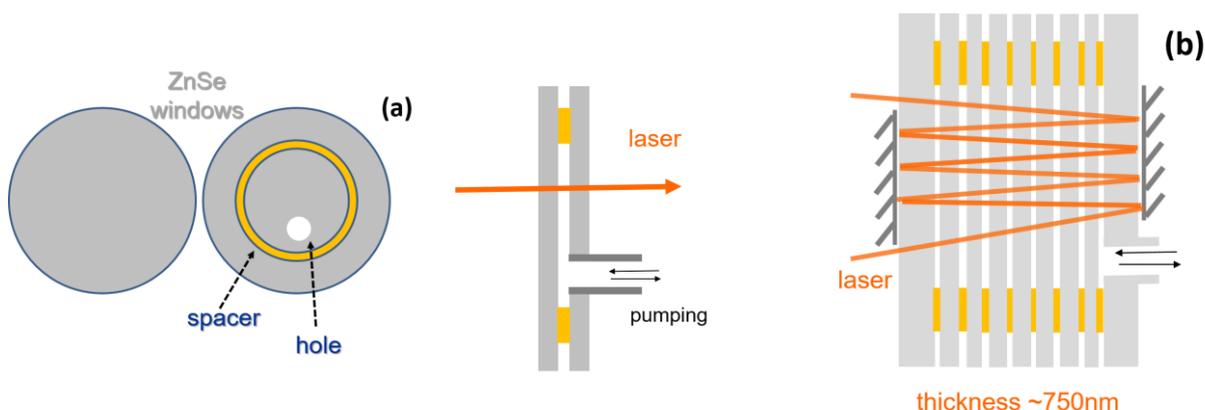

**Fig. 7.5** (a) Thin cell design with two ZnSe windows. A metallic (gold) spacer is deposited (by evaporation or sputter coating or simply placing a gold foil). The hole on one of the windows allows for pumping and filling with molecules. (b) A schematic diagram of a multi-compartment, multi-pass cell for acetylene molecules. The design should be able to increase the signal to noise ration by a few orders of magnitude.

## Acknowledgements and brief history of the experiment

For me, the idea of a performing selective reflection experiments on rovibrational molecular transitions started after discussions with Sean Tokunaga (a colleague and a very good friend) during small, after work meetings in the 'Hideout' pub. Soon we decided to confront our initial enthusiasm with the more pragmatic approach of Benoit Darquié (also a colleague and a very good friend). This is how we started asking difficult questions such as: Why is this experiment interesting? What can we measure? How much signal do we expect? and so on and so forth. Trying to answer these questions put the molecule project on the move. At the end of the summer of 2013, Benoit and I started discussing these ideas with Martial Ducloy at the Novosibirsk conference MPLP. Martial told us that some unsuccessful attempts were made in the past to measure selective reflection on iodine transitions in collaboration with P. Juncar. He also directed us towards Jose Rios Leite who came to Paris a few weeks later financed by a CAPES-COFECUB Franco-Brazilian collaboration project. Rios had also made some attempts in the past to measure evanescent wave or selective reflection spectroscopy on an $SF_6$ gas with a $CO_2$ laser. Our discussions with Martial and Rios blew a breath of enthusiasm and ideas into the project that lasts until now and for which I am truly indebted to them. Following these initial steps, selective reflection of a molecular gas started evolving into a collaboration between the group SAI and the MMTF group (Benoit Darquié and Sean Tokunaga). At first, we applied for small grants that allowed us to perform preliminary experiments in our spare time using the $CO_2$ laser set-up of the MMTF group. Eventually, we managed to buy a new QCL with money from the IISE (Institut de Sciences Experimentales) of Paris13 and the experiment moved officially under the aegis of the SAI group. Junior Lukusa Mudiayi started his PhD thesis on the subject under the wings of Daniel Bloch in September 2016. All the data presented here were taken by Junior, working closely with Isabelle Maurin, Benoit Darquié, Daniel Bloch and myself. I should also acknowledge the help of Olivier Gorceix, head of our laboratory. The project took off thanks to his support. Our initial measurements were presented in CLEO-IQEQ Munich 2017. The project seems now mature enough to write a journal article.